\documentclass[10pt,final,journal,a4paper,oneside,twocolumn]{IEEEtran}
\usepackage{amssymb}
\usepackage[tbtags,sumlimits,nointlimits,reqno]{amsmath}

\usepackage{bbm}
\usepackage{float}
\usepackage{srcltx}
\usepackage{color,psfrag}
\usepackage{epsfig}
\usepackage{color}
\usepackage{commath}
\usepackage{mathdots}
\usepackage[export]{adjustbox}
\usepackage{amsmath}
\usepackage{multicol}
\usepackage{color}
\usepackage{float}
\usepackage{mathtools, cuted}
\usepackage[makeroom]{cancel}
\usepackage{relsize}
\usepackage{bbm}
\usepackage[normalem]{ulem} 
\usepackage{cite}
\usepackage[table]{xcolor}
\usepackage{tabularx,booktabs}
\usepackage{gensymb}
\usepackage{amssymb}
\usepackage{flushend}

\makeatletter
\let\MYcaption\@makecaption
\makeatother

\usepackage[font=footnotesize]{subcaption}

\makeatletter
\let\@makecaption\MYcaption
\makeatother
\definecolor{amber}{rgb}{0.8, 0.33, 0.0}
\definecolor{ceruleanblue}{rgb}{0.16, 0.32, 0.75}

\begin{document}

\title{Millimeter-Wave Integrated Side-Fire Leaky-Wave Antenna and its Application as a Spectrum Analyzer}

\author{%
       Daniel J. King, \IEEEmembership{Student Member, IEEE}, Mohamed K. Emara, \IEEEmembership{Student Member, IEEE},\\ and Shulabh Gupta, \IEEEmembership{Senior Member, IEEE}\\      
\thanks{D. J. King, M. K. Emara and S.~Gupta are with the Department of Electronics, Carleton University, Ottawa, Ontario, Canada. (email: danielking3@cmail.carleton.ca)}
}

\maketitle

\begin{abstract}
An analog, low-profile and shielded spectrum analyzer is proposed for operation at mm-wave frequencies around the 60 GHz band based on a novel side-fire Leaky-Wave Antenna (LWA) configuration. The proposed side-fire periodic LWA is systematically developed from a conventional 3-port waveguide T-junction which is modified to a LWA unit cell with an internal matching mechanism to suppress the stop-band and enable broadside radiation based on unit cell symmetry considerations. The resulting periodic side-fire antenna radiates in the plane of the antenna, whereby the leakage power be either be allowed to radiate in free-space or kept confined inside a PPW structure. The proposed side-fire structure thus can be completely shielded useful as an analog broadband spectrum analyzer using Substrate Integrated Waveguide (SIW) Technology. Furthermore, a convex side-fire antenna is demonstrated to focus the radiated beams in the near-field of the structure to make the entire system compact. The  integrated spectrum analyzer is experimentally demonstrated between 59 GHz - 66 GHz providing 1 GHz frequency resolution. Furthermore, a simple mathematical model consisting of array of line sources is proposed to efficiently model the beam-scanning characteristics of the curved side-fire LWA in the near-field of the structure.
\end{abstract}


\section{Introduction}
Spectrum analysis is among the most fundamental signal processing operations required in engineering and forms the basis of various systems in application areas such as signal distortion measurement from audio equipment, spectrum monitoring for governmental frequency allocation of various radio services, and electromagnetic emissions testing of equipment \cite{farzaneh2009antenna,sanchez2017millimeter}. Traditional techniques include discrete Fourier transforms (DFTs) based on digital computing for Radio Frequency (RF) signals, where an analog test signal is first sampled, discretized and then digitally processed to compute its Fourier transform \cite{kawamura2014novel, bergland1969fast}. However, this approach suffers from performance degradation and challenging implementation as frequency increases and signal bandwidths become large~\cite{rappaport2017overview}. With the upcoming next generation wireless systems, novel technologies for instrumentation and communication are required particularly in the IEEE 801.11ad (60 GHz unlicensed band) mm-Wave band which are capable of handling large signal bandwidths with reduced latencies~\cite{ghosh20165g} \cite{karjalainen2014challenges}.

Consequently, alternative techniques have been explored in the literature to perform spectrum analysis using analog methods. At optics, the material dispersion is utilized to real-time separate the various spectral components of a broadband test signal, commonly utilized in Real Time Fourier Transformers (RTFTs) \cite{de2016optical}, Arrayed Waveguide Gratings (AWG) \cite{saleh2019fundamentals}, \textcolor{black}{prisms} or to use the wavelength dependent diffraction orders in periodic diffraction gratings \cite{goodman2005introduction} and Virtual Image Phased Arrays (VIPA) \cite{shirasaki1996large}, for instance. While such systems are not as flexible as a digital system, \textcolor{black}{they offer} high acquisition bandwidth and ultrafast system operation. Inspired from these optical systems, various solutions have been proposed at \textcolor{black}{RF} using both guided and radiative systems based on dispersion (frequency dependent group delay, $\tau(\omega)$) engineered Bragg gratings \cite{laso2003real}, spatial interferometers \cite{wang2018real} and leaky-wave Antennas (LWAs), for instance, exploiting their frequency scanning properties~\cite{gupta2009microwave, gupta2008crlh, gomez2010frequency, garcia20111d, martinez2011planar}. However, all of these solutions suffer from either lack of device integration or are based on unshielded microstrip implementations which are not particularly suitable for higher \textcolor{black}{mm-wave} bands (e.g. 60 GHz, for instance) leading to significant free-space radiation, not ideal for minimal electromagnetic interference.

\begin{figure*}[htbp]
   \begin{minipage}{2\columnwidth}
     \centering
	\psfrag{A}[c][c][0.6]{Port 1}
	\psfrag{C}[c][c][0.6]{Port 2}
	\psfrag{B}[c][c][0.6]{Port 3}
	\psfrag{D}[c][c][0.7]{\color{amber}\shortstack{\textsc{Simple} \\ \textsc{T-Junction}}}
	\psfrag{E}[c][c][0.7]{\color{amber}\shortstack{\textsc{Matched} \\ \textsc{T-Junction}}}
	\psfrag{F}[c][c][0.7]{\color{amber}\shortstack{\textsc{Port Switched} \\ \textsc{Simple T-Junction}}}
	\psfrag{G}[c][c][0.7]{\color{amber}\shortstack{\textsc{Port Switched} \\ \textsc{Matched T-Junction}}}
	\psfrag{H}[c][c][0.7]{\color{amber}\shortstack{\textsc{Cascaded Matched} \\ \textsc{T-Junction/Side-Fire Structure}}}
	\psfrag{J}[c][c][0.7]{$w_1$}
	\psfrag{k}[c][c][0.7]{$w_2$}
	\psfrag{L}[c][c][0.7]{$\{g,\ell\}$}
	\psfrag{M}[c][c][0.7]{$s$}
	\psfrag{N}[c][c][0.7]{$p$}
	\psfrag{O}[c][c][0.6]{$N$ Output Ports}
   	 \includegraphics[width=\columnwidth]{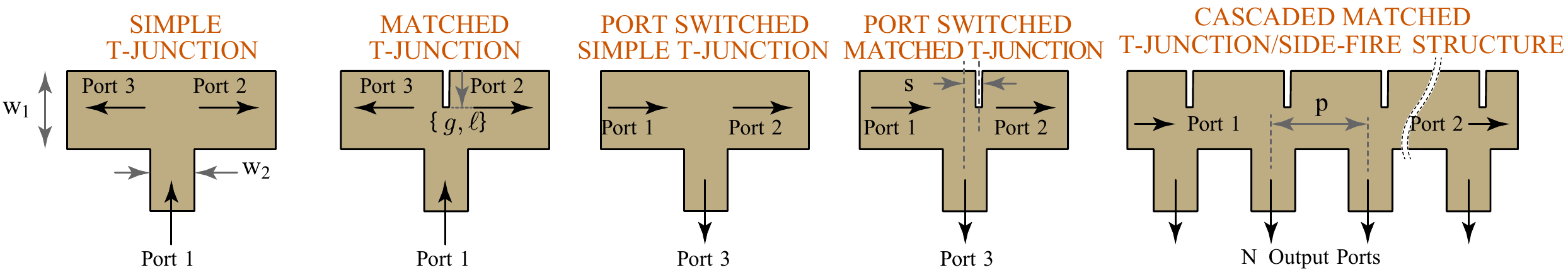}
     \subcaption{Evolution of a conventional T-junction to an $N$-port power divider.}
     \label{fig:sub1}
   \end{minipage}
      \begin{minipage}{0.5\columnwidth}
     	\centering
     	\psfrag{a}[c][c][0.7]{Frequency (GHz)}
	\psfrag{b}[c][c][0.7]{S Parameters (dB)}
	\psfrag{c}[c][c][0.5]{$S_{11}$}
	\psfrag{d}[c][c][0.5]{$S_{21}$}
	\psfrag{e}[c][c][0.5]{$S_{31}$}
    	\includegraphics[width=\columnwidth]{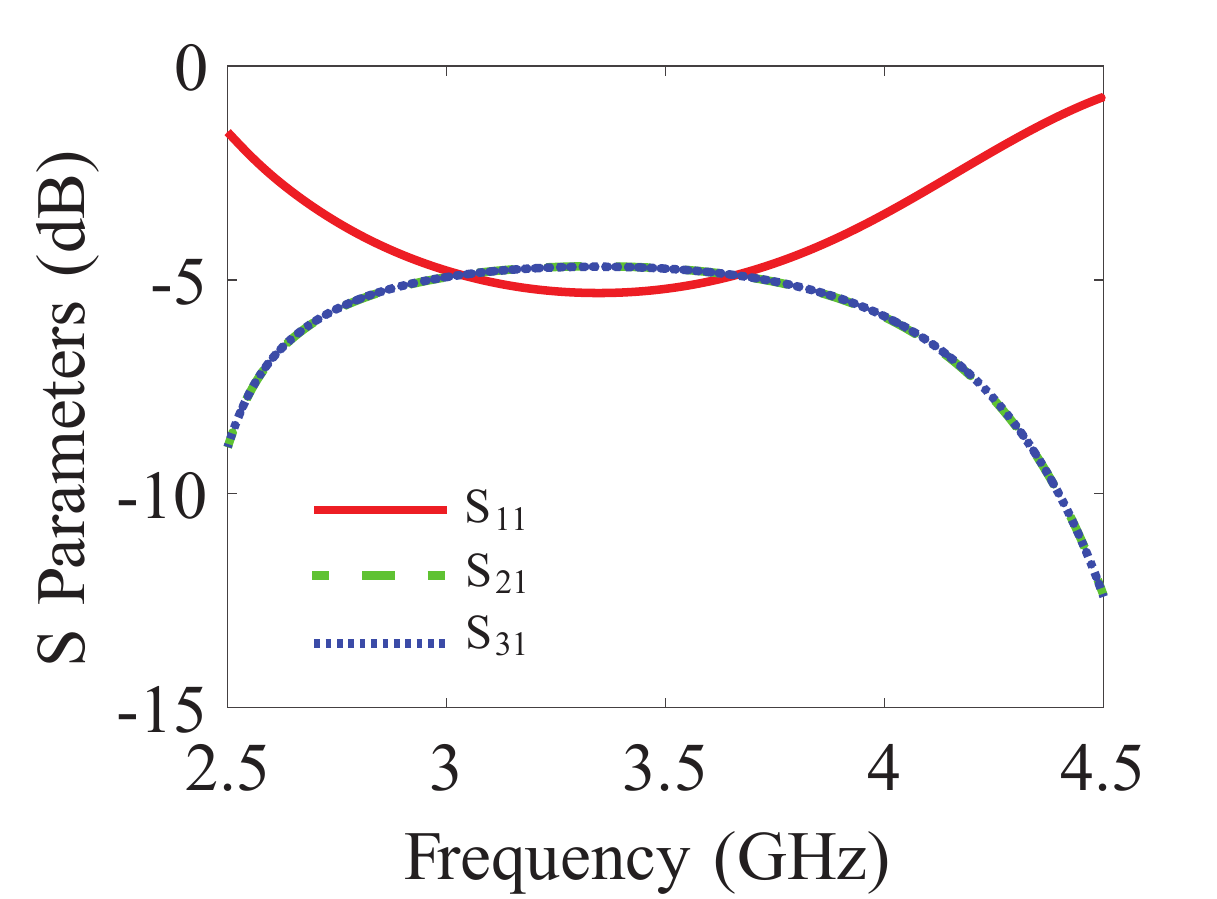}\\
     	\psfrag{a}[c][c][0.7]{$x$ (mm)}
	\psfrag{b}[c][c][0.7]{$y$ (mm)}
	\psfrag{c}[c][c][0.7]{Port 1}
	\psfrag{d}[c][c][0.7]{Port 2}
	\psfrag{e}[c][c][0.7]{Port 3}
     \includegraphics[width=\columnwidth]{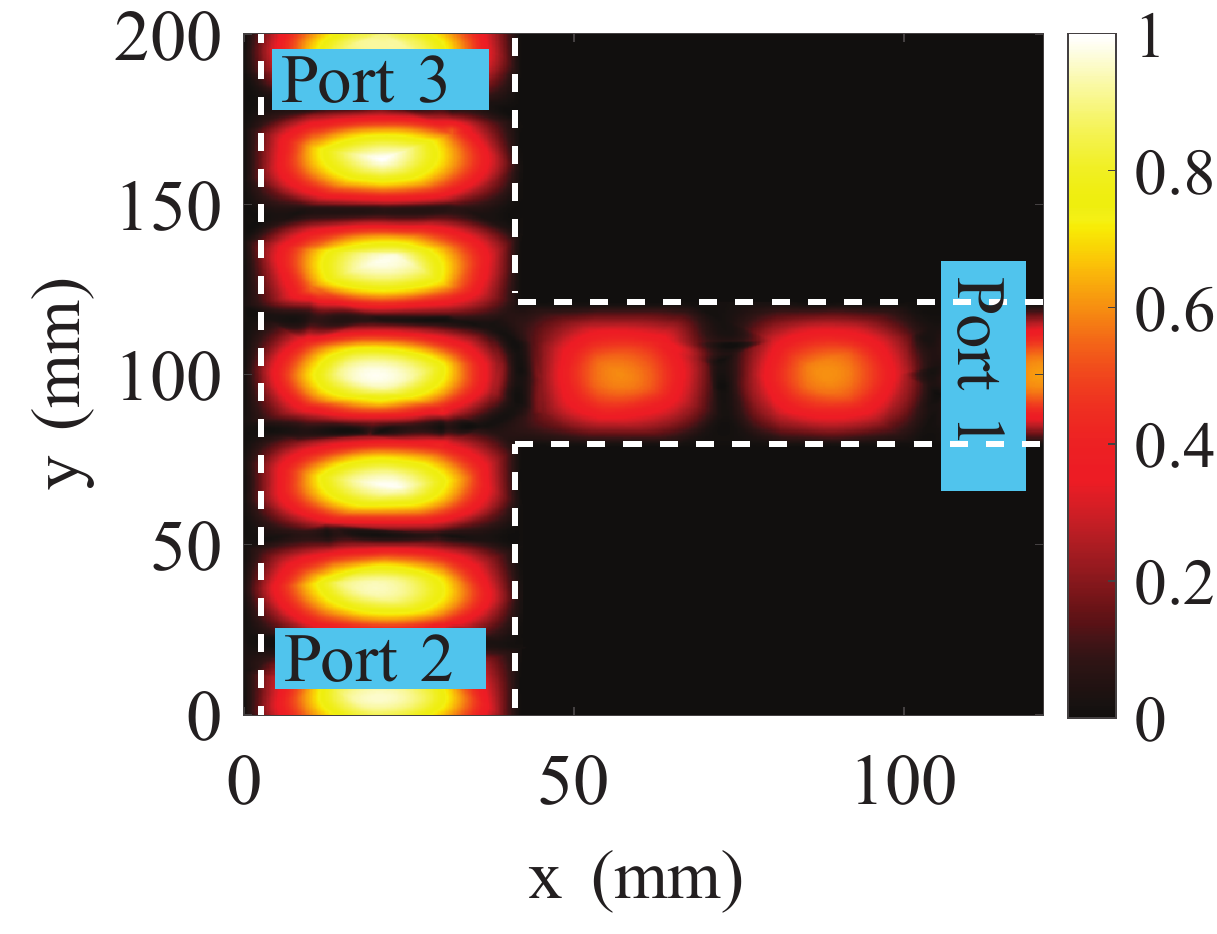}
     \subcaption{Simple T-junction.}
     \label{fig:sub1}
   \end{minipage}
      \begin{minipage}{0.5\columnwidth}
     \centering
     	\psfrag{a}[c][c][0.7]{Frequency (GHz)}
	\psfrag{b}[c][c][0.7]{S Parameters (dB)}
	\psfrag{c}[c][c][0.5]{$S_{11}$}
	\psfrag{d}[c][c][0.5]{$S_{21}$}
	\psfrag{e}[c][c][0.5]{$S_{31}$}
     \includegraphics[width=\columnwidth]{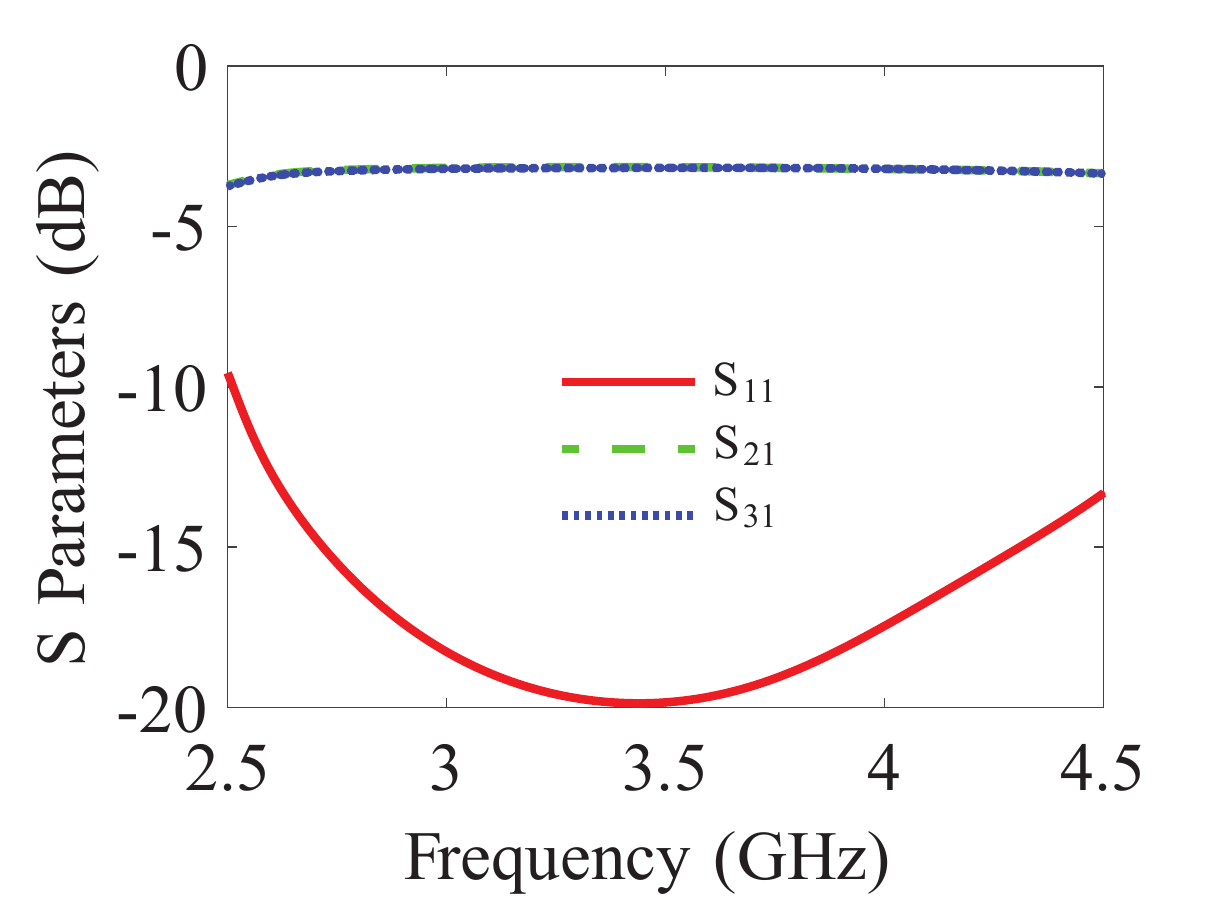}\\
     	\psfrag{a}[c][c][0.7]{$x$ (mm)}
	\psfrag{b}[c][c][0.7]{$y$ (mm)}
	\psfrag{c}[c][c][0.7]{Port 1}
	\psfrag{d}[c][c][0.7]{Port 2}
	\psfrag{e}[c][c][0.7]{Port 3}
     \includegraphics[width=\columnwidth]{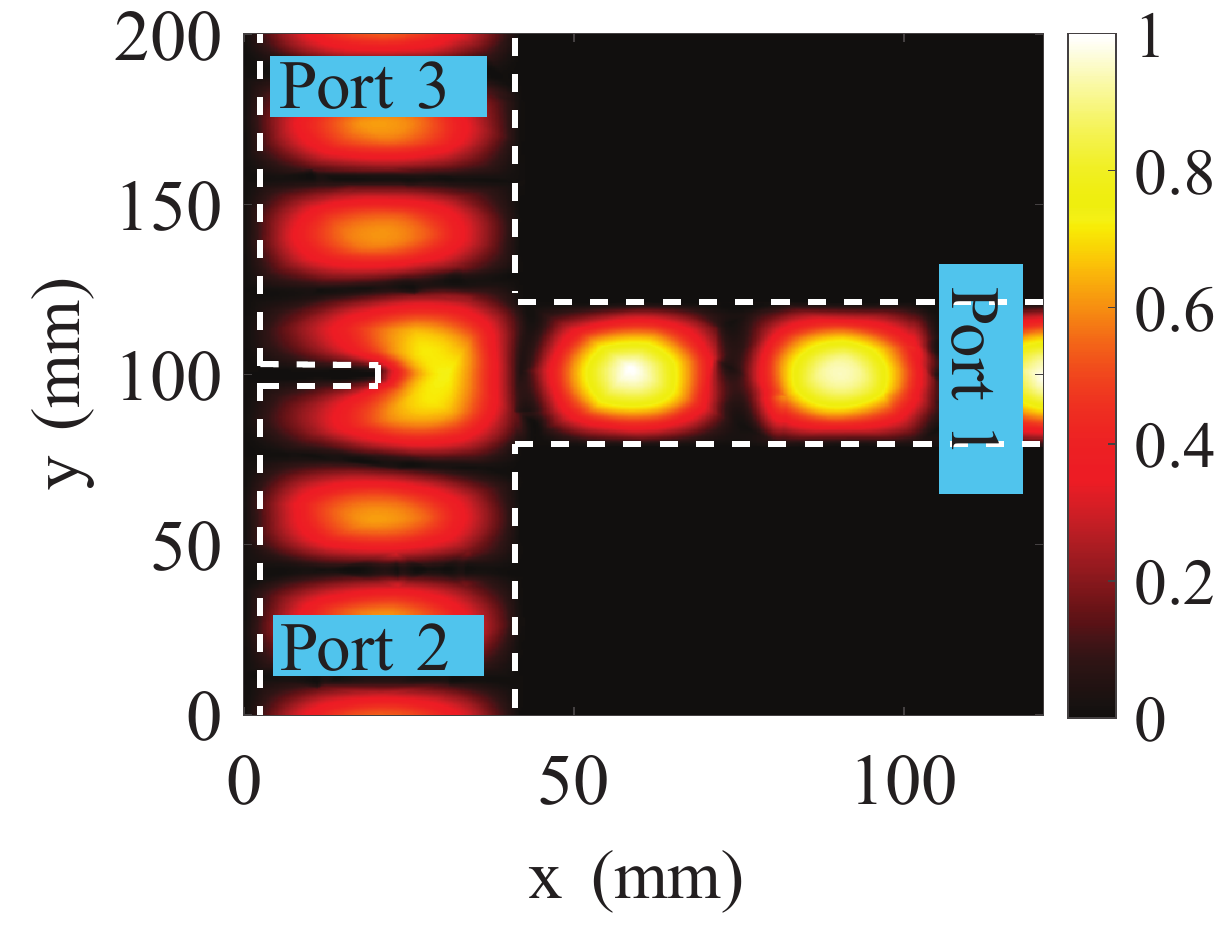}
     \subcaption{Matched T-junction.}
     \label{fig:sub1}
   \end{minipage}
      \begin{minipage}{0.5\columnwidth}
     \centering
     	\psfrag{a}[c][c][0.7]{Frequency (GHz)}
	\psfrag{b}[c][c][0.7]{S Parameters (dB)}
	\psfrag{c}[c][c][0.5]{$S_{11}$}
	\psfrag{d}[c][c][0.5]{$S_{21}$}
	\psfrag{e}[c][c][0.5]{$S_{31}$}
     \includegraphics[width=\columnwidth]{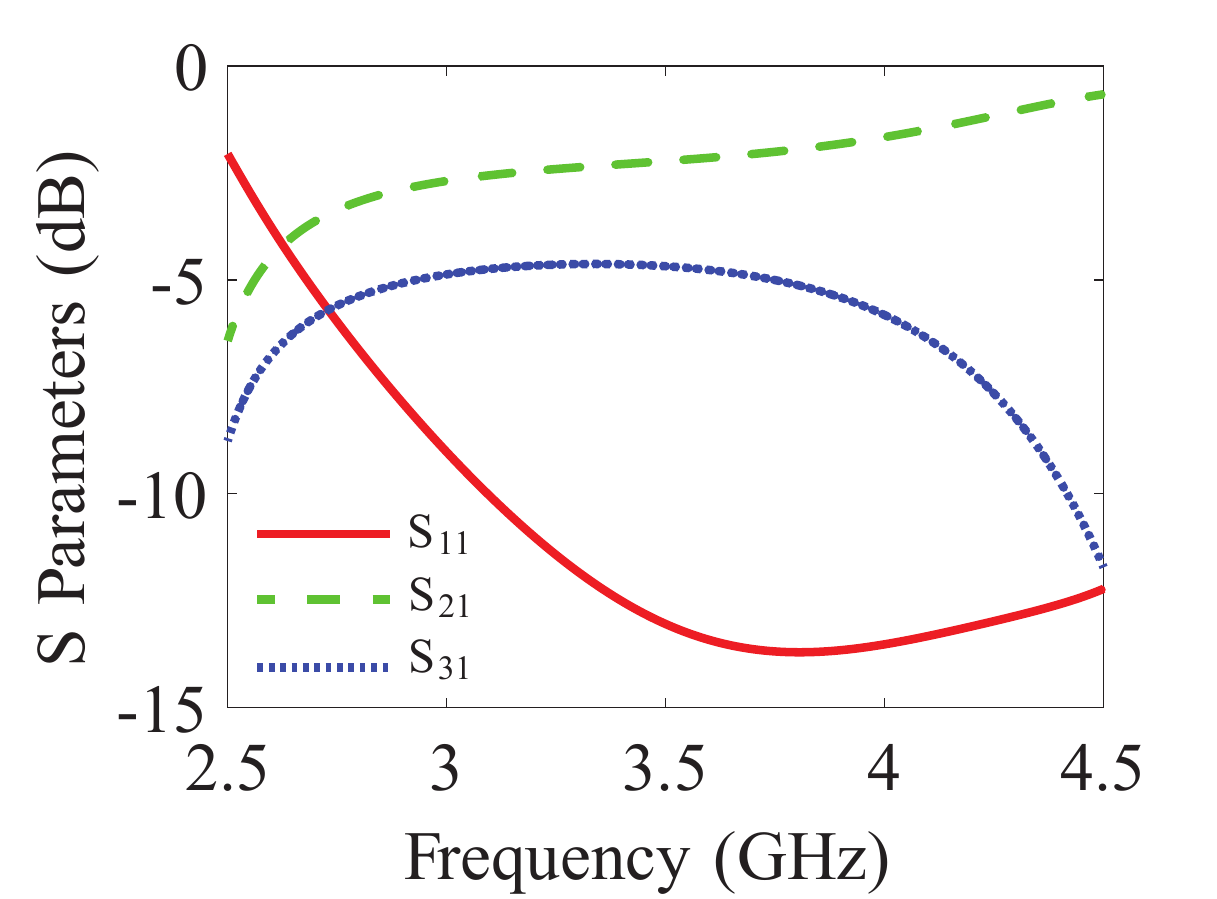}\\
     	\psfrag{a}[c][c][0.7]{$x$ (mm)}
	\psfrag{b}[c][c][0.7]{$y$ (mm)}
	\psfrag{c}[c][c][0.7]{Port 1}
	\psfrag{d}[c][c][0.7]{Port 2}
	\psfrag{e}[c][c][0.7]{Port 3}
     \includegraphics[width=\columnwidth]{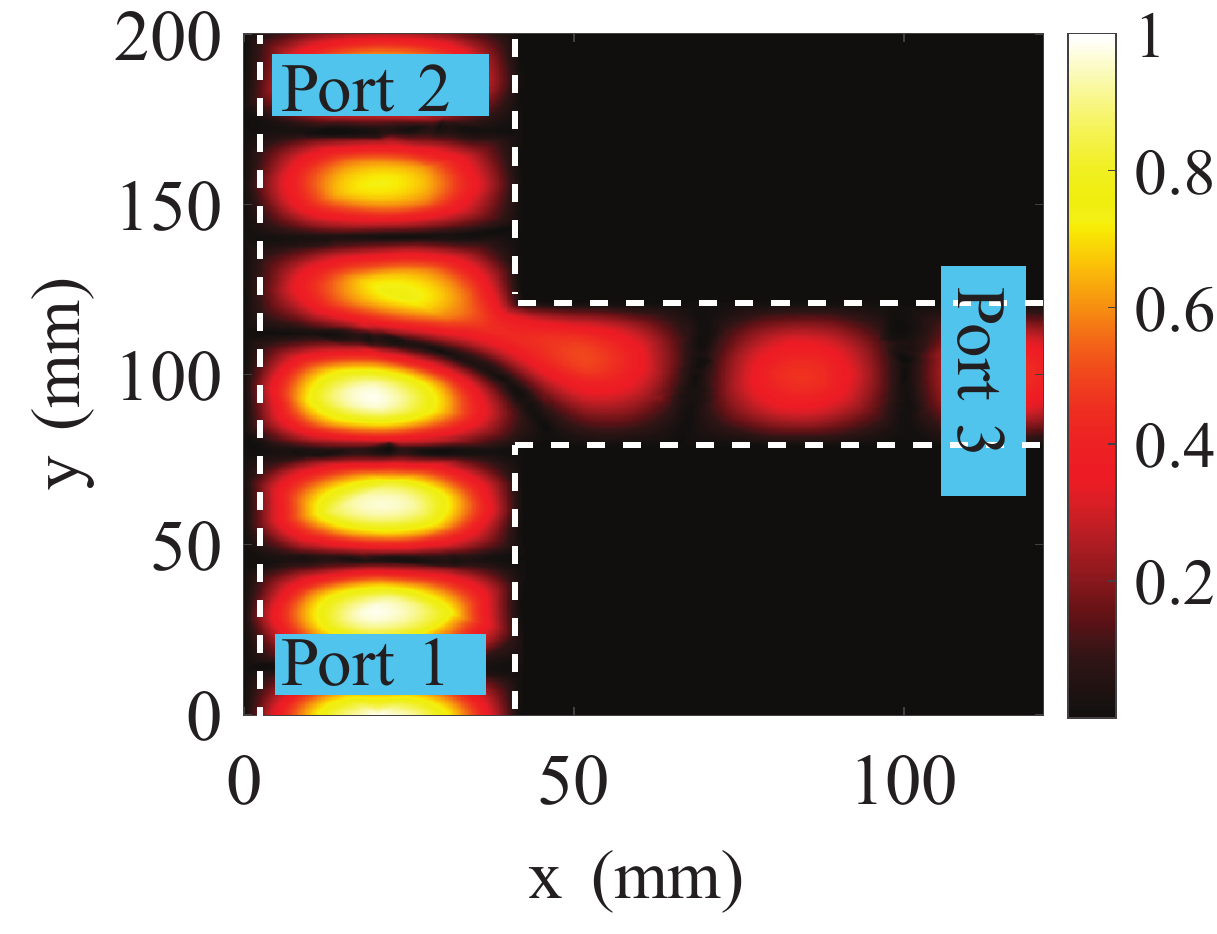}
     \subcaption{Port switched simple T-junction.}
     \label{fig:sub1}
   \end{minipage}
      \begin{minipage}{0.5\columnwidth}
     \centering
     	\psfrag{a}[c][c][0.7]{Frequency (GHz)}
	\psfrag{b}[c][c][0.7]{S Parameters (dB)}
	\psfrag{c}[c][c][0.5]{$S_{11}$}
	\psfrag{d}[c][c][0.5]{$S_{21}$}
	\psfrag{e}[c][c][0.5]{$S_{31}$}
     \includegraphics[width=\columnwidth]{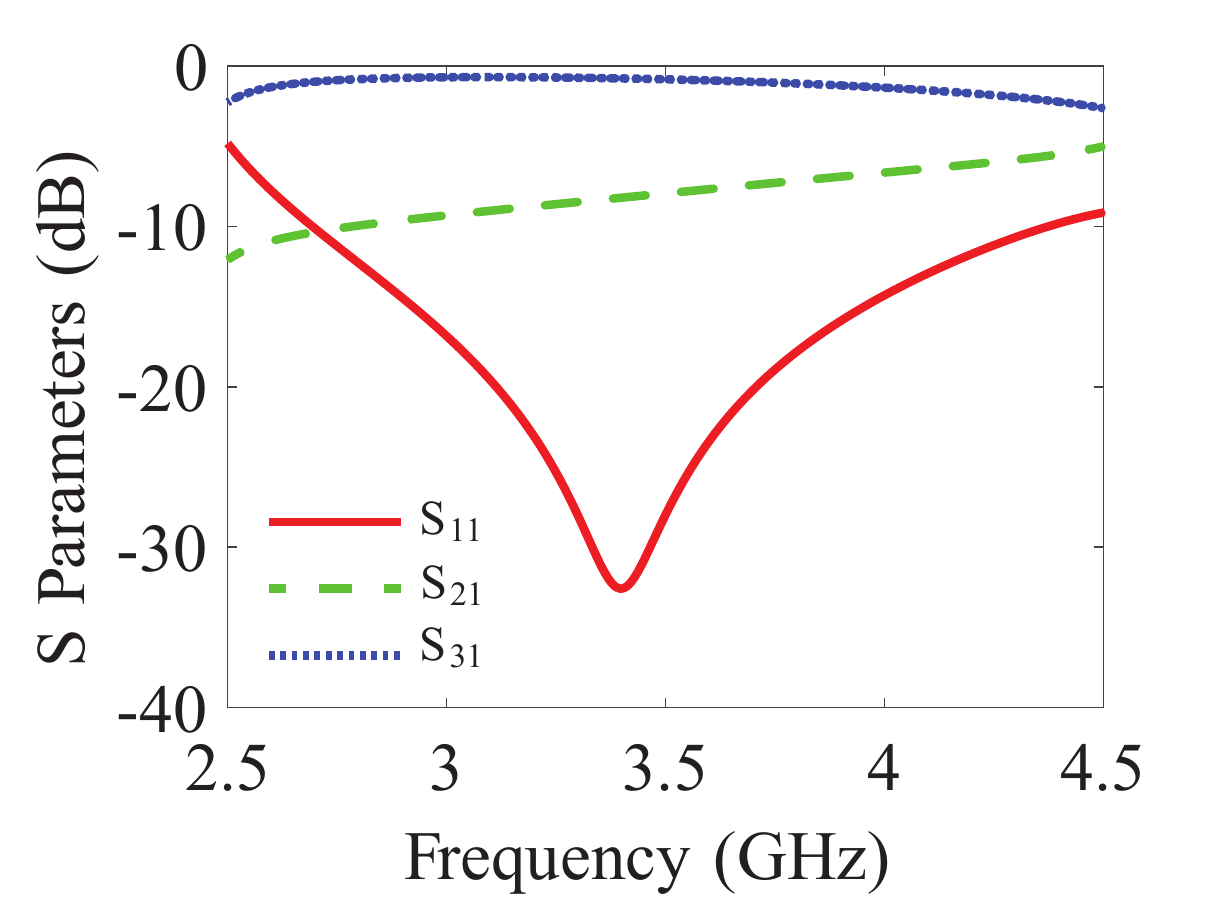}\\
     	\psfrag{a}[c][c][0.7]{$x$ (mm)}
	\psfrag{b}[c][c][0.7]{$y$ (mm)}
	\psfrag{c}[c][c][0.7]{Port 1}
	\psfrag{d}[c][c][0.7]{Port 2}
	\psfrag{e}[c][c][0.7]{Port 3}
     \includegraphics[width=\columnwidth]{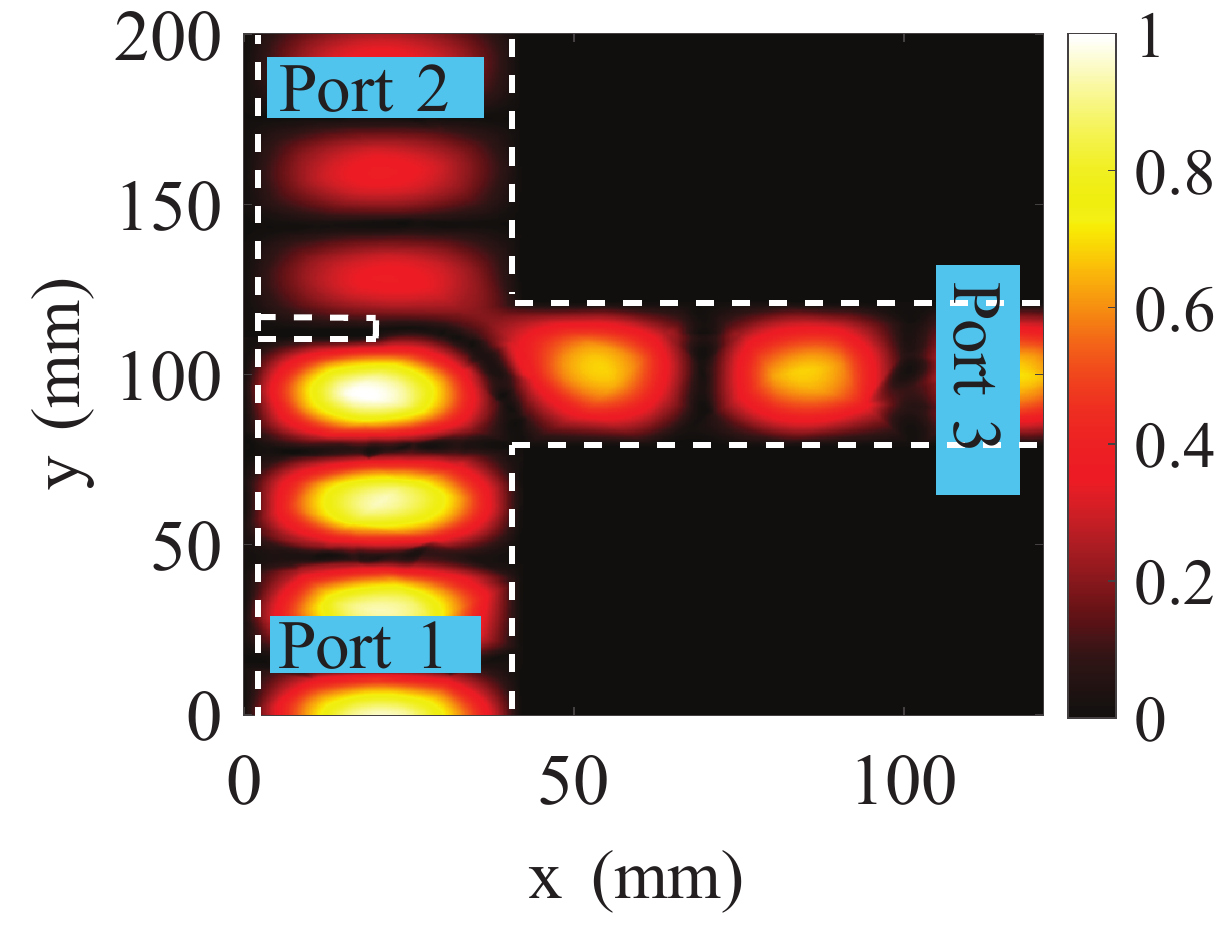}
     \subcaption{Port switched matched T-junction.}
     \label{fig:sub1}
   \end{minipage}\\[1em]
   \begin{minipage}{2\columnwidth}
     \centering
	\psfrag{a}[c][c][0.7]{$x$ (mm)}
	\psfrag{b}[c][c][0.7]{$y$ (mm)}
	\psfrag{c}[c][c][0.5]{Port 1}
	\psfrag{d}[c][c][0.5]{Port 2}
	\psfrag{A}[c][c][0.5]{\color{white}$f = 3.7$~GHz}
	\psfrag{f}[c][c][0.5]{P}
	\psfrag{f}[c][c][0.7]{Normalized $\abs{\mathbf{E}}$ (dB)}
 	\includegraphics[width=0.24\columnwidth]{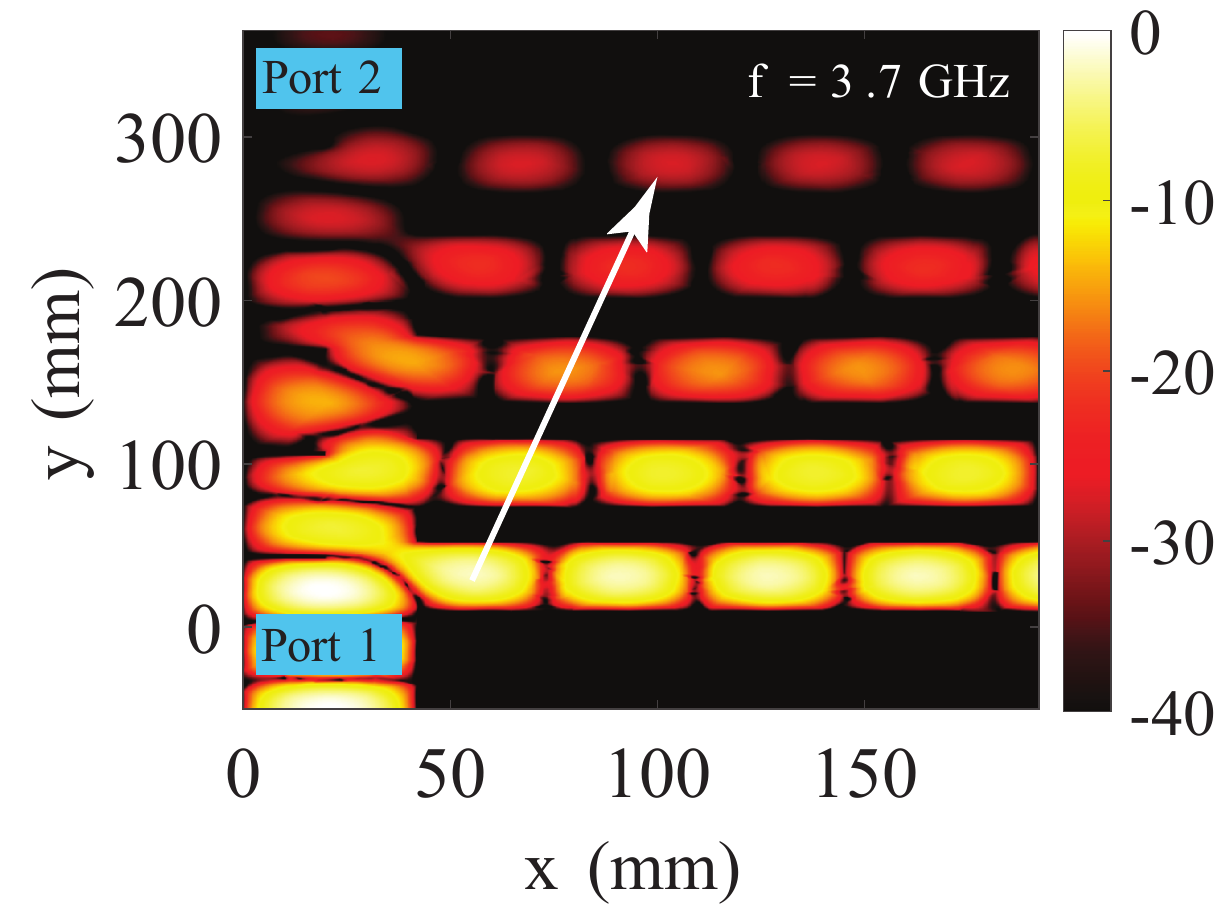}
	\psfrag{a}[c][c][0.7]{$x$ (mm)}
	\psfrag{b}[c][c][0.7]{$y$ (mm)}
	\psfrag{c}[c][c][0.5]{Port 1}
	\psfrag{d}[c][c][0.5]{Port 2}
	\psfrag{A}[c][c][0.5]{\color{white}$f = 4$~GHz}
	\psfrag{f}[c][c][0.7]{Normalized $\abs{\mathbf{E}}$ (dB)}
   	 \includegraphics[width=0.24\columnwidth]{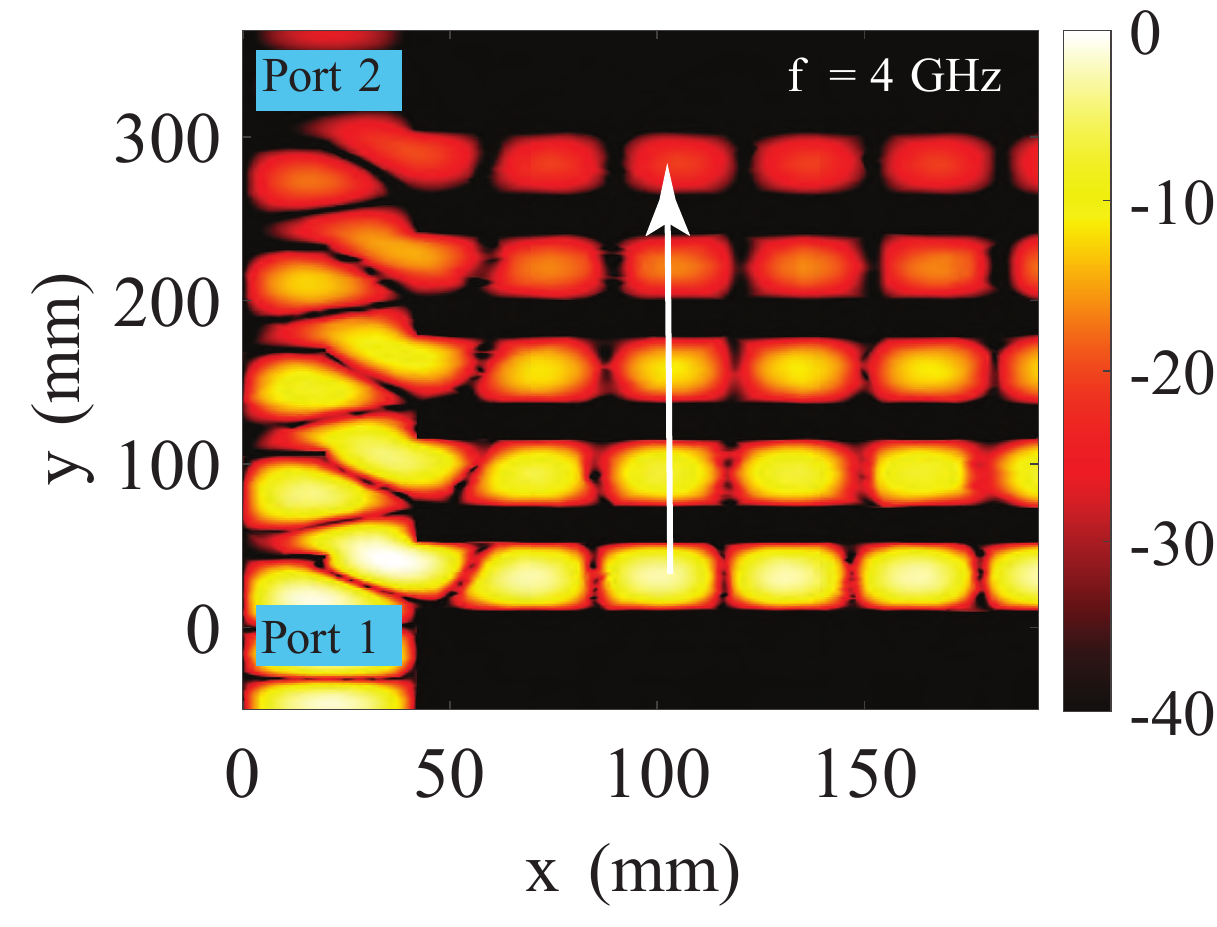}
	\psfrag{a}[c][c][0.7]{$x$ (mm)}
	\psfrag{b}[c][c][0.7]{$y$ (mm)}
	\psfrag{c}[c][c][0.5]{Port 1}
	\psfrag{d}[c][c][0.5]{Port 2}
	\psfrag{A}[c][c][0.5]{\color{white}$f = 4.3$~GHz}
	\psfrag{f}[c][c][0.7]{Normalized $\abs{\mathbf{E}}$ (dB)}
   	 \includegraphics[width=0.24\columnwidth]{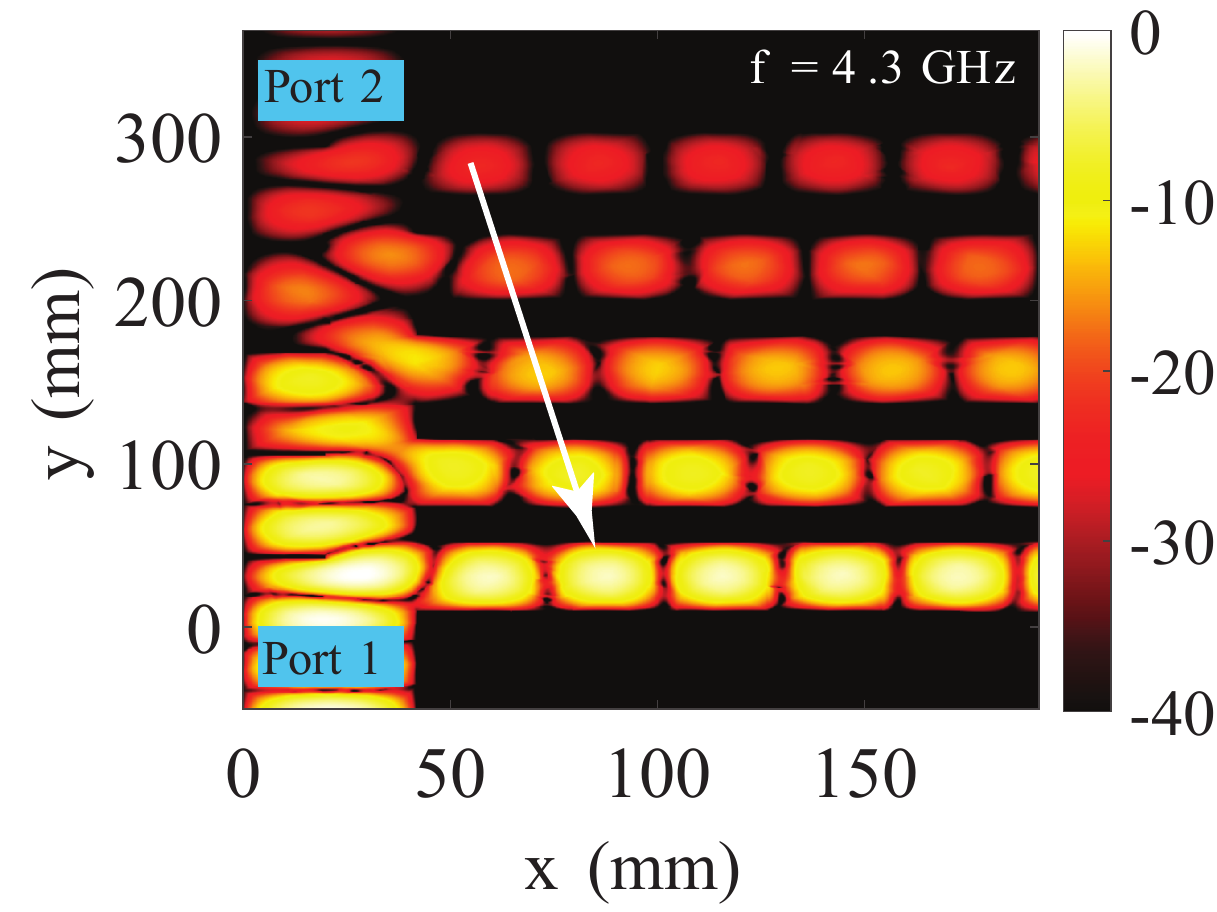}
         \centering
	\psfrag{a}[c][c][0.7]{Frequency (GHz)}
	\psfrag{b}[c][c][0.7]{$S_{11}$ (dB)}
	\psfrag{c}[c][c][0.5]{No Notch}
	\psfrag{d}[c][c][0.5]{Notch}
   	 \includegraphics[width=0.24\columnwidth]{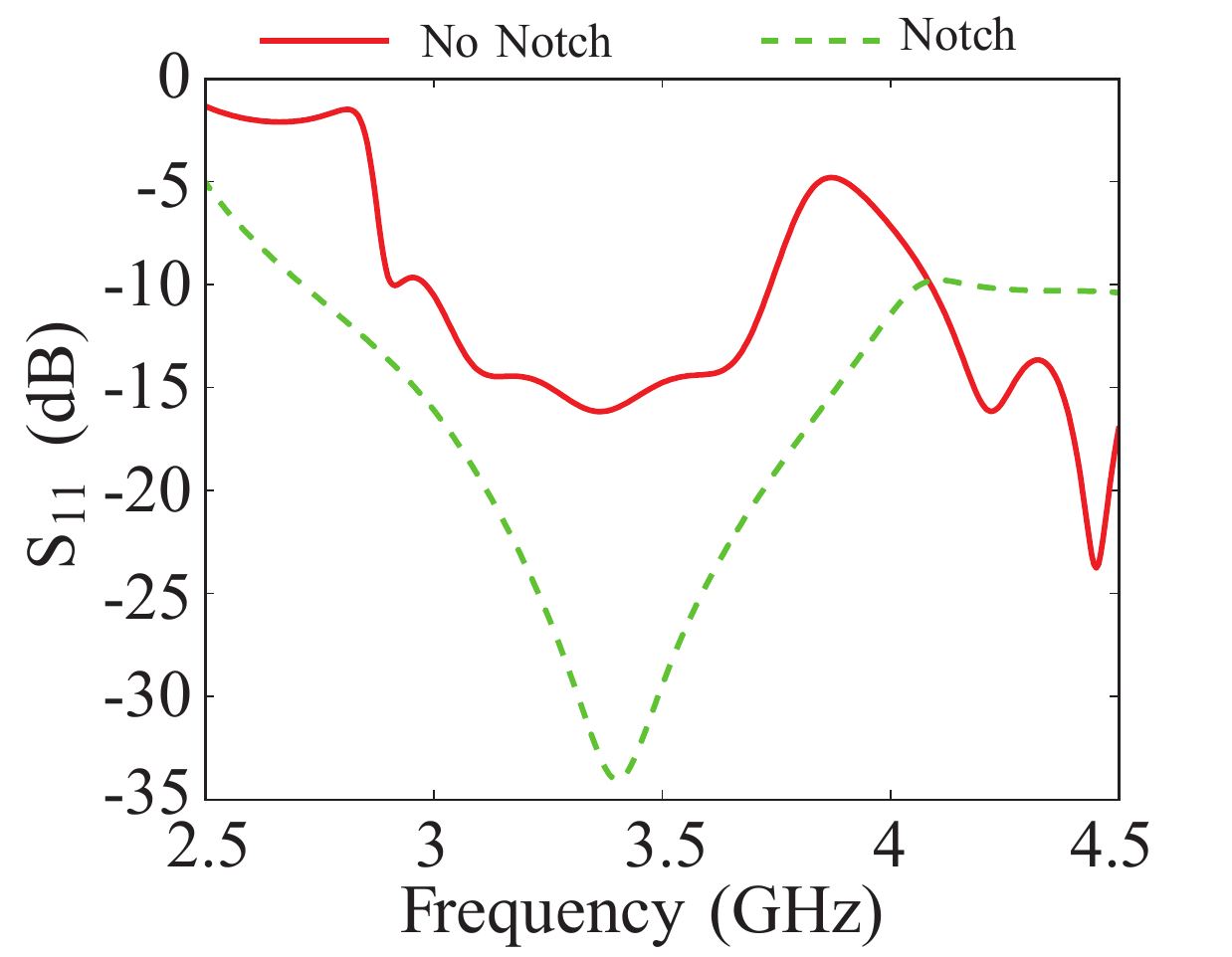}
     \subcaption{Cascaded matched T-junction.}
     \label{fig:sub2}
   \end{minipage}
	\caption{Principle of constructing an $N-$port power divider starting from a conventional 3-port waveguide T-junction. Various design parameters are: $s=12$ mm, $w_1=w_2=42$ mm, $p=63.32$ mm, notch length $\ell=20$ mm, and notch width $g=2$ mm. The fields are plotted at $f = 4$~GHz for (b-e) }  \label{Fig:TjunctionPrinciple}
\end{figure*}

Recently, an idea of a novel mm-wave near-field real-time spectrum analyzer (RTSA) based on an integrated side-fire LWA has been proposed in \cite{king2019millimeter}. Here this idea is further explored in details and experimentally demonstrated. The subsequent contributions of this works are \textcolor{black}{two-fold}: 1) It is based on a novel side-fire LWA using integrated waveguide periodic apertures on its side walls, providing a typical frequency scanned power leakage \emph{along the plane of the antenna} as opposed to radiation normal to the antenna plane in conventional designs. The LWA unit cell is designed with a self-matching mechanism that \textcolor{black}{suppresses} the stop-band characteristics of periodic structures, thereby enabling seamless frequency scanning from \textcolor{black}{the} backward to forward region through broadside. 2) Its application to broadband spectrum analysis at \textcolor{black}{mm-wave} frequency bands. In this work, a detailed design analysis of a side-fire antenna and its application to spectrum analysis is presented along with full experimental demonstrations around the 60 GHz band. The system is the first of its kind at mm-wave frequencies, and is fully integrated inside a low profile \textcolor{black}{parallel-plate waveguide} (PPW) with no radiation to free-space, eliminating any electromagnetic interference (EMI) with neighbouring instruments. The system is further designed to operate in the near-field region by introducing a curved shape of the structure resulting in compact size and low-profile compatible with a standard \textcolor{black}{printed circuit board} (PCB) process. 

This paper is organized as follows. Section II describes the principle of \textcolor{black}{the} proposed side-fire leaky-wave antenna inspired from a \textcolor{black}{cascaded} matched T-junction cell \textcolor{black}{with} an asymmetric notch (or inductive post in fabrication). Through cascading the T-junctions, the output ports can be opened for radiation forming a periodic LWA. Detailed eigenmode analysis is further provided to highlight important features of the structure. Section III shows how the side-fire LWA can \textcolor{black}{be} used as a real-time spectrum analyzer through the property of spatial-spectral decomposition. By introducing curvature, a compact size for the spectrum analyzer can be achieved. Furthermore, a simple analytical method is used to model the beam-scanning property of this structure providing further insight. Section IV provides experimental results showing the fabricated structures and the corresponding measurement results \textcolor{black}{demonstrating} the successful operation of the side-fire LWA for \textcolor{black}{spectral} decomposition. Finally, conclusions are provided in Sec. V.

\section{Proposed Side-Fire Leaky-Wave Antenna (LWA)}

\textcolor{black}{A} LWA is typically a 2-port traveling-wave structure that gradually radiates EM energy to free-space as the wave propagates along it \cite{jackson2012leaky}. It is characterized by a complex propagation constant $\gamma(\omega)=\alpha(\omega) + j\beta(\omega)$, where $\alpha$ is the \textcolor{black}{leakage per unit length} along the structure and $\beta$ is the \textcolor{black}{phase} constant. LWAs can be considered as series-fed phased arrays, and are typically broadband and directive in nature. While $\alpha$ (and the physical aperture) controls the beamwidth of radiation, $\beta$ provides the necessary frequency dependent phase shifts to scan the radiation beams. The Majority of LWAs are periodic in nature with near-sub-wavelength periodicities, where they radiate from the $n=-1$ space harmonics which fall within the fast-wave region with both backward and forward radiation \cite{yang2010full, Sakakibara_slots, jackson2012leaky}. On the other hand, uniform LWA structures based on metamaterial principles feature deep sub-wavelength periodicities and enable full-space frequency scanning using fundamental mode directly \cite{Caloz_CRLH, paulotto2008full}.

The majority \textcolor{black}{of LWAs} provide frequency-scanned peak radiation normal (in the $y-z$ plane) to the plane of the antenna (say $x-y$ plane, with antenna along $y-$axis for instance). An exception is a rectangular waveguide based LWA, where a continuous slot is opened enabling peak radiation from the sides, in the $x-y$ plane \cite{collin1969antenna}. Such a structure however is restricted to operate in the forward region only with no broadside radiation capability. Now, our objective here is to design a \textcolor{black}{LWA}, which is integrated into the subtrate, and radiates in the $x-y$ plane with radiation in \textit{both forward and backward regions including broadside}. In this section, we will present a step by step procedure to show how such a structure may be devised starting from a conventional waveguide T-junction.

\subsection{Conventional Waveguide T-Junctions}

The microwave and \textcolor{black}{mm-wave} T-junction is an important and commonly used component in microstrip circuits such as filters, amplifiers, ring hybrids, and power dividers \cite{pozar2005microwave}. Depending on the symmetry of the junction, power can be split in various ways. The S matrix can be written as

\begin{equation}
   [S] = \begin{bmatrix} 
   		 S_{11} & S_{12} & S_{13}  \\
         S_{21} & S_{22} & S_{23}  \\
         S_{31} & S_{32} & S_{33}  \\
         \end{bmatrix}.
\end{equation}

\noindent A conventional matched power splitter has $\abs{S_{11}} = \abs{S_{22}} = \abs{S_{33}} = 0$, and $\abs{S_{12}} = \abs{S_{13}} = \abs{S_{23}} = 1/2$, with reciprocity \cite{pozar2005microwave}. However, this assumes frequency invariance, which is untrue in practice. Impedance mismatch further degrades performance. A simple waveguide T-junction operating in its \textcolor{black}{fundamental} TE$_{10}$ mode with equal power split is shown in Fig.~\ref{Fig:TjunctionPrinciple}(a), with its typical response shown in Fig.~\ref{Fig:TjunctionPrinciple}(b) \textcolor{black}{exhibiting} strong reflection characteristics. This impedance mismatch can be improved through the introduction of a notch located in an optimized region of the junction, acting as a matching element~\cite{arndt1987optimized,hirokawa1991analysis}. Fig.~\ref{Fig:TjunctionPrinciple}\textcolor{black}{(b)} along with its typical response is shown in Fig.~\ref{Fig:TjunctionPrinciple}(c) with a significantly improved broadband matching. In actual fabrication, this notch is manifested as an inductive post. 

Let us switch the input and \textcolor{black}{output} ports of a conventional T-junction, so that the input port now becomes an output port, and one of the output ports becomes the new input, as shown in Fig.~\ref{Fig:TjunctionPrinciple}(a), indicated with \emph{Port-Switched Simple T-junction}, with its corresponding response shown in Fig.~\ref{Fig:TjunctionPrinciple}(d). Naturally, the power balance is disturbed between the two output ports and poor matching is still present. Next, let us introduce the notch to match this section, where by it is seen that by an optimum design and placement of the notch, an optimal matching is achieved as shown in Fig.~\ref{Fig:TjunctionPrinciple}(e), i.e. \emph{Port-Switched Matched T-junction}. This configuration must be compared with a matched T-junction, where the notch was placed symmetrically with respect to the two output ports. For a Port-Switched Matched T-junction, a longitudinal offset $s$ is needed which introduce\textcolor{black}{s} a \emph{transverse} asymmetry in the structure, where the transverse plane is $y=0$. Now by controlling the width of Port 3, power flow in this section can be controlled.

\begin{figure*}[htbp]
\centering
      \begin{minipage}{0.95\columnwidth}
     \centering
	\psfrag{a}[c][c][0.6]{Port 1}
	\psfrag{b}[c][c][0.6]{Port 2}
	\psfrag{x}[c][c][0.6]{$x$}
	\psfrag{y}[c][c][0.6]{$y$}
	\psfrag{c}[c][c][0.7]{$\theta (\omega )$}
	\psfrag{d}[c][c][0.7]{Leaky-Wave Region}
	\psfrag{e}[c][c][0.7]{Radiation Boundary}
	\psfrag{f}[c][c][0.7]{Apertures}
\includegraphics[width=\columnwidth]{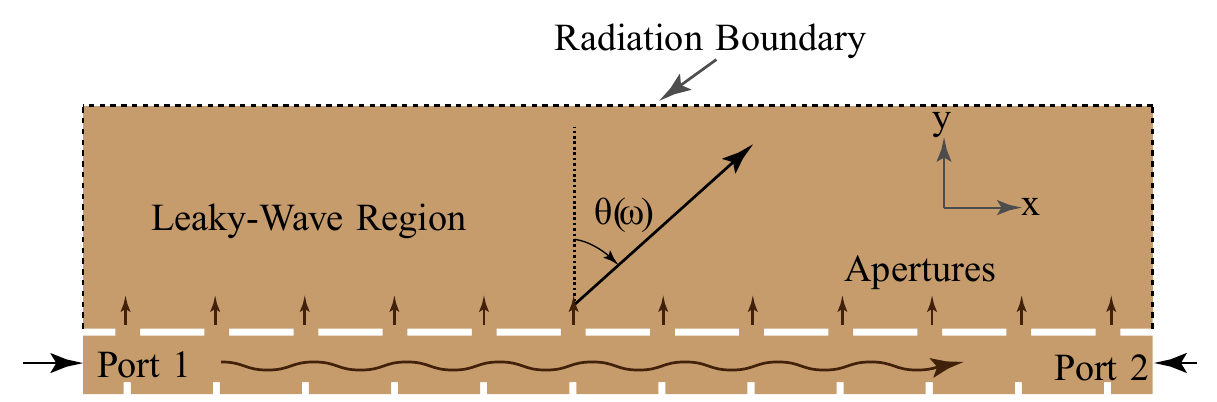}
\par\bigskip
     %
	\psfrag{a}[c][c][0.7]{$s$}
	\psfrag{b}[c][c][0.7]{$L$}
	\psfrag{c}[c][c][0.7]{$\phi = \beta p$}
	\psfrag{d}[c][c][0.7]{$t$}
	\psfrag{e}[c][c][0.7]{$a$}
	\psfrag{f}[c][c][0.7]{$w_1$}
	\psfrag{g}[c][c][0.7]{$p$}
	\psfrag{h}[c][c][0.7]{\shortstack{Free-space Impedance \\ $Z=377~\Omega$}}
	\psfrag{i}[c][c][0.7]{Floquet Boundaries}
	\psfrag{j}[c][c][0.7]{$s$}
	\psfrag{k}[c][c][0.7]{$w_2$}
	\psfrag{l}[c][c][0.7]{$\ell_n$}
\includegraphics[width=\columnwidth]{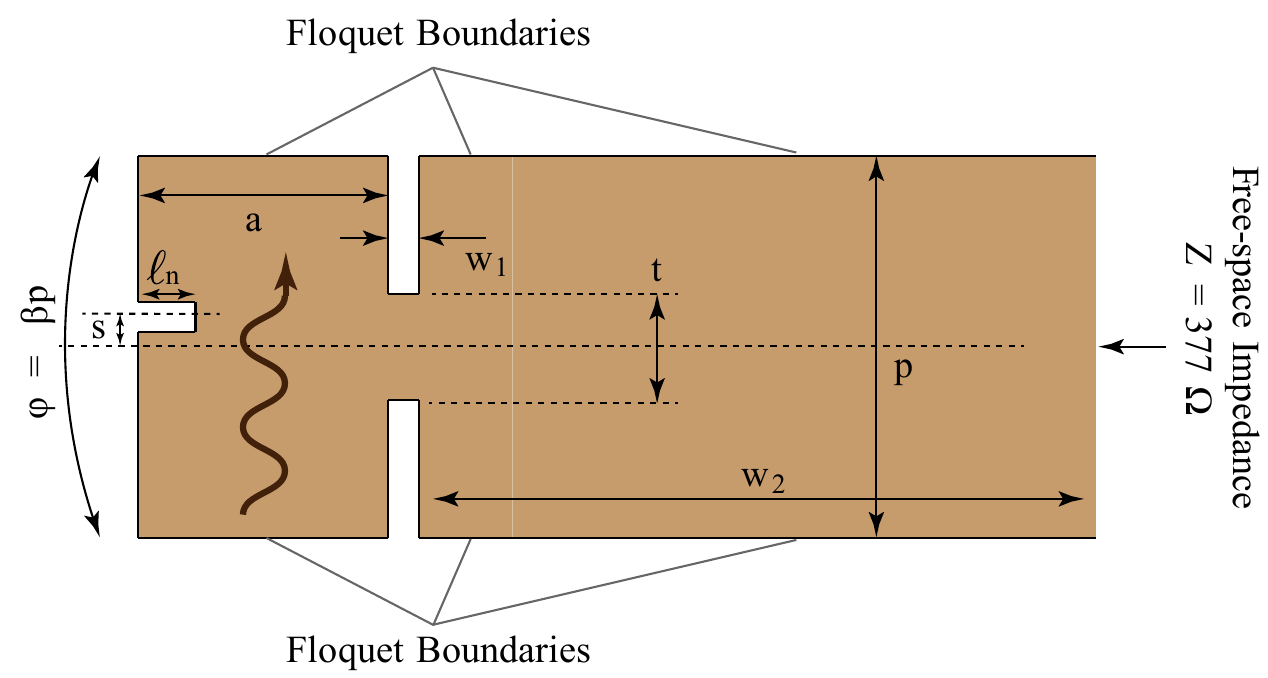}
     \subcaption{}
   \end{minipage}\hfill
\begin{minipage}{1.05\columnwidth}
	\psfrag{a}[c][c][0.7]{$\phi = \beta p$ ($\degree$)}
	\psfrag{b}[c][c][0.7]{Frequency (GHz)}
	\psfrag{c}[c][c][0.7]{Optimal asymmetry}
	\psfrag{d}[c][c][0.7]{Full symmetry}
	\psfrag{e}[c][c][0.7]{Suboptimal asymmetry}
\includegraphics[width=\columnwidth]{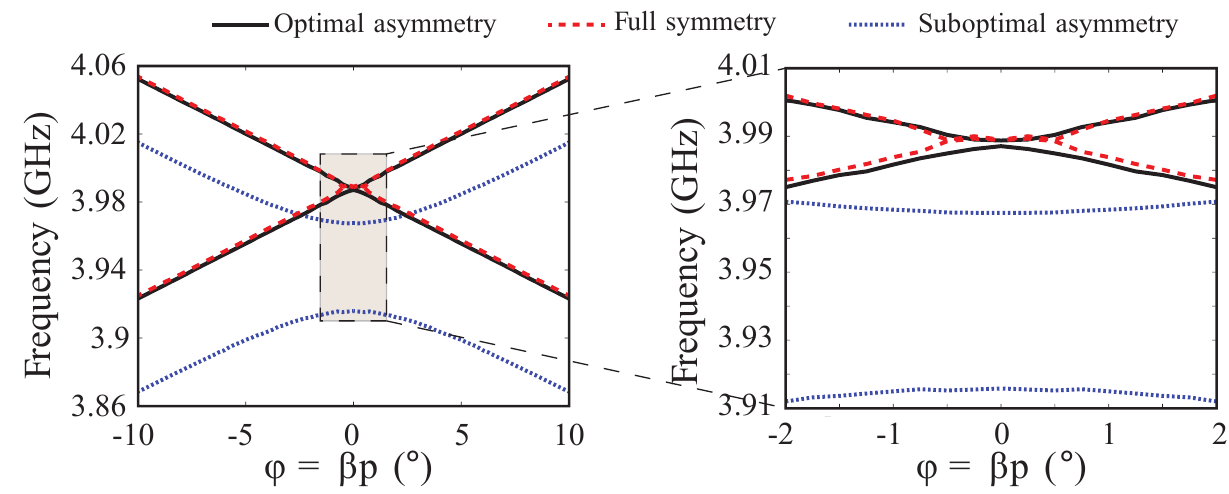}%
     \subcaption{}
     	\psfrag{a}[c][c][0.7]{Frequency (GHz)}
	\psfrag{b}[c][c][0.7]{S-Parameters (dB)}
	\psfrag{c}[c][c][0.7]{Peak $G_\theta(\phi, \theta=90^\circ)$~(dB)}
	\psfrag{h}[c][c][0.5]{\shortstack{Stopband\\ Region}}
\includegraphics[width=\columnwidth]{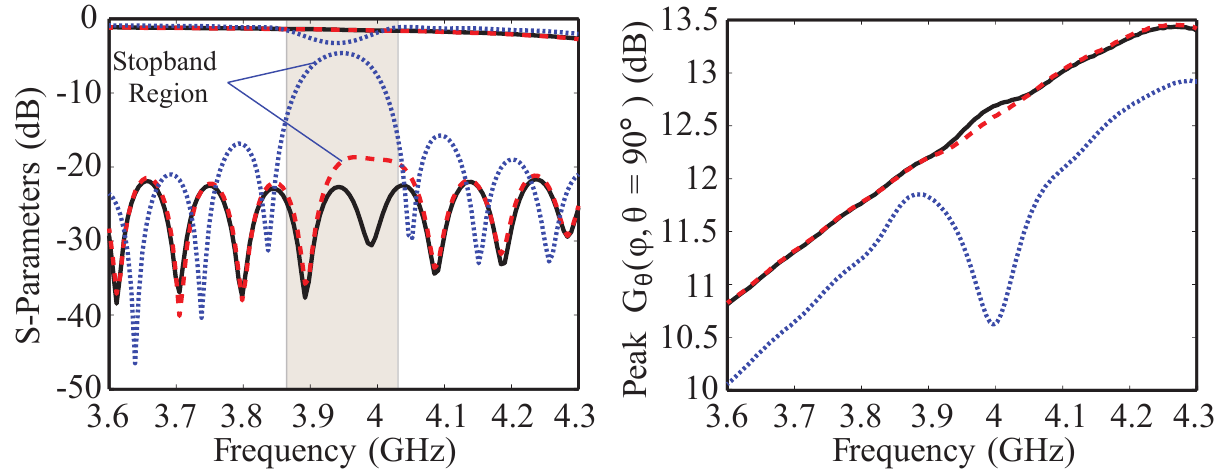}
     \subcaption{}
\end{minipage}
\begin{minipage}{2\columnwidth}
	\psfrag{a}[c][c][0.7]{$x$~(mm)}
	\psfrag{b}[c][c][0.7]{$y$~(mm)}
	\psfrag{A}[c][c][0.7]{\color{white} \shortstack{\textsc{Backward Radiation}\\ $f = 3.7$~GHz}}
	\psfrag{B}[c][c][0.7]{\color{white} \shortstack{\textsc{Broadside Radiation}\\ $f = 4.0$~GHz}}
	\psfrag{C}[c][c][0.7]{\color{white} \shortstack{\textsc{Forward Radiation}\\ $f = 4.3$~GHz}}
\includegraphics[width=\columnwidth]{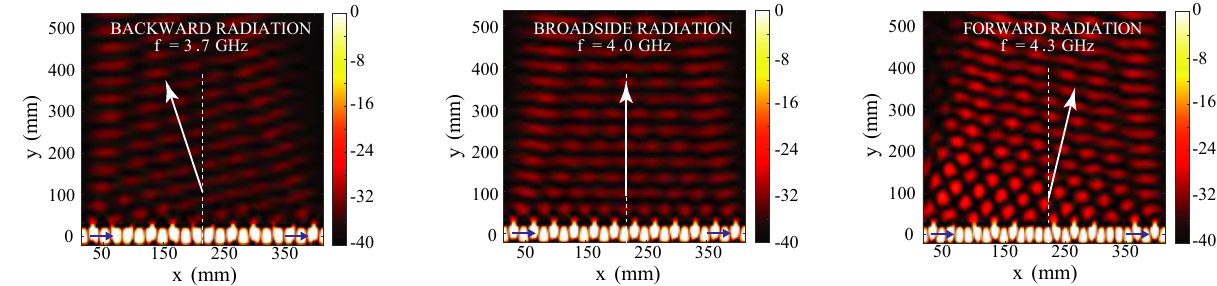}%
     \subcaption{}
\end{minipage}
\caption{The proposed side-fire Leaky-Wave Antenna (LWA) constructed using the $N-$port power divider of Fig.~\ref{Fig:TjunctionPrinciple}. a) A driven-mode LWA structure consisting of 12 cells and the unit cell used for eigenmode analysis in FEM-HFSS, b) the corresponding eigenfrequencies showing three unit cell configurations. c) The driven mode S-parameters and the peak Gain vs frequency. d) Near-fields $E_z$ showing the radiated wavefronts in the backward and forward regions, along with that at broadside. Here $p=64$ mm, $t=17.78$ mm, $a=42$ mm, $w_1=5.08$ mm, and $w_2=152.4$ mm. For optimal asymmetry: $\ell_n=8.64$ mm, and $s=0.32$ mm. For full symmetry: $\ell_n=8.76$ mm, and $s=0$ mm. For suboptimal asymmetry: $\ell_n=2.54$ mm, and $s=0.34$ mm. } \label{Fig:SidefireLWA}
\label{figabc}
\end{figure*}

\subsection{Side-Fire LWA}

The port-switched matched T-junction now acts as a power splitter by splitting the input from port 1 to output ports 2 and 3, respectively, with identical port impedances at port 1 and 2. This represents a fundamental building unit which is self-matched due to optimal design of the notch element. This allows a series cascade of this unit to form an $N$ element periodic structure with $N+2$ ports and a spatial period $p$, which can now be seen simply as a $1:N$ power divider, as illustrated in Fig.~\ref{Fig:TjunctionPrinciple}(a). Since each unit cell element is self-matched using a transversally asymmetric placed notch, the resulting periodic structure also exhibits a broadband matching, whereby its stopband characteristics can be easily controlled. The typical reflection and transmission characteristics for $1:5$ power divider are shown in Fig.~\ref{Fig:TjunctionPrinciple}(f), where a broadband matching is clearly evident. Due to the periodicity and the traveling-wave nature of this structure, the phase of the side ports (2-7) are frequency dependent. For instance, when $p \approx \lambda_g$, all the wavefronts that exit these ports are in phase, as shown in the Fig.~\ref{Fig:TjunctionPrinciple}(f). We will refer to this frequency $f_0$ as the \emph{broadside frequency of the structure}. For frequencies, where $p> \lambda_g$ ($f > f_0$), there is a progressive phase delay leading to a phase tilt, which will later will shown to correspond to \emph{forward radiation}. Finally, for $p< \lambda_g$ ($f < f_0$), there is a progressive phase advance corresponding to \emph{backward radiation}. These wavefront \textcolor{black}{advances} and delays, leading to a phase tilts are further shown in Fig.~\ref{Fig:TjunctionPrinciple}(f) for two specific frequencies.

The cascaded $1:N$ power divider can now be allowed to have its outputs open to free space readily forming a periodic LWA with beam-scanning capability. Thanks to the self-matched transversally asymmetric unit cell based on a notch, a seamless transition from \textcolor{black}{the} backward to the forward region is possible including broadside radiation. This resulting structure \textcolor{black}{represents} a periodic LWA consisting of $N$ unit cells with radiating apertures, allowing free-space radiation from the side walls of a rectangular waveguide operating in its fundamental TE$_{10}$ mode. Since this structure is compatible with \textcolor{black}{substrate integrated waveguide} (SIW) technology, it can now be readily integrated inside a dielectric slab, where the radiation from the aperture array can either be coupled into free-space or confined between top and bottom conductors of the dielectric forming a \textcolor{black}{PPW}, as shown in Fig.~\ref{Fig:SidefireLWA}(a). If the leaky-wave region outside the main waveguide is confined inside a PPW, the entire structure remains shielded and completely enclosed.

A cell for such a periodic antenna has a stop band at broadside that requires suppression~\cite{mallahzadeh2015periodic}. This behavior can be examined through considering the eigenmodes of the unit cell~\cite{baccarelli2019open,otto2013importance}. The cell schematic is shown in Fig.~\ref{Fig:SidefireLWA}(a). Depending on the dimensions and location of the notch, various dispersion relations of the unit cell may be computed, depending on the symmetries of the unit cell \cite{otto2013importance, otto2014transversal}. The eigenmode analysis assumes Floquet (periodic) boundaries on the two faces of the unit cell along the direction of wave propagation. Let us consider the case, when the leaky-wave region is confined inside a PPW, which is terminated with the medium impedance at the far end of the waveguide. While the width of the main waveguide, $a$, the period $p$, the aperture size $t$ and width $w_1$ are kept constant, the length $\ell_n$ and offset $s$ of the notch is varied to study its traveling wave characteristics. The FEM-HFSS computed eigenmodes for various cases of notch configurations is shown in Fig.~\ref{Fig:SidefireLWA}(b), which shows regions of $\beta<0$ corresponding to backward radiation, and $\beta>0$ for \textcolor{black}{foward radiation} (for \textcolor{black}{positively} sloped branches of the dispersion curves). As expected from the works of Otto et. al. \cite{otto2013importance, otto2014transversal}, a perfectly closed stopband requires an optimal transversal asymmetry in the unit cell. For sub-optimal asymmetries a large stop-band is present, which prohibits any broadside radiation along $\theta=0^\circ$. A fully symmetric unit cell on the other exhibits a frequency balanced condition, but is still a non-optimal condition which is manifested in the flat phase region around the broadside frequency. Ultimately, at the optimal set of notch parameters, the eigenmodes crossover at a single point as shown in shown in Fig.~\ref{Fig:SidefireLWA}(b), and the reflection at this frequency corresponding to broadside is expected to be optimal~\cite{guglielmi1993broadside}.

To further confirm these eigenmode results, a driven model is analyzed in FEM-HFSS consisting of $N=12$ cells, whose computed frequency dependent transmission/reflection response and far-field peak gain is shown in Fig.~\ref{Fig:SidefireLWA}(c). As expected, the cell with the optimal asymmetry provides an excellent match throughout the frequency band compared to a \textcolor{black}{suboptimally asymmetric} case, while the symmetric unit cell still exhibits a noticeable increase in reflection around broadside \textcolor{black}{in spite} of an overall good matching response \cite{otto2014transversal}. Fig.~\ref{Fig:SidefireLWA}(d) finally shows the radiating wavefronts at three different frequencies, clearly showing the wave propagation along backward, broadside and forward directions. Therefore, the proposed side-fire structure represents a fully integrated periodic LWA with complete control over its dispersion relation and the subsequent closure of the stop-band to enable broadside radiation. Exploiting its integrated nature and fully shielded configuration, we will next describe an integrated analog spectrum analyzer for a 60 GHz mm-Wave frequency band.

\section{Proposed mm-Wave Spectrum Analyzer}

\subsection{Principle}

Leaky-wave antennas have the property of frequency scanning. For a periodic leaky-wave antenna, the aperture spacing provides a phase shift between locations of radiation. This spacing is electrically frequency dependent, and therefore provides a phase gradient which enables beam steering. The beam scanning law is described by~\cite{jackson2012leaky}

\begin{equation}\label{Eq:Scanning}
\theta(\omega ) = \text{sin}^{-1}\left(\frac{\beta(\omega )}{k_0}\right),
\end{equation} 

\noindent where $k_0$ is the free space wavenumber, $\beta(\omega)$ is the guided propagation constant, and $\theta(\omega)$ is the beam angle measured from broadside. A real-time spectrum analyzer can be formed using the process of spatial-spectral frequency decomposition, similar to that used in optical prisms~\cite{gupta2009microwave, gupta2008crlh}. Accordingly, a broadband time domain test signal sent at the input port of a LWA would have its frequency components radiated at particular angles described by \eqref{Eq:Scanning} with appropriate magnitudes. This enables one to deduce the presence of the frequency components of the signal and their respective magnitudes by inspecting the corresponding far-field radiation patterns. Such a determination of the spectral components of an unknown test signal based on spatial-spectral frequency decomposition principle, is purely analog and thus real-time in nature.

The radiated signals from the LWA are typically measured in the \textcolor{black}{far-field} regime of the radiation, which is approximately located at distances greater than $2d^2/ \lambda_0$, where $d$ is the largest dimension of the antenna aperture, and $\lambda_0$ is the free space wavelength. For typical mm-wave leaky-wave antennas which are several wavelengths long, this distance readily exceeds 5~m when operated around 60 GHz. Therefore, to perform spectral analysis using a typical mm-wave LWA, one would need large space to make the measurement. However, this measurement would not be protected from the environment, and thus undesired signal interference would occur between neighbouring devices. Thus shielding is necessary, which makes the proposed side-fire LWA configuration of Fig.~\ref{Fig:SidefireLWA} particularly attractive. However, the far-field requirement of large system size is still a challenge, since for far-field distances of several meters, the PCB would be impractically large. A potential solution is then to add a physical curvature to the LWA aperture to focus the far-field radiation in the near-field region of the antenna, which is proposed next.

\subsection{Curved Side-Fire LWA}

A solution to the issue of large far-field distances where the beam is found is by introducing a curvature to the antenna~\cite{josefsson2006conformal}. In the case of parabolic reflectors, a parabolic surface allows for a beam focal point where the radiation is concentrated. Depending on the exact surface curvature, the focal location can change, subject to sufficiently small aberrations. \cite{balanis2016antenna, stutzman2012antenna}. In this sense a convex lens with a close focal point can solve the far distance problem. This principle can be easily applied to the side-fire LWA configuration of Fig.~\ref{Fig:SidefireLWA}, where instead of a linear array, we devise a curved array with desired frequency dependent beam focussing characteristics in its near-fields. Such a structure may behave like a lens, allowing the beam location to be within the range of a typical PCB lengths, while preserving the spatial-spectral decomposition information needed for spectral analysis. 

\begin{figure}[htbp]
	\centering
       \begin{subfigure}[b]{0.85\columnwidth}
         \centering
	\psfrag{A}[c][c][0.7]{$(x_1, y_1)$}
	\psfrag{B}[c][c][0.7]{$(x_2, y_2)$}
	\psfrag{C}[c][c][0.7]{\textsc{Source} $(x', y')$}
	\psfrag{D}[c][c][0.7]{\textsc{Observation} $(x, y)$}
	\psfrag{E}[c][c][0.7]{$2\alpha$}
	\psfrag{x}[c][c][0.7]{$x$}
	\psfrag{y}[c][c][0.7]{$y$}
	\psfrag{F}[c][c][0.7]{$r$}
	\psfrag{G}[c][c][0.7]{$\theta$}
	\psfrag{H}[c][c][0.7]{$\rho$}
   	 \includegraphics[width=\linewidth]{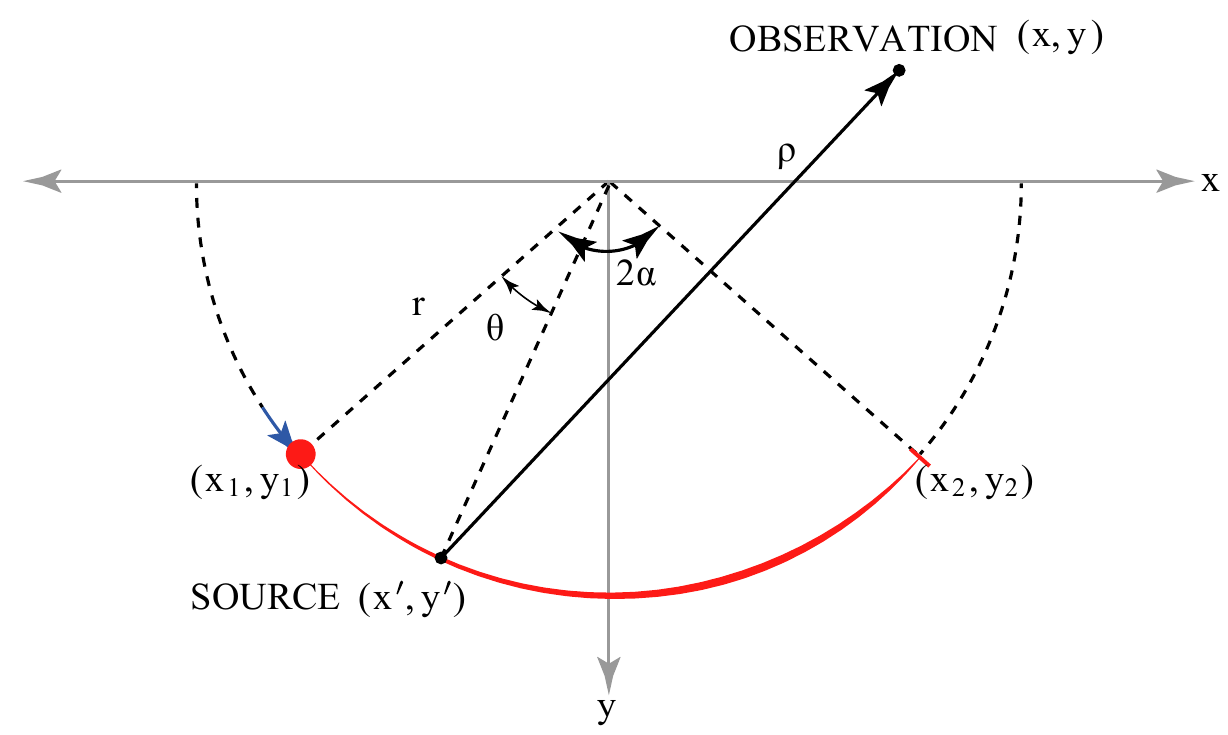}\caption{}
     \end{subfigure} 
       \begin{subfigure}[b]{0.493\columnwidth}
         \centering
	\psfrag{a}[c][c][0.7]{Observation Distance (mm)}
	\psfrag{b}[c][c][0.7]{Normalized $|\mathbf{E}|$}
	\psfrag{c}[c][c][0.5]{57~GHz}
	\psfrag{d}[c][c][0.5]{58~GHz}
	\psfrag{e}[l][c][0.5]{Analytical,~\eqref{Eq:AnalyticalField}}
	\psfrag{f}[l][c][0.5]{FEM-HFSS}
 	\includegraphics[width=\linewidth]{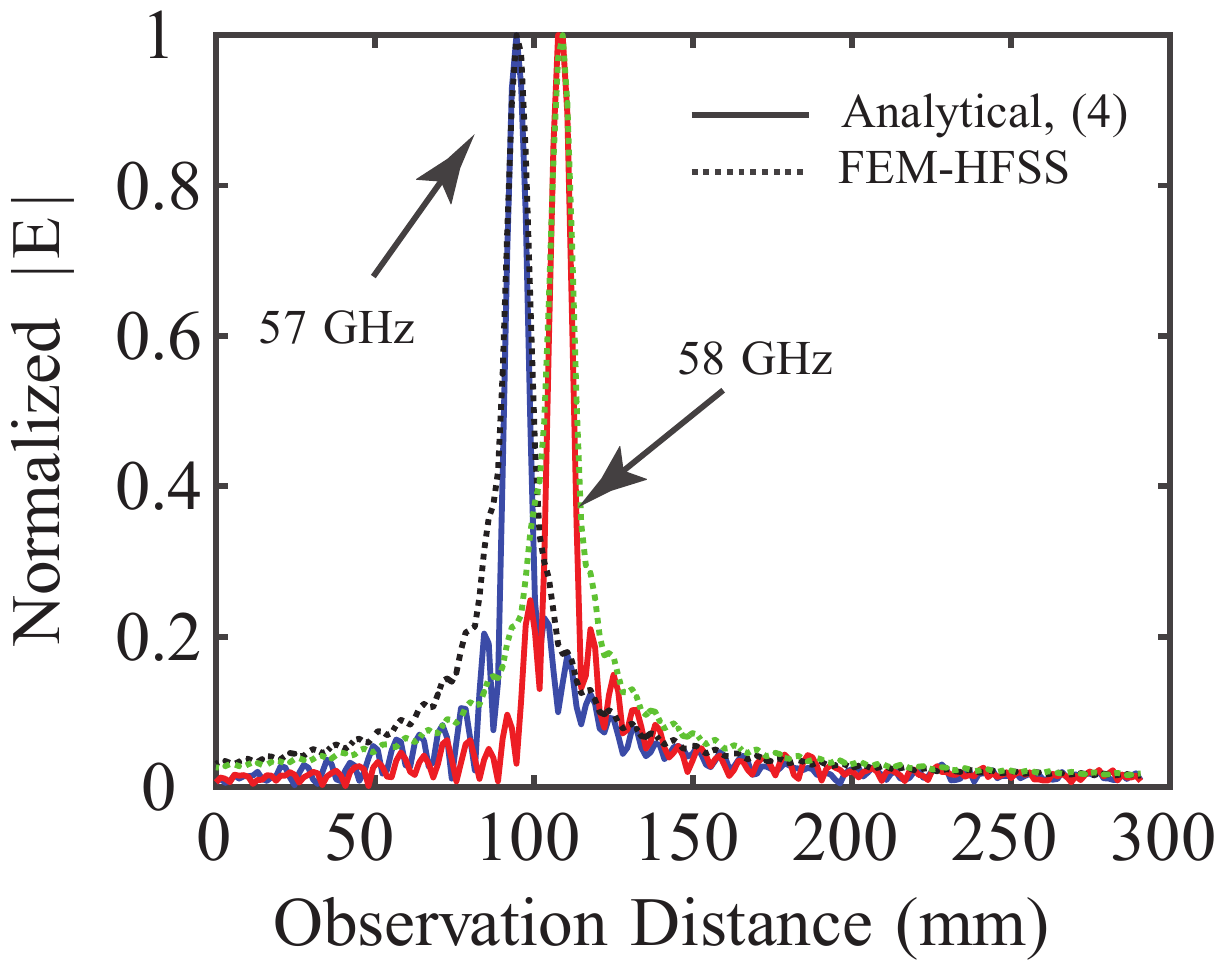}\caption{}
    	\end{subfigure}  
	     \begin{subfigure}[b]{0.493\columnwidth}
         \centering
	\psfrag{a}[c][c][0.7]{Observation Distance (mm)}
	\psfrag{b}[c][c][0.7]{Normalized $|\mathbf{E}|$}
	\psfrag{c}[c][c][0.5]{59~GHz}
	\psfrag{d}[c][c][0.5]{60~GHz}
	\psfrag{e}[l][c][0.5]{Analytical,~\eqref{Eq:AnalyticalField}}
	\psfrag{f}[l][c][0.5]{FEM-HFSS}
   	 \includegraphics[width=\linewidth]{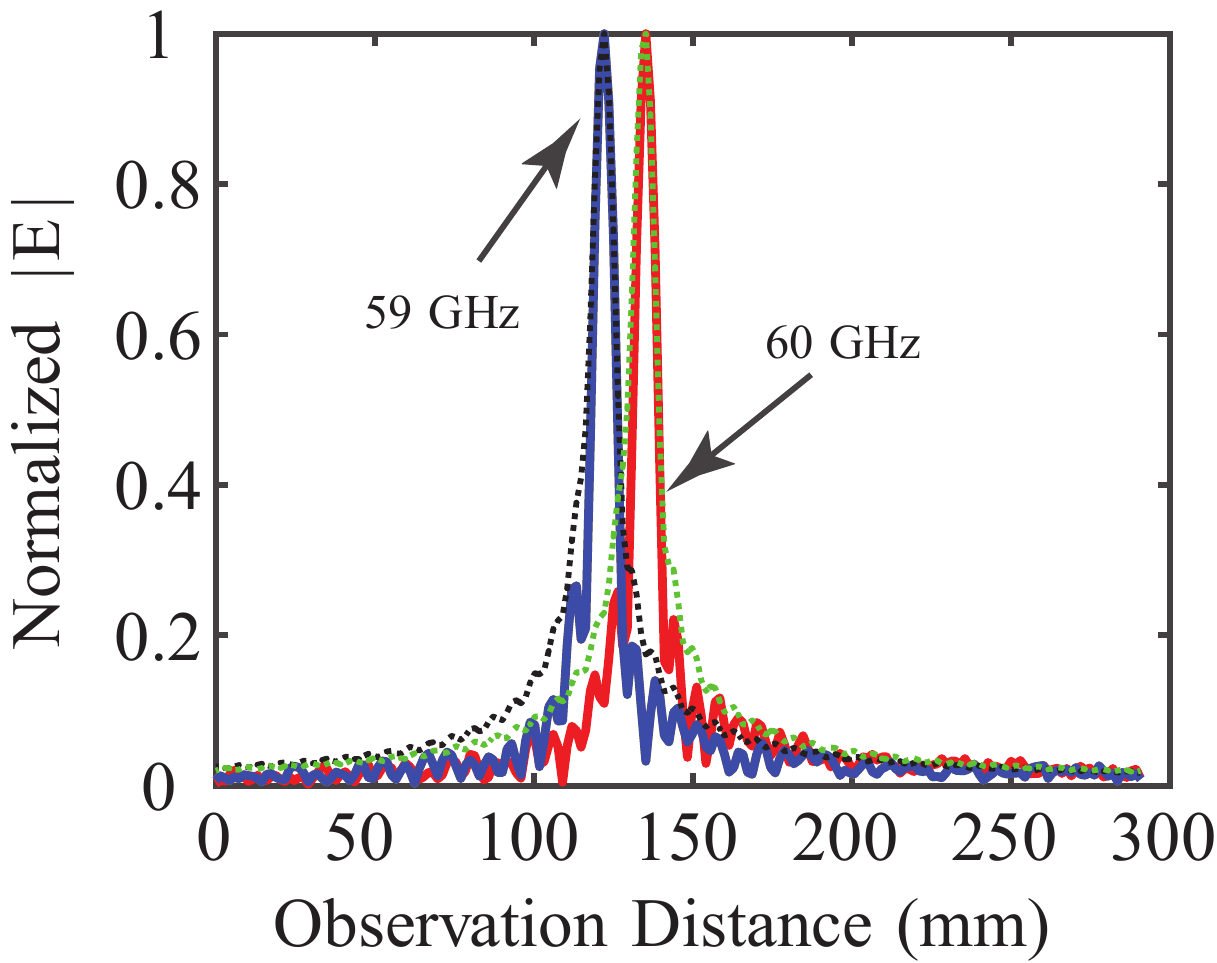}\caption{}
     \end{subfigure}  
 	     \begin{subfigure}[b]{0.493\columnwidth}
         \centering
	\psfrag{a}[c][c][0.7]{Observation Distance (mm)}
	\psfrag{b}[c][c][0.7]{Normalized $|\mathbf{E}|$}
	\psfrag{c}[c][c][0.5]{61~GHz}
	\psfrag{d}[c][c][0.5]{62~GHz}
	\psfrag{e}[l][c][0.5]{Analytical,~\eqref{Eq:AnalyticalField}}
	\psfrag{f}[l][c][0.5]{FEM-HFSS}
   	 \includegraphics[width=\linewidth]{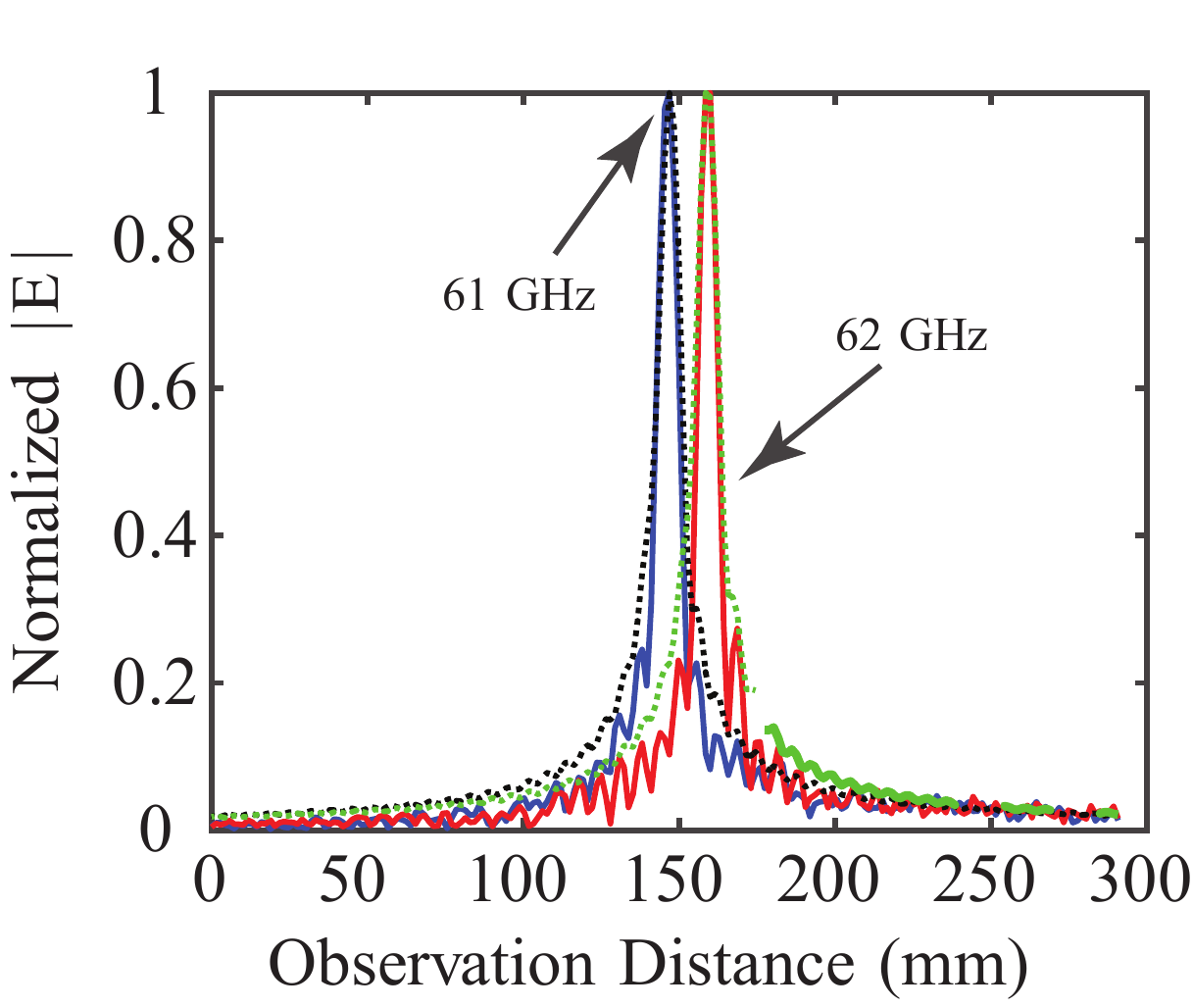}\caption{}
     \end{subfigure}  
 	     \begin{subfigure}[b]{0.493\columnwidth}
         \centering
	\psfrag{a}[c][c][0.7]{Observation Distance (mm)}
	\psfrag{b}[c][c][0.7]{Normalized $|\mathbf{E}|$}
	\psfrag{c}[c][c][0.5]{63~GHz}
	\psfrag{d}[c][c][0.5]{64~GHz}
	\psfrag{e}[l][c][0.5]{Analytical,~\eqref{Eq:AnalyticalField}}
	\psfrag{f}[l][c][0.5]{FEM-HFSS}
   	 \includegraphics[width=\linewidth]{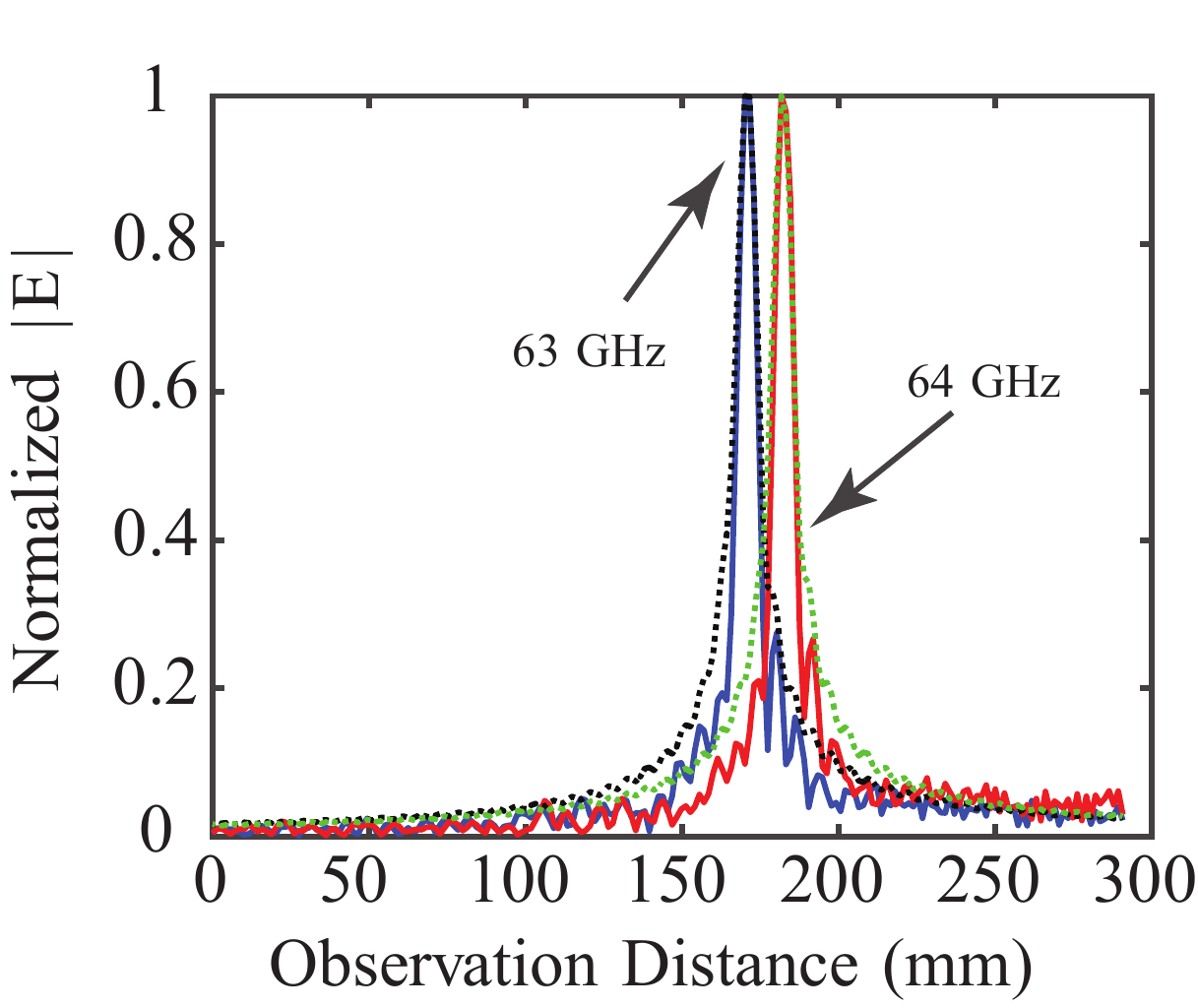}\caption{}
     \end{subfigure}  
	\caption{Analytical method to compute the near-fields of a side-fire antenna. a) Illustration of a curved side-fire antenna showing the source and observation points. (b)-(e) Comparison of electric fields calculated along a line at the focal point using the wavenumber computed from FEM-HFSS eigenmode analysis with full-wave electric fields along a line at the focal distance ($r=56$ cm).}\label{Fig:Analytical}
\end{figure} 

\begin{figure}[htbp]
	\centering              
	     \begin{subfigure}[b]{1\columnwidth}
         \centering
	\psfrag{a}[c][c][0.7]{Matched Boundary, $Z = \eta_0/\sqrt{\epsilon_r}$}
	\psfrag{b}[c][c][0.7]{Port 1}
	\psfrag{c}[c][c][0.7]{Port 2}
	\psfrag{d}[c][c][0.7]{$2d$}
	\psfrag{e}[c][c][0.7]{$s_3$}
	\psfrag{f}[c][c][0.7]{$\ell_1$}
	\psfrag{g}[c][c][0.7]{$d$}
	\psfrag{i}[c][c][0.7]{$s_1$}
	\psfrag{j}[c][c][0.7]{$\ell_2$}
	\psfrag{k}[c][c][0.7]{$s_2$}
   	 \includegraphics[width=\linewidth]{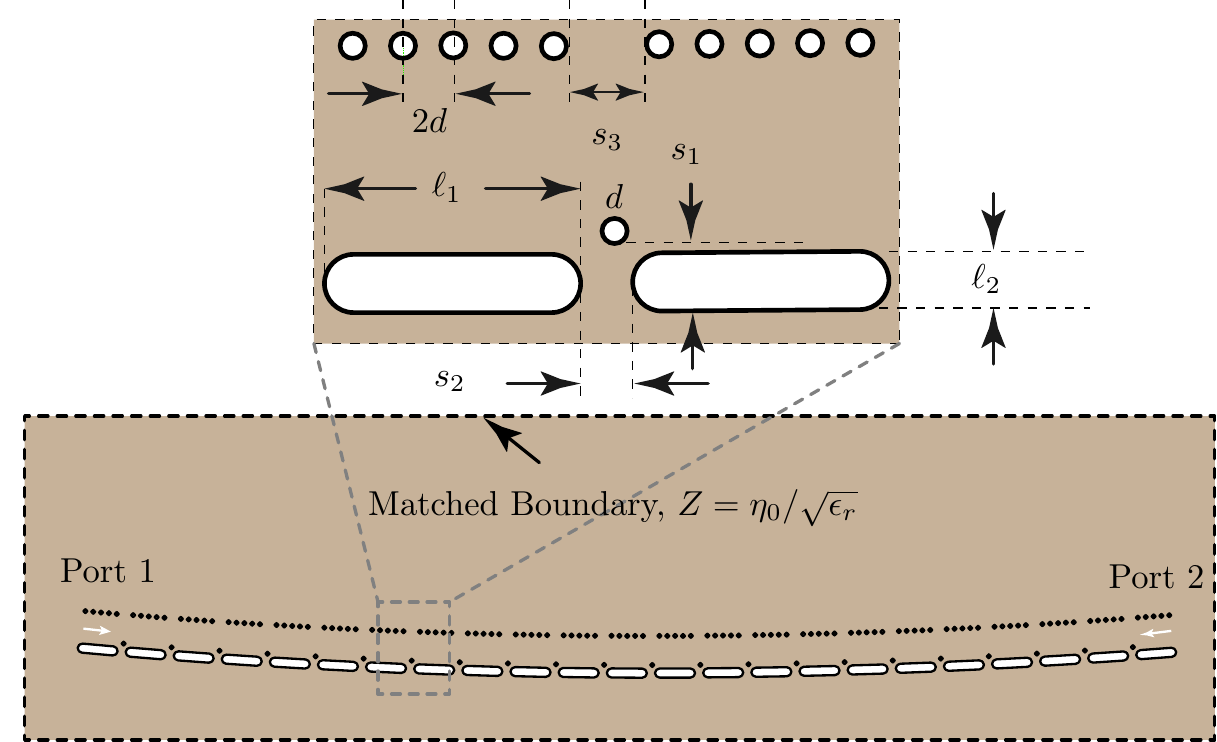}\caption{}
     \end{subfigure} 
 	     \begin{subfigure}[b]{0.493\columnwidth}
         \centering
	\psfrag{a}[c][c][0.7]{Frequency (GHz)}
	\psfrag{b}[c][c][0.7]{$S_{11}$ (dB)}
	\psfrag{c}[c][c][0.7]{Optimal}
	\psfrag{d}[c][c][0.7]{Suboptimal}
   	 \includegraphics[width=\linewidth]{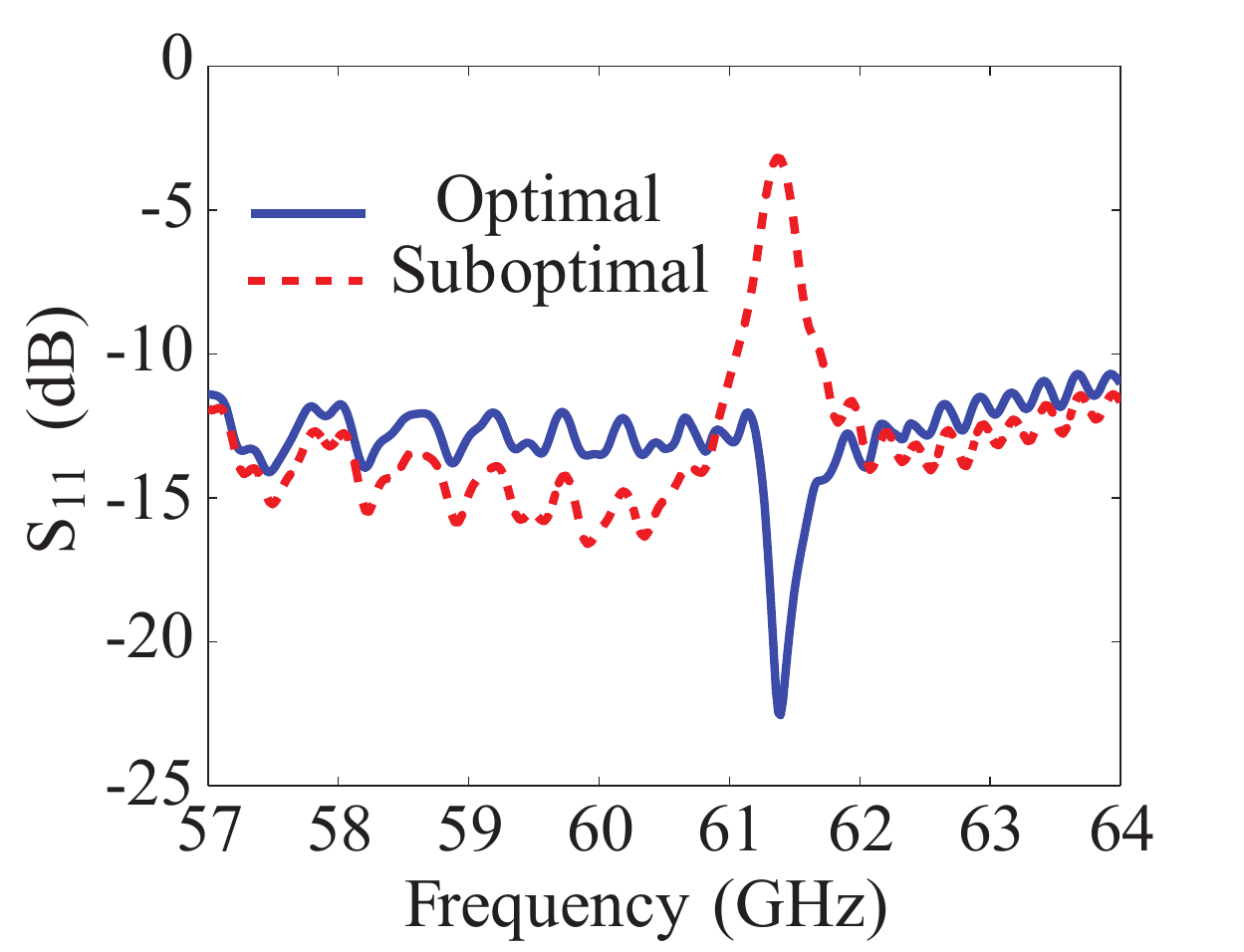}\caption{}
     \end{subfigure}
 	     \begin{subfigure}[b]{0.493\columnwidth}
    \psfrag{a}[c][c][0.7]{Frequency (GHz)}
	\psfrag{b}[c][c][0.7]{$S_{21}$ (dB)}
	\psfrag{c}[c][c][0.7]{Optimal}
	\psfrag{d}[c][c][0.7]{Suboptimal}
   	 \includegraphics[width=\linewidth]{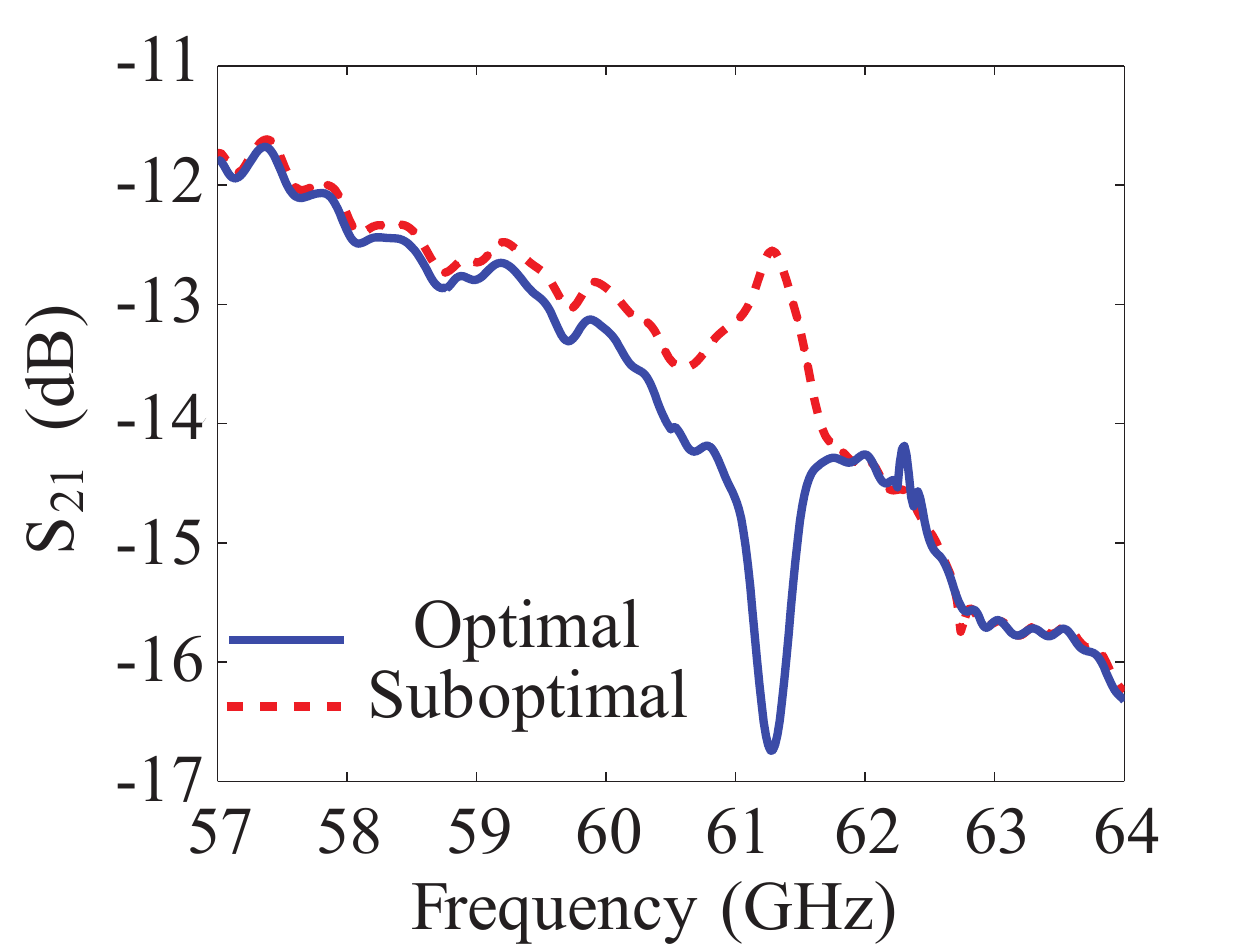}\caption{}
     \end{subfigure}
	\caption{Integrated Side-Fire LWA design. (a) FEM-HFSS cell model of the curved LWA based on substrate integration, showing the unit cell design parameters. Here $d=0.34$ mm, $s_3=1.08$ mm, $\ell_1=3.46$ mm, $\ell_2=0.8$ mm, $\epsilon_r=2.2$ mm, and $s_1=1.1$ mm.  (b) $S_{11}$ and (c) $S_{21}$ of the curved antenna with optimal via $s_2=0.1$ mm compared to suboptimal fully symmetric via $s_2=0$ mm.}\label{Fig:AnlyVShfss}
\end{figure}

With a determined antenna structure and periodic aperture producing a specified per unit length propagation constant $\beta(\omega)$, a numerical model to calculate the radiation and focussing properties is needed to understand and model this structure. Consider a uniform circular LWA as shown in Fig.~\ref{Fig:Analytical}(a), of an angular \textcolor{black}{span} $2\alpha$ and a radius $r$, excited from the left. The field variation along the $z-$axis is assumed to be zero, i.e. a 2D problem is assumed. Each point on this LWA can be assumed to be a \textcolor{black}{line} source $J_z(x', y';\omega)$ which is radiating with a certain phase accumulated along the LWA. Each \textcolor{black}{source} \textcolor{black}{coordinate} follows the following parametric curve
\begin{equation}
x'^2 + y'^2 = r^2.
\end{equation}
\noindent The total radiated electric field in the $x-y$ plane can be analytically expressed as \cite{gomez2010frequency}
\begin{equation}\label{Eq:AnalyticalField}
\mathbf{E}(x,y) = \int_{-r\sin\alpha}^{r\sin\alpha} J_z(x', y')G(x', y';x,y)dx'~\hat{z}
\end{equation}
\noindent where $G(x', y';x,y)$ is the Green's function. This integral represents the contribution of each \textcolor{black}{source} at the observation point $(x,y)$ along the entire extent of the antenna. For a 2D problem \textcolor{black}{we have} a Hankel's function of the second kind, given by
\begin{equation}\label{Eq:Green}
G(x', y';x,y) = -\frac{\omega \mu_0}{4}H_0^{(2)}(k\rho),
\end{equation}
\noindent with
\begin{equation}
\rho(x', y';x,y) = \sqrt{(x-x')^2+(y - y')^2}~\hat{\rho},
\end{equation} 
\noindent where $\rho$ represents the vector from the source to the observation point, as illustrated in Fig.~\ref{Fig:Analytical}(a). Finally, the position dependent current source is assumed to take the following form:
\begin{align}\label{Eq:Current}
J_z(x',y') \approx e^{-j\beta s}
\end{align} 
\noindent where the length $s = r\theta$ represents the arc length from $\{(x_1,y_1) - (x', y')\}$ and can be easily shown to be
\begin{align}
s  = r\cos^{-1}\left\{1 - \frac{(x'-x_1)^2 + (y'-y_1)^2}{2r^2}\right\}
\end{align} 
\noindent Therefore, substituting \eqref{Eq:Green}, \eqref{Eq:Current} inside \eqref{Eq:AnalyticalField} leads to the analytical form of the field anywhere in the observation area.

Next, an integrated spectrum analyzer based on a side-fire LWA configuration of Fig.~\ref{Fig:SidefireLWA} with a 1 GHz spatial resolution operating from 57-64 GHz was designed and simulated in FEM-HFSS, as shown in Fig.~\ref{Fig:AnlyVShfss}(a). To be compatible with standard PCB fabrication, the vertical solid walls of Fig.~\ref{Fig:SidefireLWA} are realized using a metallized via array as typically done in SIW implementations, and a curvature is introduced. An isolated via is used to emulate a notch and its precise placement is obtained to close the stop-band following the analysis of Fig.~\ref{Fig:SidefireLWA} to obtain the \textcolor{black}{required} asymmetry. Fig.~\ref{Fig:AnlyVShfss}(b,c), further shows the simulated S-parameters for the case of an optimal and \textcolor{black}{suboptimal} asymmetry, clearly showing broadband matching within the bandwidth of interest, i.e. 57-64 GHz.

Next, an analytical model is built for this side-fire configuration, where its propagation constant $\beta(\omega)$ corresponding to the $n=-1$ space harmonic is approximated as

\begin{equation}\label{Eq:beta}
\beta(\omega) \approx \sqrt{\left(\frac{\omega}{c}\right)^2\epsilon_r - \left(\frac{\pi}{a}\right)^2} - \frac{2\pi}{p}.
\end{equation}

\noindent This analytical result closely approximates the full-wave computed dispersion relation extracted from \textcolor{black}{an eigenmode} simulation of a single cell, and is thus used \textcolor{black}{due} to its simplicity to estimate beam-scanning of the structure in the near-field region of the antenna. Using  \eqref{Eq:Green}, \eqref{Eq:Current} and \eqref{Eq:beta} in \eqref{Eq:AnalyticalField}, the radiated near-fields were calculated along a line at the focal distance $r$, and compared with those obtained using \textcolor{black}{the} full-wave HFSS model, as shown in Fig.~\ref{Fig:Analytical}(b-d) for several frequencies. An excellent match is seen for all frequencies which successfully validates the model. While the HFSS model is rigorously correct, it is electrically large and computationally expensive. The analytical model on the other hand, captures the fundamental behavior of the structure using a much simpler model based on array of line sources, while providing a reasonably good accuracy. It thus represents an ideal method for an efficient design and fast analysis of a side-fire LWA based spectrum analyzer.

\newcolumntype{C}{>{\centering\arraybackslash}X}
\setlength{\extrarowheight}{1pt}

\begin{figure}[htbp]
	\centering
          \begin{subfigure}[b]{0.95\columnwidth}
         \centering
   	 \includegraphics[width=\linewidth]{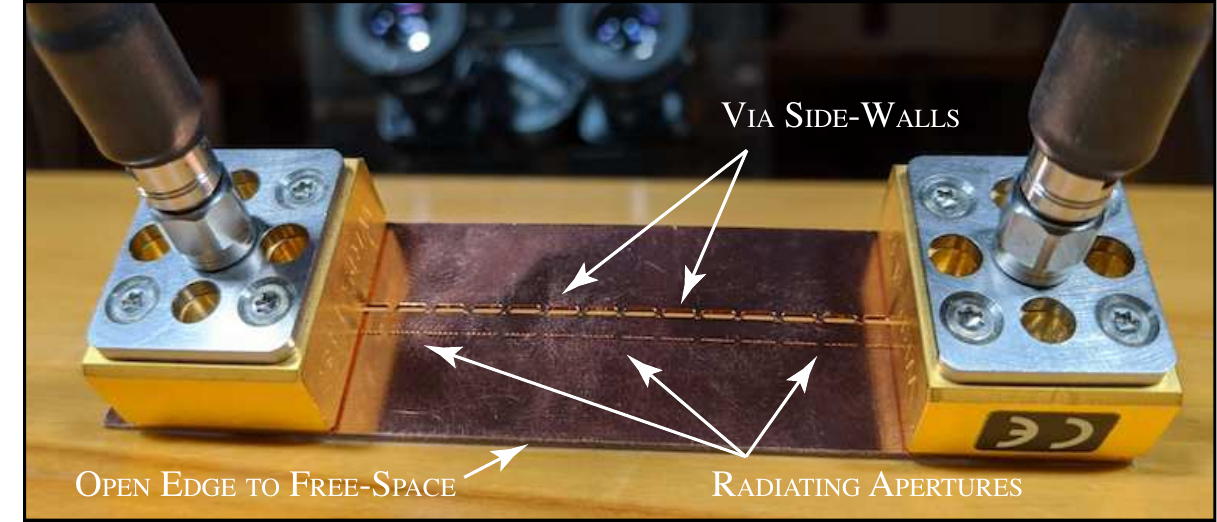}\caption{}
     \end{subfigure} 
 	     \begin{subfigure}[b]{0.493\columnwidth}
         \centering
\psfrag{a}[c][c][0.7]{Frequency (GHz)}
\psfrag{b}[c][c][0.7]{$S_{11}$ (dB)}
\psfrag{c}[c][c][0.50]{Measured}
\psfrag{d}[c][c][0.5]{Arlon}
\psfrag{e}[c][c][0.5]{Characterized}
 	\includegraphics[width=\linewidth]{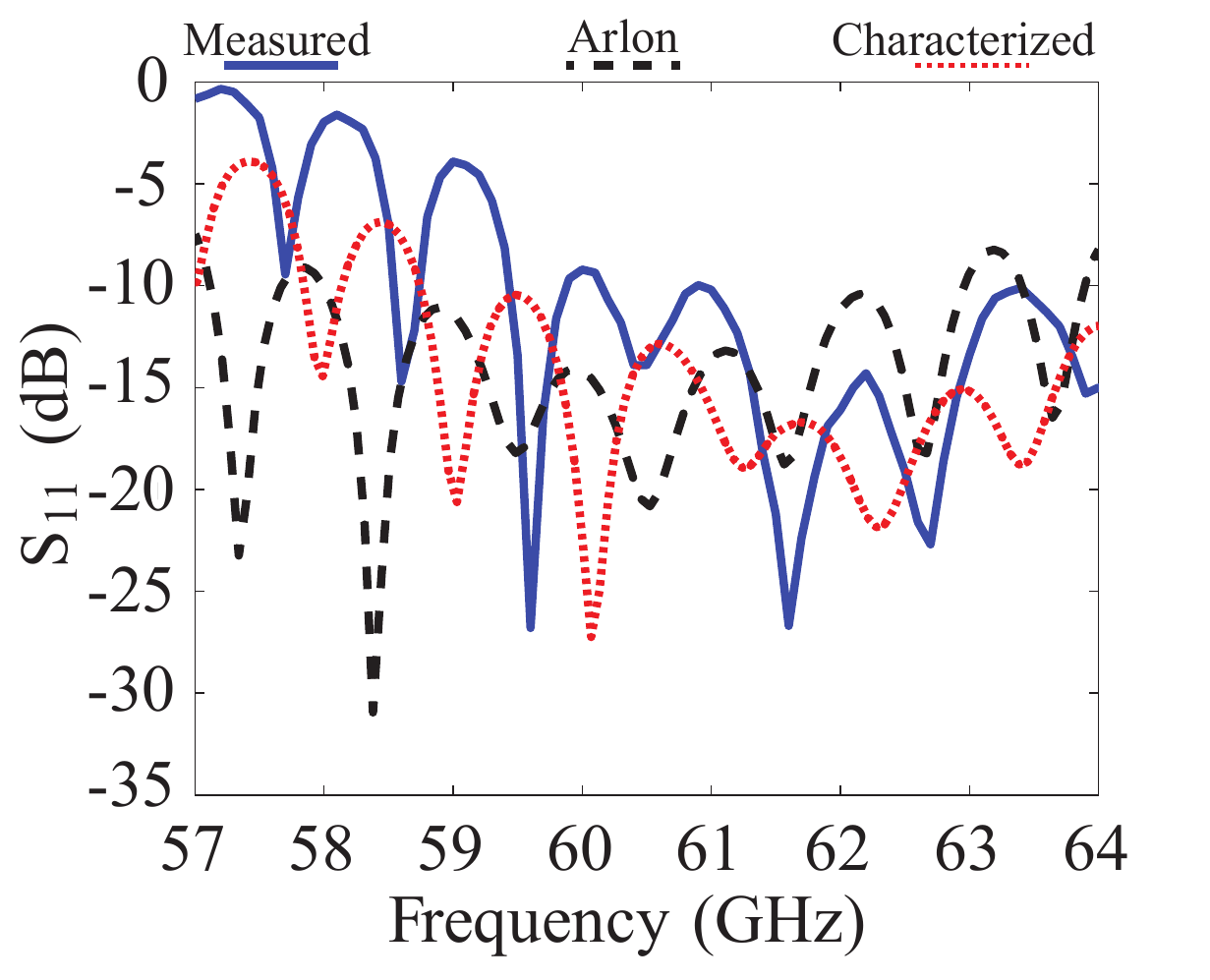}\caption{}
    	\end{subfigure}   
     \begin{subfigure}[b]{0.493\columnwidth}
         \centering
\psfrag{a}[c][c][0.7]{Frequency (GHz)}
\psfrag{b}[c][c][0.7]{$S_{21}$ (dB)}
\psfrag{c}[c][c][0.5]{Measured}
\psfrag{d}[c][c][0.5]{Arlon}
\psfrag{e}[c][c][0.5]{Characterized}
 	\includegraphics[width=\linewidth]{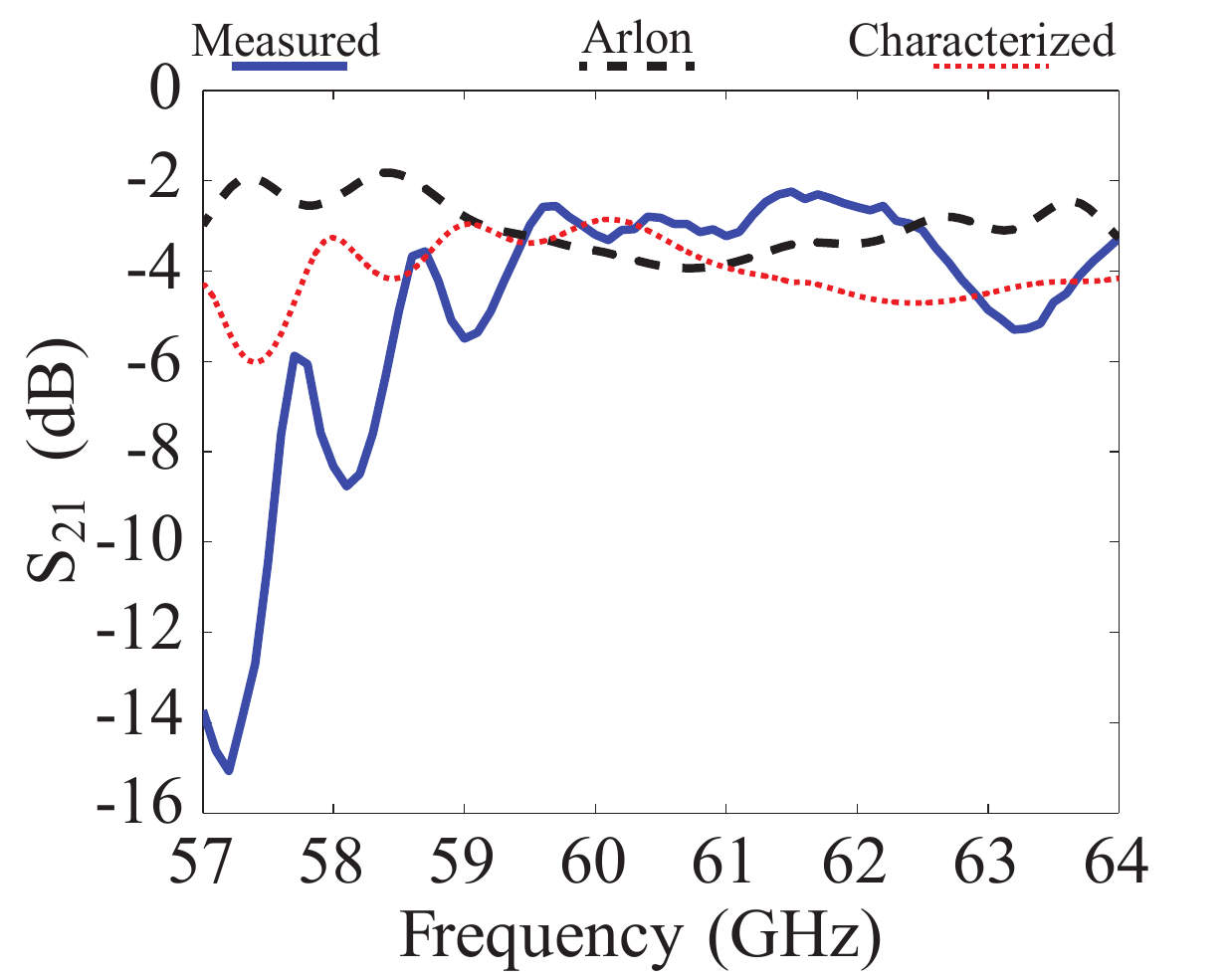}\caption{}
    	\end{subfigure}
    	\begin{subfigure}[b]{0.493\columnwidth}
         \centering
\psfrag{a}[c][c][0.7]{Frequency (GHz)}
\psfrag{b}[c][c][0.7]{Gain (dB)}
\psfrag{c}[c][c][0.7]{Characterized}
\psfrag{d}[c][c][0.7]{Arlon}
\psfrag{e}[c][c][0.5]{$f_{0,\text{des.}} = 58.9$~GHz}
\psfrag{f}[c][c][0.5]{$f_{0,\text{exp.}} = 60.7$~GHz}
\psfrag{g}[c][c][0.5]{shift $ \Delta f \approx 1.6$~GHz}
 	\includegraphics[width=\linewidth]{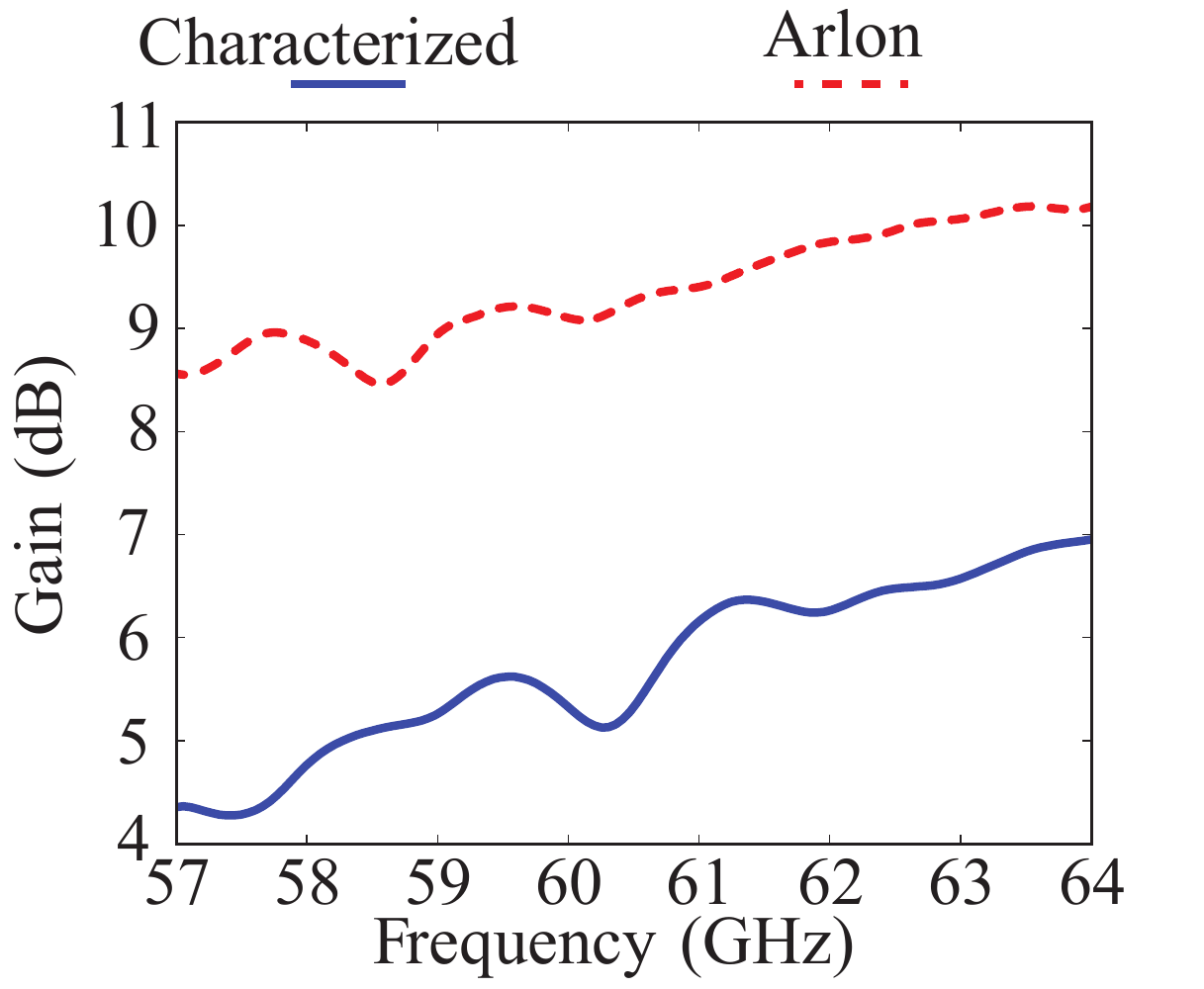}\caption{}
    	\end{subfigure}    
     \begin{subfigure}[b]{0.493\columnwidth}
         \centering
\psfrag{a}[c][c][0.7]{$\phi$ ($\degree$)}
\psfrag{b}[c][c][0.7]{Gain (dB)}
\psfrag{c}[c][c][0.5]{57 GHz}
\psfrag{d}[c][c][0.5]{64 GHz}
 	\includegraphics[width=\linewidth]{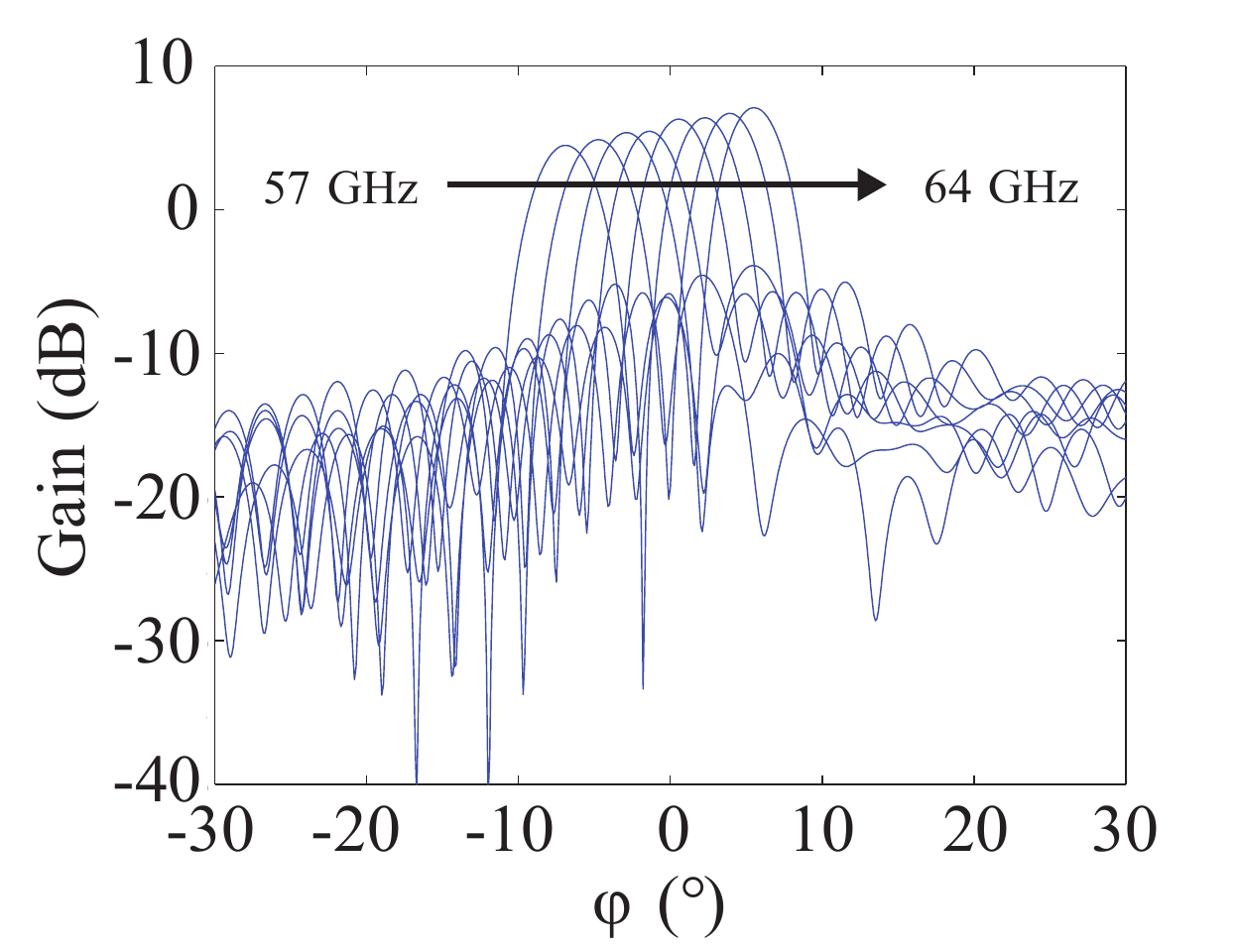}\caption{}
    	\end{subfigure}         
	\caption{Experimental demonstration of a straight side-fire LWA radiating into free-space. (a) Fabricated antenna prototype ($N=15$ cells). (b) $S_{11}$ of the antenna in FEM-HFSS with Arlon and the resonator characterized material compared to the VNA measurement. Here the stopband is closed at 61 GHz (broadside). (c) $S_{21}$ of the antenna in FEM-HFSS with Arlon and the resonator characterized material compared to the VNA measurement. (d) Simulated gain as a function of frequency for the antenna with the characterized material and Arlon DiClad 880. (e) Radiation patterns of the simulated antenna with characterized material in 1 GHz steps over the band of interest.}\label{Fig:StraightSideFire}
\end{figure}

\section{Experimental Demonstration}

In this section, two experimental demonstrations will be presented: 1) A straight integrated side-fire LWA radiating into free space and, b) a curved side-fire antenna radiating inside an integrated PPW acting as a mm-\textcolor{black}{wave} spectrum analyzer.

\begin{figure*}[htbp]%
\centering
\begin{subfigure}{1.2\columnwidth}
\centering
\psfrag{a}[c][c][0.7]{Port 3}
\psfrag{b}[c][c][0.7]{Port 10}
\psfrag{c}[c][c][0.7]{Port 1}
\psfrag{d}[c][c][0.7]{Port 2}
\psfrag{e}[c][c][0.7]{$2\alpha$}
\psfrag{f}[c][c][0.7]{$\Delta y = 57$~cm}
\psfrag{g}[c][c][0.7]{$\Delta x = 33$~cm}
\psfrag{h}[c][c][0.7]{$\theta_h$}
\psfrag{i}[c][c][0.7]{$a_{ap}$}
\psfrag{j}[c][c][0.7]{$a$}
\psfrag{k}[c][c][0.7]{$b$}
\psfrag{l}[c][c][0.7]{$L_1$}
\psfrag{m}[c][c][0.7]{$L_2$}
\psfrag{o}[c][c][0.7]{$g$}
\psfrag{p}[c][c][0.7]{$L_4$}
\psfrag{q}[c][c][0.7]{$a$}
\psfrag{r}[c][c][0.7]{$L_3$}
\psfrag{A}[c][c][0.7]{\shortstack{Arlon DiClad 880\\ $t=0.762$~mm,~$\epsilon_r = 2.2$,~$\tan\delta = 0.0009$}}
\includegraphics[width=\columnwidth]{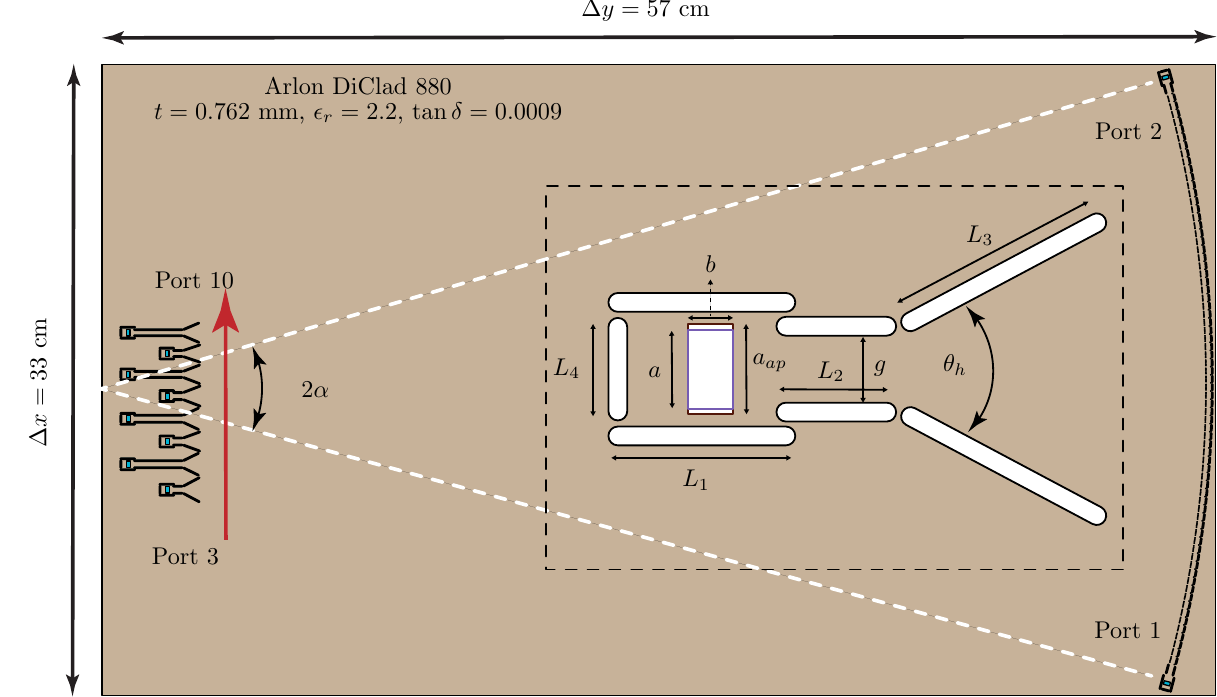}%
\caption{}%
\label{subfigb}%
\end{subfigure}\hfill%
\begin{subfigure}{.7\columnwidth}
\centering
	\psfrag{a}[c][c][0.7]{Port Number}
	\psfrag{b}[c][c][0.7]{Frequency (GHz)}
	\psfrag{c}[c][c][0.85]{Transmission, $|S_{n,1}|$ (dB)}
\includegraphics[width=\columnwidth]{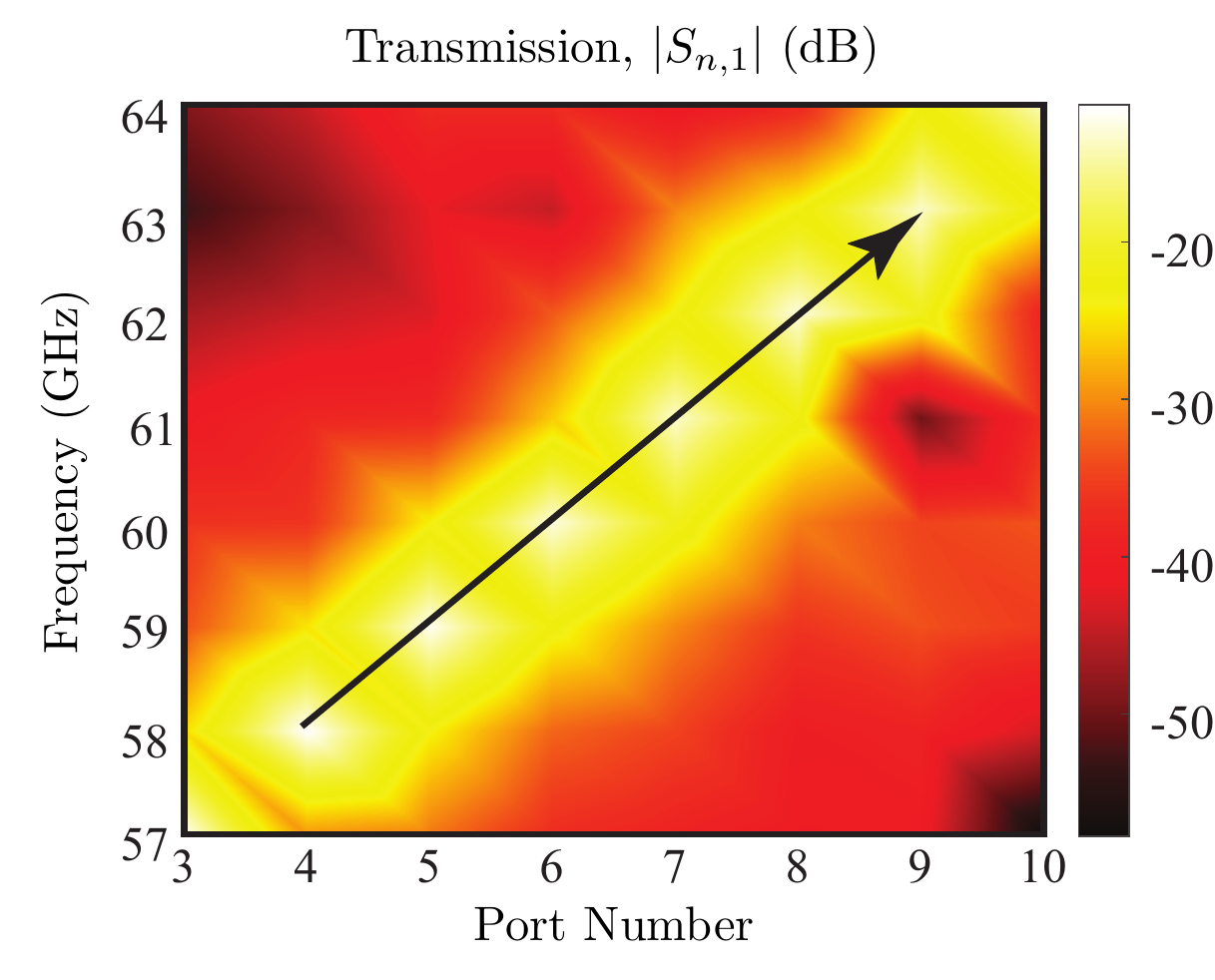}%
\caption{}%
\label{subfigb}%
\end{subfigure}\hfill%
\begin{subfigure}{2\columnwidth}
\centering
\psfrag{a}[c][c][0.7]{$x$ (mm)}
\psfrag{b}[c][c][0.7]{$y$ (mm)}
\psfrag{A}[c][c][0.7]{\color{white}$n=0$}
\psfrag{B}[c][c][0.7]{\color{white}$n=-2$}
\psfrag{C}[c][c][0.7]{\color{white}$n=-1$}
\psfrag{D}[l][c][0.7]{\color{white}{$f=57$~GHz}}
\psfrag{E}[l][c][0.7]{\color{white}{$f=59$~GHz}}
\psfrag{F}[l][c][0.7]{\color{white}{$f=61$~GHz}}
\psfrag{G}[l][c][0.7]{\color{white}{$f=63$~GHz}}
\includegraphics[width=\columnwidth]{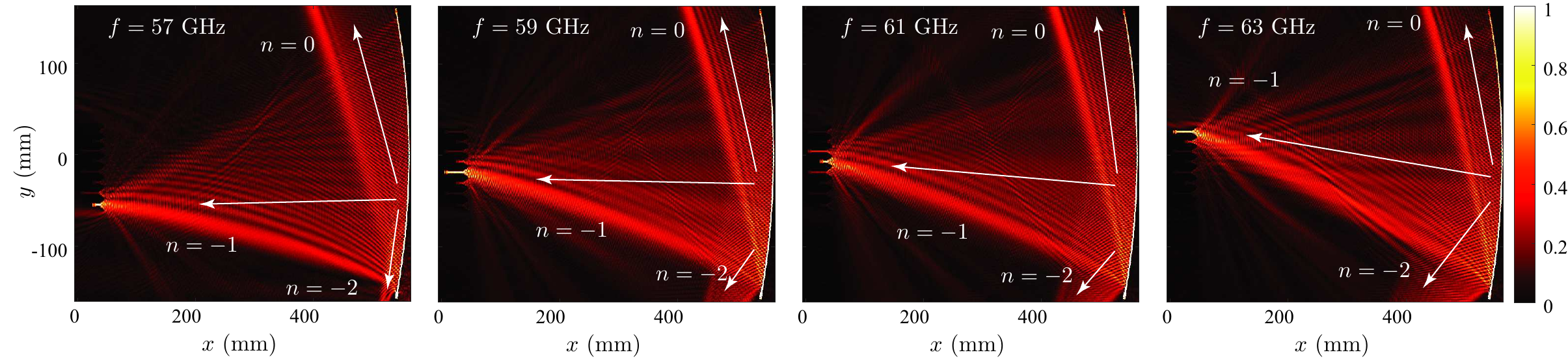}%
\caption{}%
\label{subfiga}%
\end{subfigure}\hfill%
\caption{FEM-HFSS modeling of the integrated side-fire LWA based spectrum analyzer. (a) Schematic model showing an array of integrated horn receivers with $\Delta x=570$ mm and $\Delta y=323$ mm, and the curved side-fire LWA consisting of $N=72$ unit cells. (b) Interpolated transmission into the receiving horn ports. (c) Normalized $|E_z|$ plots from HFSS of the spectrum analyzer showing the beam focusing. All horns and aperture dimensions shown in Tab.~1.}\label{Eq:FabSA}
\label{figabc}
\end{figure*}

\subsection{Straight Side-Fire LWA}

\textcolor{black}{Fig}.~\ref{Fig:StraightSideFire}(a) shows a photograph of a 15 cell straight side-fire LWA operating around the 60 GHz band, following the configuration of Fig.~\ref{Fig:AnlyVShfss}(a), except with a straight cell. The matching via is optimized in simulations to close the stop-band using the nominal permittivity of the dielectric specified by the manufacturer. A transition is further designed to excite the LWA antenna using a standard WR-15 waveguide, as shown in the input and output ends of the LWA structure. One side of the antenna is left open to free-space with a width of the PPW controlling the amount of radiation along the $-y$ direction. Fig.~\ref{Fig:StraightSideFire}(b-c) shows the measured S-parameters compared with FEM-HFSS. While good matching is observed in measurements as a result of optimal via placement, a slight increase in the reflection is also observed along with appreciable mismatch with the simulated response. This mismatch between simulation and measurement can be attributed to the the unknown material properties at 60 GHz, and the tolerances in fabrication, \textcolor{black}{especially} considering that the response is particularly sensitive to \textcolor{black}{the} matching via location.

\textcolor{black}{To} estimate the material properties, \textcolor{black}{a straight waveguide resonator} was also fabricated, which \textcolor{black}{was} then used to estimate the effective dielectric constants of the Arlon DiClad 880 material used in fabrication. The resonator results are presented in the Appendix, and the extracted dielectric permittivity and loss tangent are found to be $\approx 2.08$ and $0.003$, respectively, compared to the design values of $\epsilon_r=2.2$ and $\tan\delta = 0.0009$. The fabricated straight side-fire LWA \textcolor{black}{was then} re-simulated with the new material parameters, and the new S-parameters are shown in Fig.~\ref{Fig:StraightSideFire}(b-c), improving the agreement with the measurements. Importantly, a frequency shift of almost 1.6~GHz is observed, with a new broadside frequency of $f_0 = 60.7$~GHz (observed in HFSS with peak radiation at $\theta=0^\circ$). The resulting frequency dependent gain and the simulated radiation patterns are shown in Fig.~\ref{Fig:StraightSideFire}(d-e) showing the seamless radiation from backward to forward regions including broadside at 60.7 GHz, as expected. Due to an actual higher loss tangent of the dielectric, the gain is seen to be significantly dropped with a slight drop at broadside, while maintaining a monotonically increasing trend, as expected.

\subsection{Integrated Spectrum Analyzer}

Next, an integrated spectrum analyzer based on a side-fire LWA will be demonstrated. Fig.~\ref{Eq:FabSA}(a) shows the general schematic of the proposed system. A curved side-fire LWA of Fig.~\ref{Fig:AnlyVShfss} with larger number of cells and input/output transitions to standard WR-15 waveguides is placed on the far right of the dielectric layer, where Port 1 is intended as the input to the system. the leakage from the side apertures is maintained inside the PPW structure, due to the beam-scanning property, the spatial-spectral decomposition is achieved in the $x-y$ plane, with broadside along $-x$ and $x<0$ and $x>0$ as the backward and forward radiation regions, respectively. The system is designed to operate between 57-64 GHz, with a frequency resolution of 1~GHz. 

\begin{figure}[htbp]
	\centering 
       \begin{subfigure}[b]{0.49\columnwidth}
         \centering
	\psfrag{a}[c][c][0.7]{$r$ (cm)}
	\psfrag{b}[c][c][0.7]{Scan Range (mm)}
 	\includegraphics[width=\linewidth]{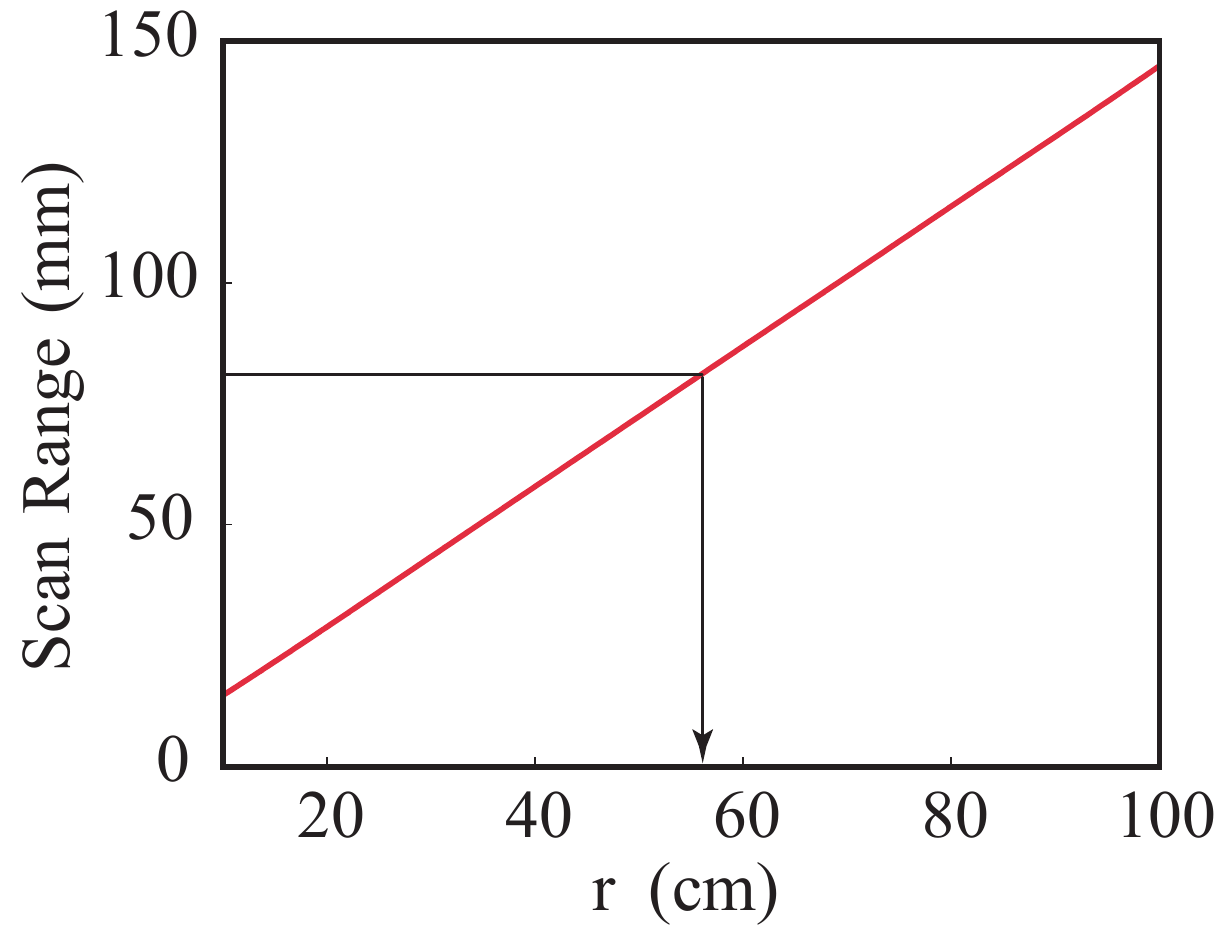}\caption{}
    	\end{subfigure}  
 	     \begin{subfigure}[b]{0.465\columnwidth}
         \centering
	\psfrag{a}[c][c][0.7]{$2\alpha$ ($\degree$)}
	\psfrag{b}[c][c][0.7]{3-dB Beamwidth (mm)}
	\psfrag{c}[c][c][0.7]{Horn Beamwidth}
   	 \includegraphics[width=\linewidth]{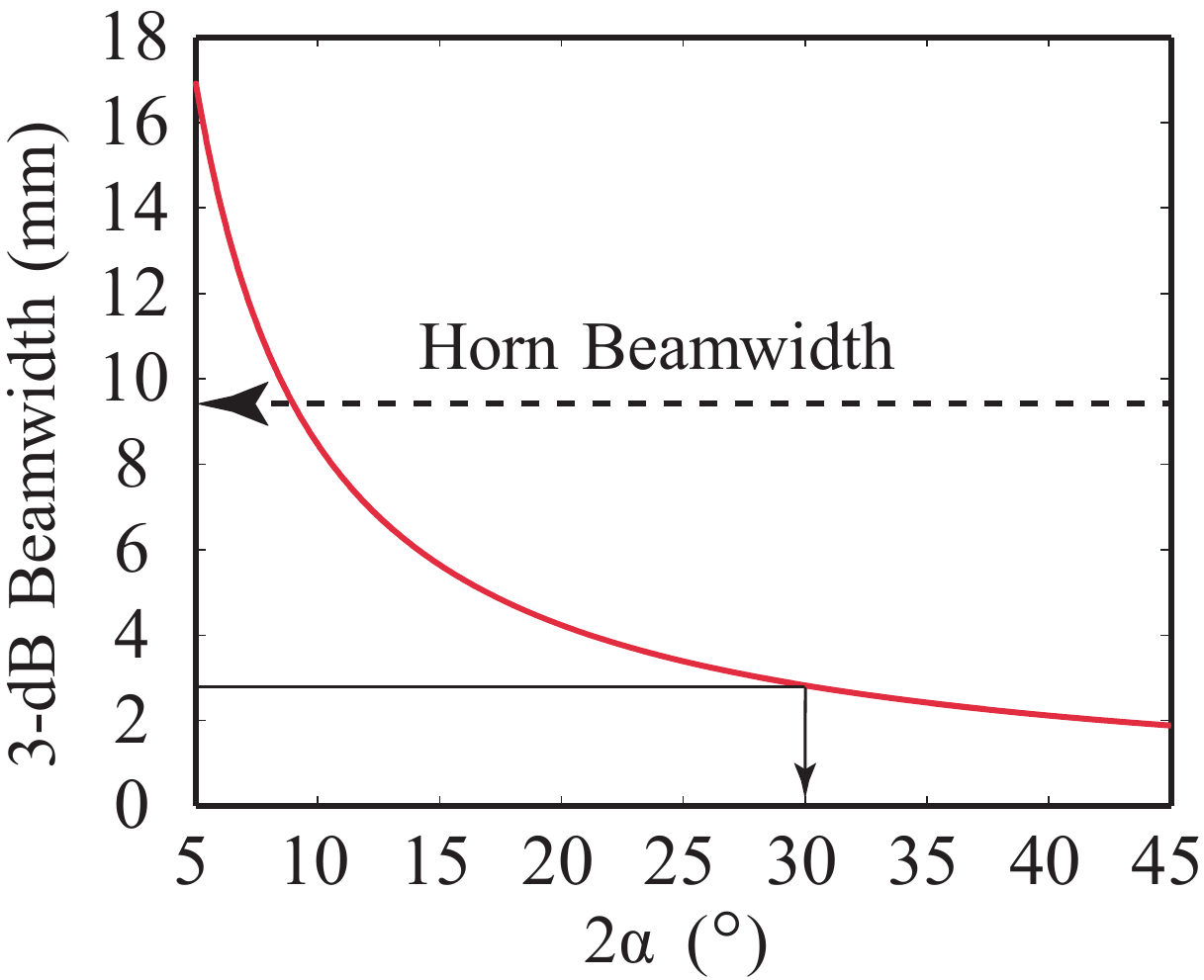}\caption{}
     \end{subfigure}
	\caption{Procedure to determine the radius of curvature of the curved side-fire LWA and the arc length $2\alpha$ to achieve a frequency resolution of 1 GHz within the operating bandwidth of 57 GHz - 64 GHz. (a) Beam scan variation with $r$ at $f=60.7$ GHz. b) Beamwidth as a function of angular span of the LWA for $r=56$ cm.}\label{Fig:parameterization}
\end{figure}

To capture the power radiated along different angles, an integrated horn array is chosen for the sake of a proof of concept experiment (detailed dimensions are provided in Tab.~I). To achieve a 1 GHz resolution over the 57-64 GHz band, 8 horns for each 1 GHz frequency step are needed where the transmission in the form of S-parameters can be measured. The receivers are waveguide fed, and thus must be spaced to allow room for \textcolor{black}{WR-15} flanges, which have a diameter of 19.05 mm. The unknown physical dimensions to achieve this resolution are: the horn spacing, radius of curvature and the angular span of the side-fire antenna. To determine these parameters, a simple parametric analysis of Fig.~\ref{Fig:Analytical}(a) is performed. Fig.~\ref{Fig:parameterization}(a) shows the maximum scan range (using peak amplitudes of the focussed beam) over 57-64 GHz for each $r$ at a fixed frequency $f$, for instance. Knowing the required separation between the horn receivers of $\approx 12$~mm, $r$ is set at $56$ cm. Next, Fig.~\ref{Fig:parameterization}(b) shows the proper antenna arc length for a given radius of 56 cm to produce a sufficiently narrow beam, less than the 3-dB beamwidth of the horn. It was then set at $2\alpha = 30^\circ$. This would then allow 1 GHz resolution while preventing electromagnetic coupling to neighbouring receivers which could compromise the spectral decomposition.

\begin{figure}[htbp]
\centering
\psfrag{a}[c][c][0.7]{$\beta$ (rad/m)}
\psfrag{b}[c][c][0.7]{Frequency (GHz)}
\psfrag{c}[c][c][0.7]{$n=0$}
\psfrag{d}[c][c][0.7]{$n=-1$}
\psfrag{e}[c][c][0.7]{$n=-2$}
\psfrag{f}[c][c][0.7]{\shortstack{Light-line\\ $\beta c/2\pi\sqrt{\epsilon_r}$}}
\psfrag{g}[c][c][0.7]{Broadside, $f_0$}
\psfrag{h}[c][c][0.7]{Cut-off region}
\psfrag{i}[c][c][0.7]{Evanescent}
\psfrag{j}[c][c][0.7]{57-64 GHz}
\includegraphics[width=0.8\columnwidth]{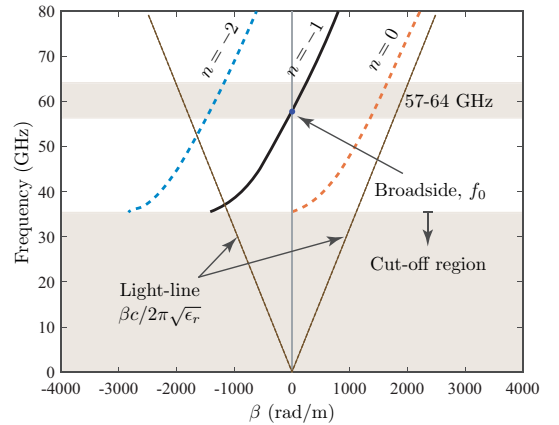}
\caption{Dispersion diagram of the conformal leaky-wave antenna illustrating the issue of multi-beam radiation within the fast-wave region.}\label{Fig:DispersionDiag}
\end{figure}

Fig.~\ref{Eq:FabSA}(b) shows the FEM-HFSS simulated power transmission between the input port 1 and the various horn receiver ports. A linear frequency scanning can be seen across the ports, with clear separation between the power received by adjacent ports. Direct incidence of the beam on a port results in a transmission of more than -15 dB. Neighbouring ports in such a case would have less than -30 dB in transmission. To this end, a nominal frequency resolution of 1 GHz is readily shown. Fig.~\ref{Eq:FabSA}(c) further shows the near-field distribution inside the PPW for various frequencies, which are being received by different horn receivers demonstrating the frequency scanning behavior of the side-fire \textcolor{black}{antenna}. In addition, extra radiation beams are also observed in the forward region, which are analogous to grating lobes in regular antenna arrays.

To understand the nature of these grating lobes, the dispersion diagram of the side-fire LWA must be inspected as shown in Fig.~\ref{Fig:DispersionDiag}. For the designed substrate permittivity and antenna dimensions (curvature, width, aperture period), more than one fast-wave harmonic appears in the radiative region. Since radiation is in the substrate, the slope of the light line is increased by \textcolor{black}{a factor} of $1/\sqrt{\epsilon_r}$ compared to that in free-space. As a result, the fundamental mode does not leave the radiative region since it is parallel to the light line. Therefore, the fundamental mode \emph{always} remain inside the fast-wave region and radiates in the forward direction. At the same time, other spatial harmonics, particularly $n=\{-1,-2\}$ harmonics, also radiate within the band of interest~\cite{paulotto2008full} \cite{yang2010full}. They appear as  extra beams in the full-wave simulated fields of Fig.~\ref{Eq:FabSA}(c), where $n=-1$ are the ones \textcolor{black}{captured} by the horn receivers. While these spurious grating lobes represent unwanted power loss through radiation, the dominant power is still found to be contained in the $n=-1$ harmonic sufficient for the current purpose of spectrum measurement.

\begin{figure}[htbp]
	\centering
          \begin{subfigure}[b]{0.95\columnwidth}
         \centering
 	\includegraphics[width=\linewidth]{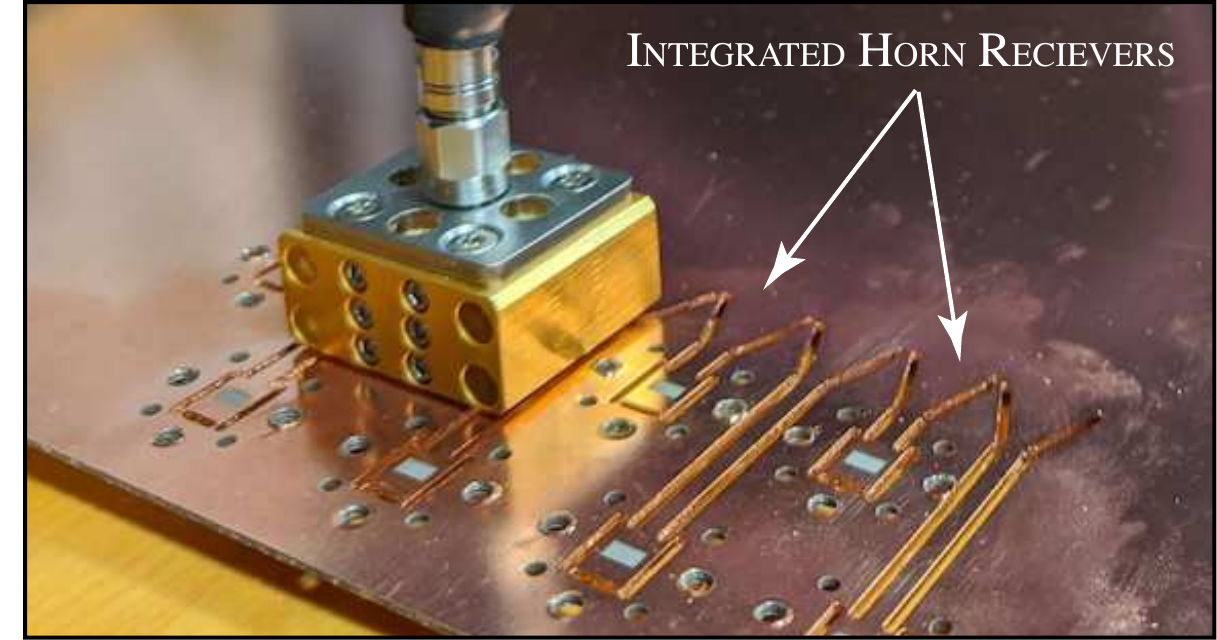}\caption{}\hfill
    	\end{subfigure}  
     %
     \begin{subfigure}[b]{0.95\columnwidth}
         \centering
   	 \includegraphics[width=\linewidth]{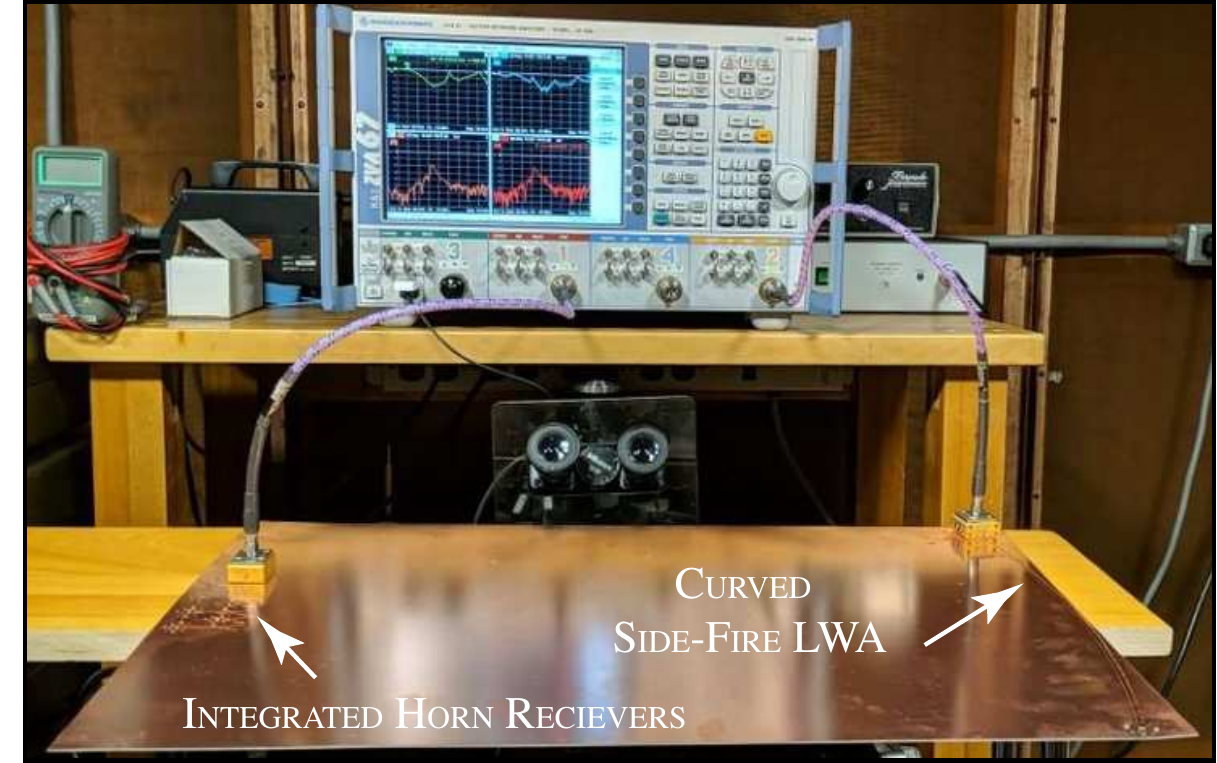}\caption{}
     \end{subfigure}
               \begin{subfigure}[b]{0.493\columnwidth}
        \centering
        \psfrag{a}[c][c][0.7]{Frequency (GHz)}
		\psfrag{b}[c][c][0.7]{$S_{11}$ (dB)}
		\psfrag{c}[c][c][0.5]{Measured}
		\psfrag{d}[c][c][0.5]{Arlon}
		\psfrag{e}[c][c][0.5]{Characterized}
 	\includegraphics[width=\linewidth]{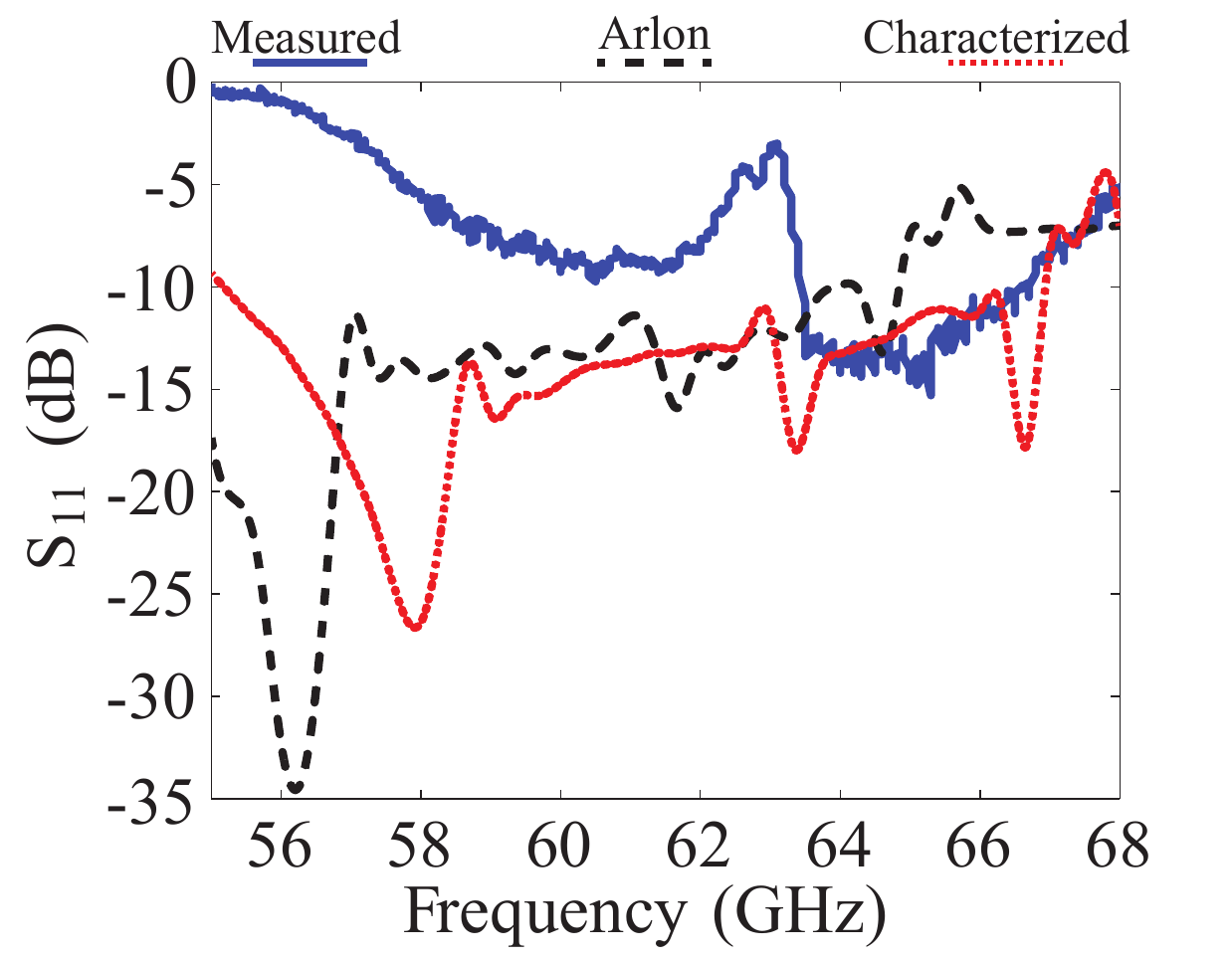}\caption{}
    	\end{subfigure}  
	     \begin{subfigure}[b]{0.493\columnwidth}
         \centering
        \psfrag{a}[c][c][0.7]{Frequency (GHz)}
		\psfrag{b}[c][c][0.7]{$S_{21}$ (dB)}
		\psfrag{c}[c][c][0.5]{Measured}
		\psfrag{d}[c][c][0.5]{Arlon}
		\psfrag{e}[c][c][0.5]{Characterized}
   	 \includegraphics[width=\linewidth]{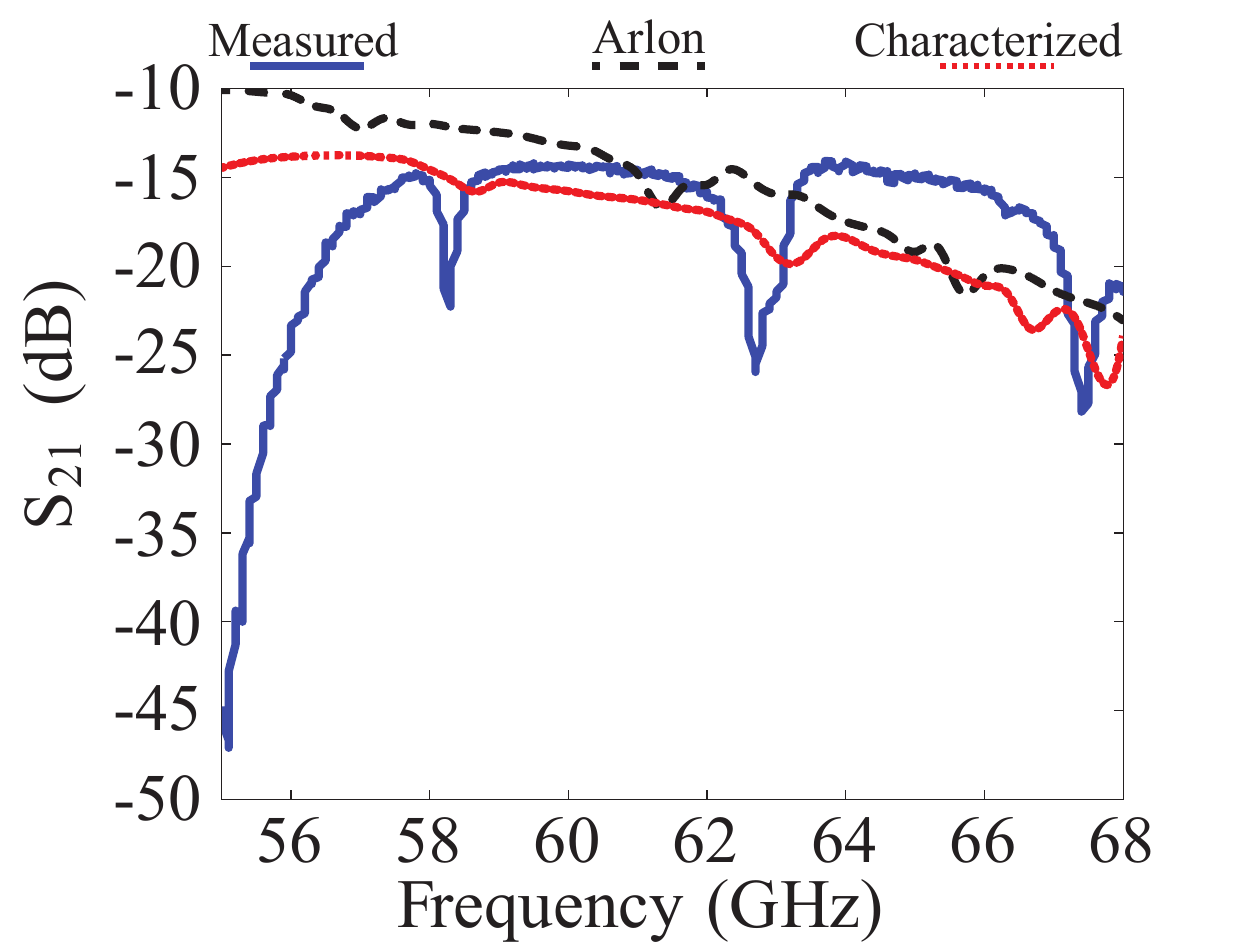}\caption{}
     \end{subfigure} 
	\caption{Experimental setup and results of the fabricated spectrum analyzer. (a) Zoomed in view of the receiver horn array during S-parameter measurement. b) S-Parameter measurement of the VNA. (c) Measured $S_{11}$ and (d) $S_{21}$ of the spectrum analyzer compared to simulation with Arlon DiClad 880, and in simulation fitted material parameters (Appendix A).}\label{Fig:SAPrototype}
\end{figure}

\begin{figure*}[htbp]%
\centering
\begin{subfigure}{.493\columnwidth}
\centering
\psfrag{a}[c][c][0.7]{Frequency (GHz)}
\psfrag{b}[c][c][0.7]{S$_{31}$ (dB)}
\psfrag{c}[l][c][0.5]{Measured}
\psfrag{d}[l][c][0.5]{HFSS-Arlon}
\psfrag{e}[l][c][0.5]{HFSS-Charcterized}
\includegraphics[width=\columnwidth]{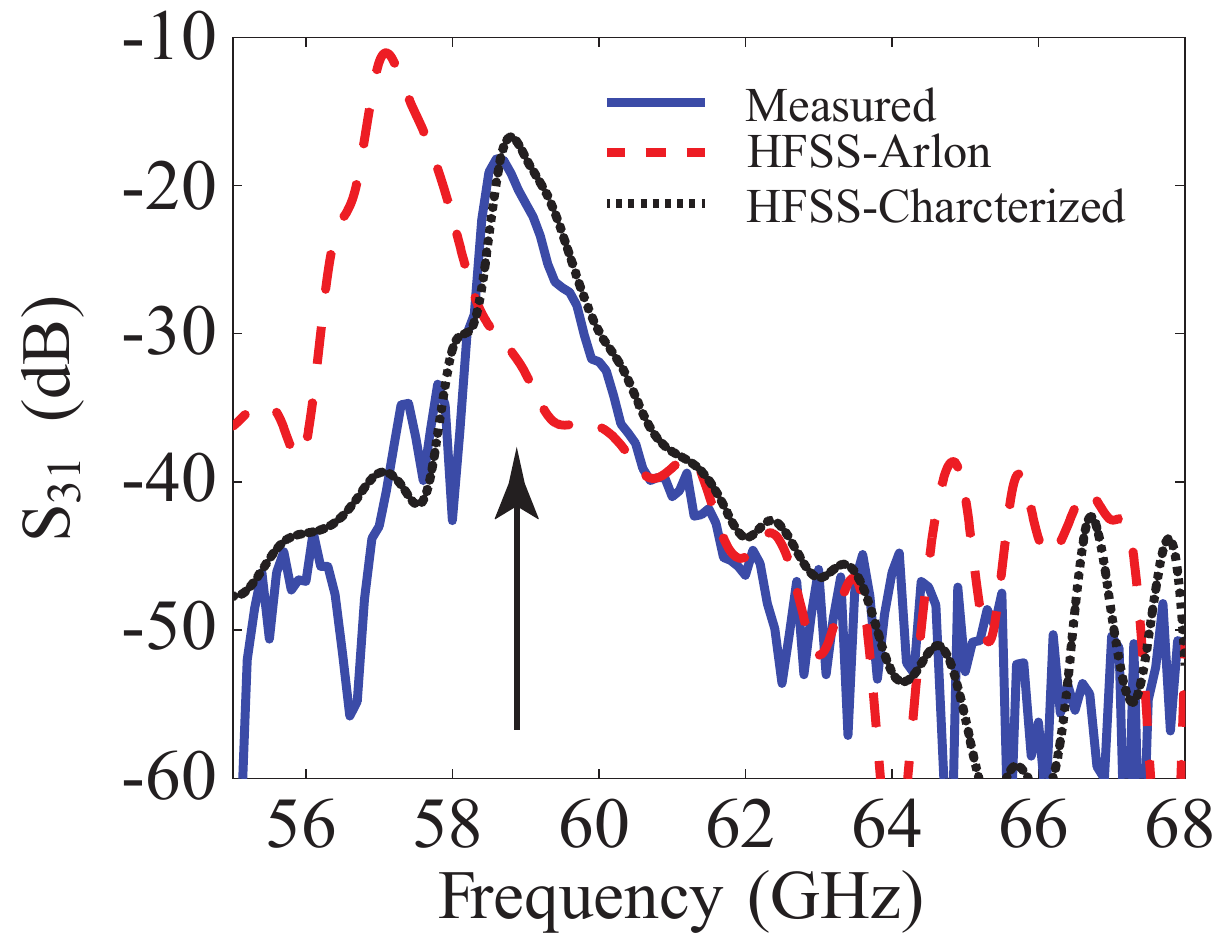}%
\caption{}%
\label{subfigb}%
\end{subfigure}\hfill%
\begin{subfigure}{.493\columnwidth}
\centering
\psfrag{a}[c][c][0.7]{Frequency (GHz)}
\psfrag{b}[c][c][0.7]{S$_{41}$ (dB)}
\includegraphics[width=\columnwidth]{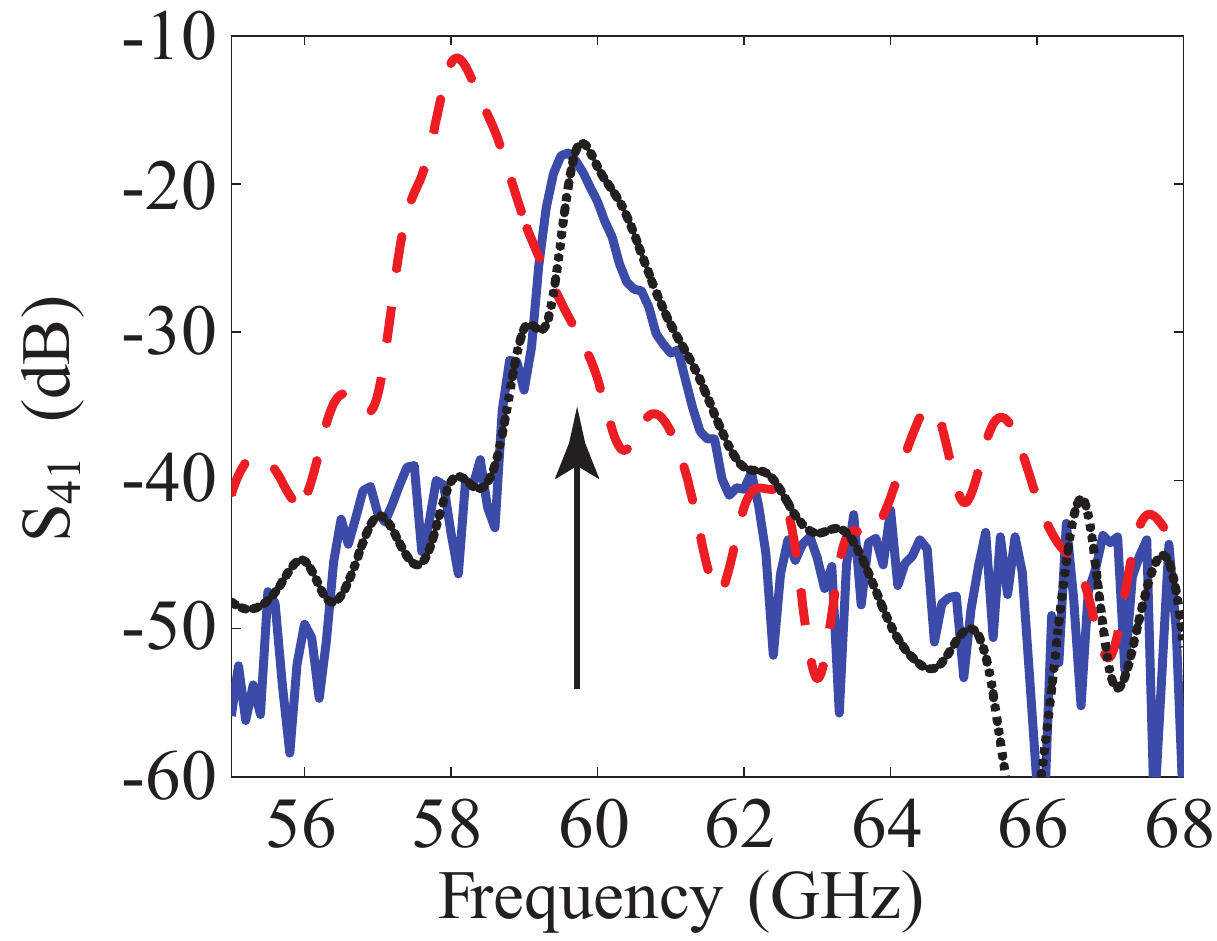}%
\caption{}%
\label{subfigb}%
\end{subfigure}\hfill%
\begin{subfigure}{.493\columnwidth}
\centering
	\psfrag{a}[c][c][0.7]{Frequency (GHz)}
	\psfrag{b}[c][c][0.7]{S$_{51}$ (dB)}

\includegraphics[width=\columnwidth]{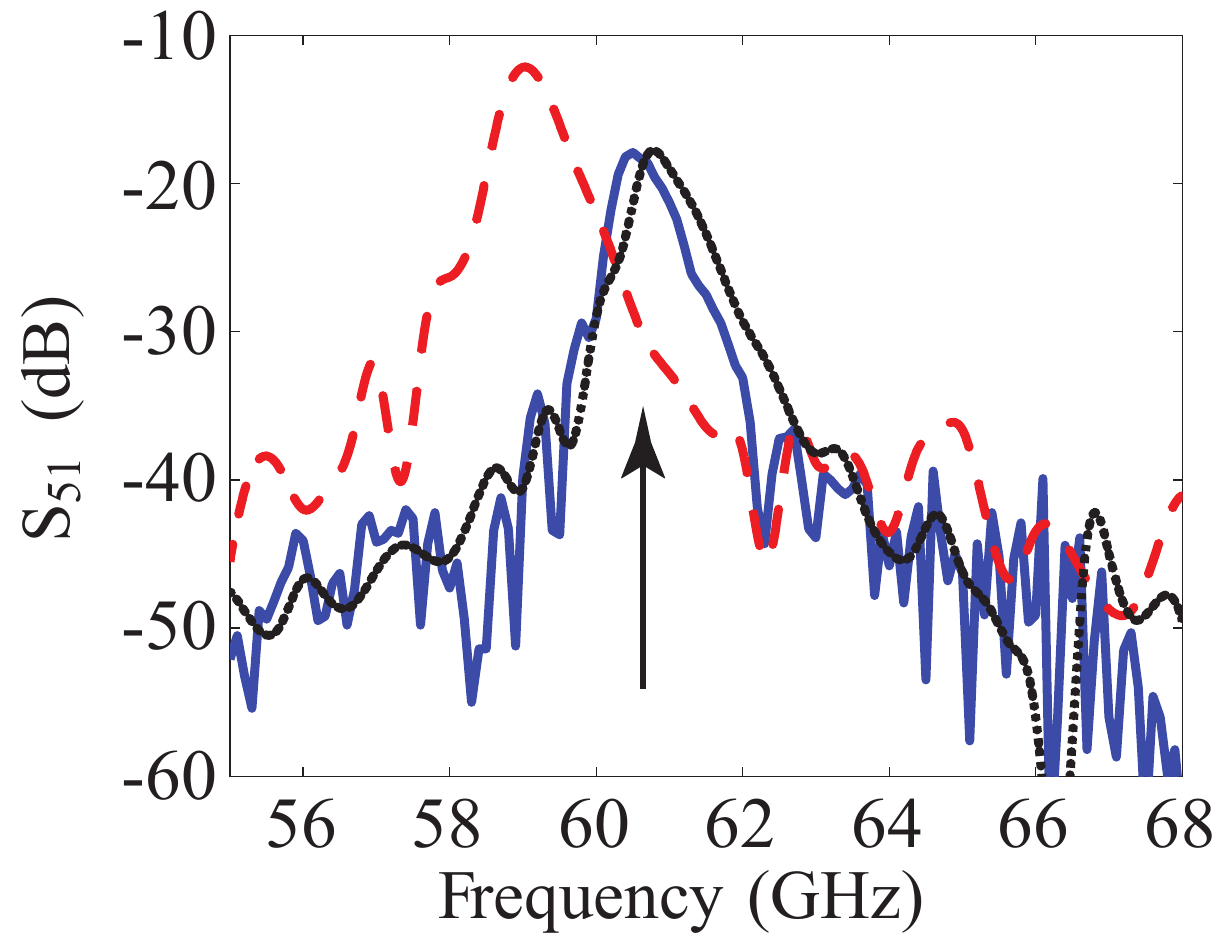}%
\caption{}%
\label{subfigb}%
\end{subfigure}\hfill%
\begin{subfigure}{0.493\columnwidth}
\centering
\psfrag{a}[c][c][0.7]{Frequency (GHz)}
\psfrag{b}[c][c][0.7]{S$_{61}$ (dB)}
\includegraphics[width=\columnwidth]{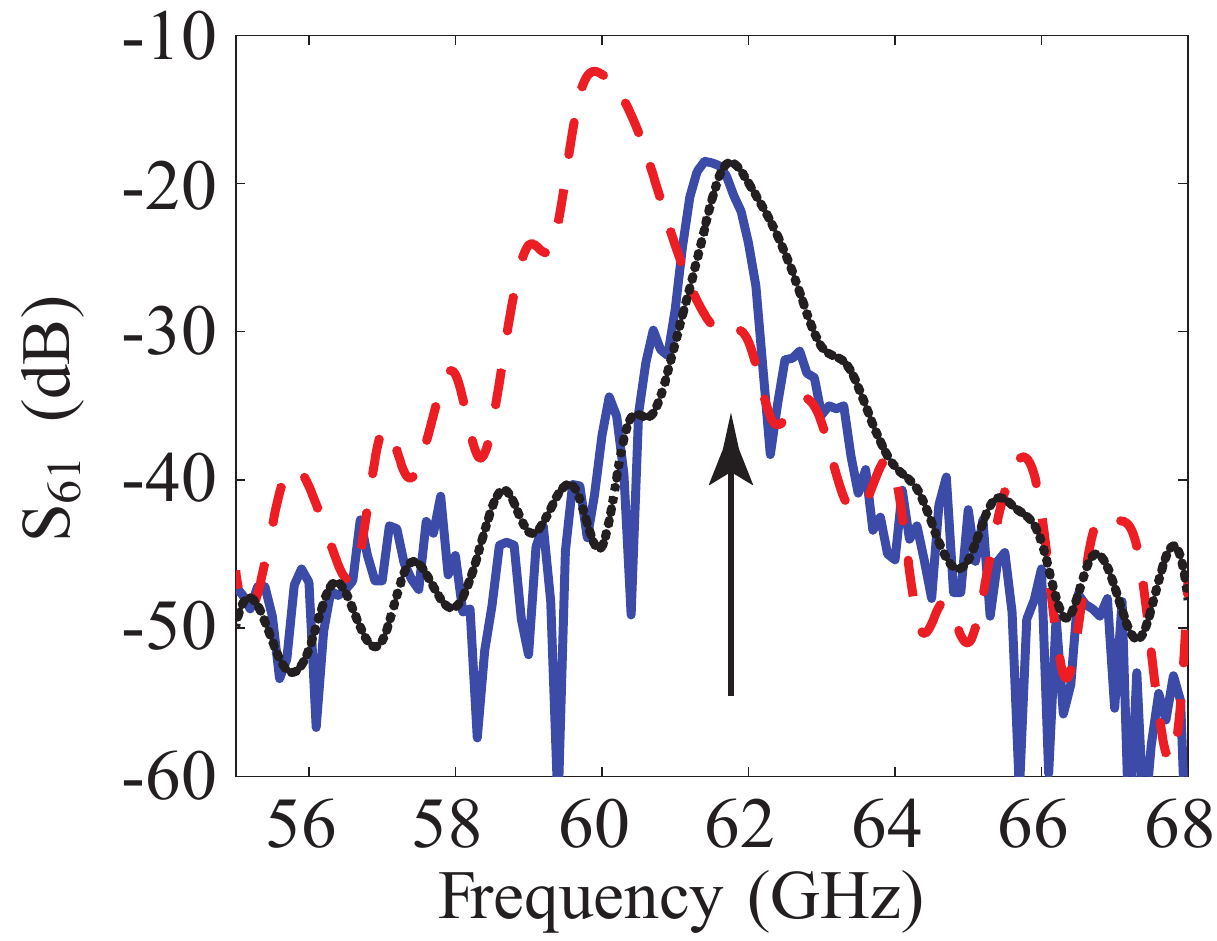}%
\caption{}%
\label{subfiga}%
\end{subfigure}\hfill%
\begin{subfigure}{.493\columnwidth}
\centering
\psfrag{a}[c][c][0.7]{Frequency (GHz)}
\psfrag{b}[c][c][0.7]{S$_{71}$ (dB)}
\includegraphics[width=\columnwidth]{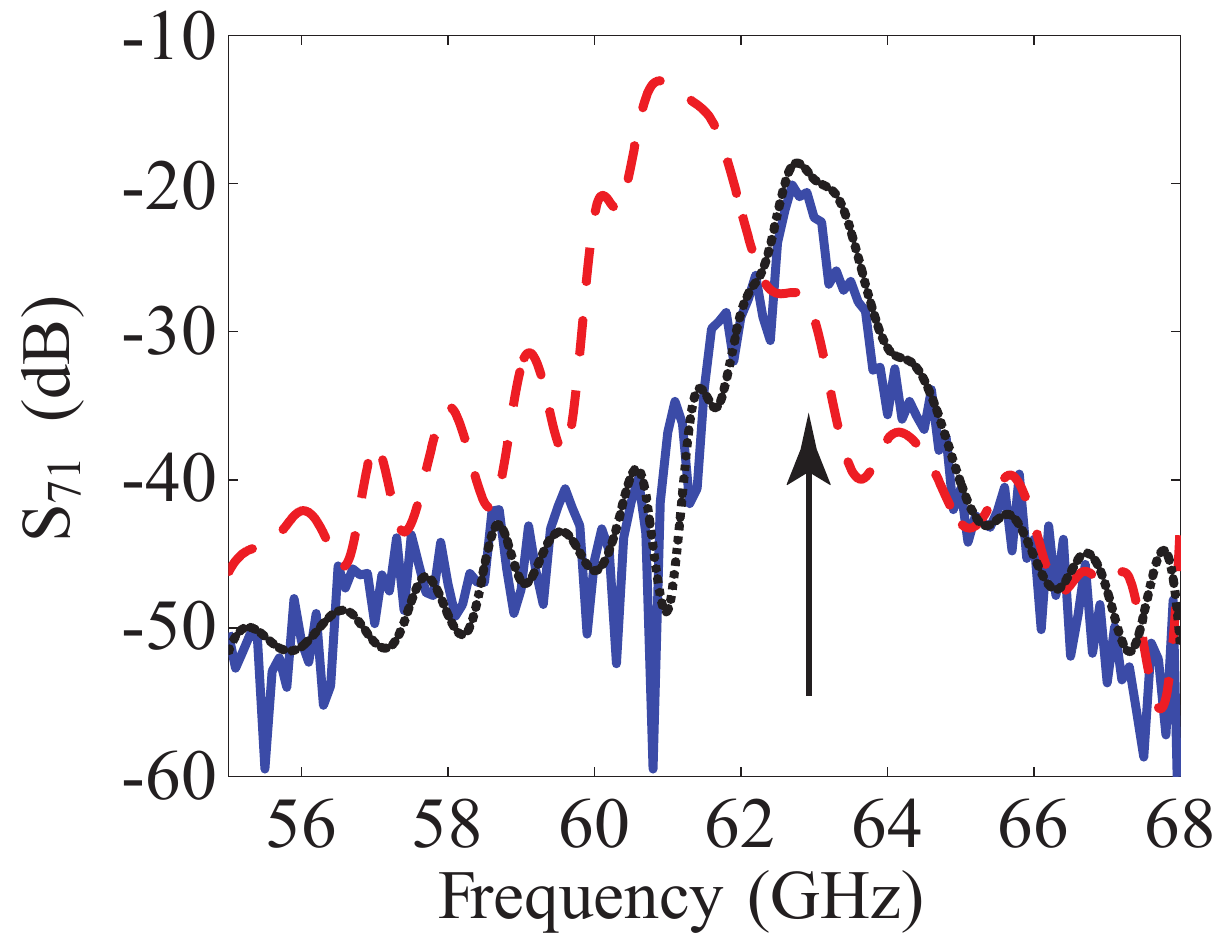}%
\caption{}%
\label{subfigb}%
\end{subfigure}\hfill%
\begin{subfigure}{.493\columnwidth}
\centering
\psfrag{a}[c][c][0.7]{Frequency (GHz)}
\psfrag{b}[c][c][0.7]{S$_{81}$ (dB)}
\includegraphics[width=\columnwidth]{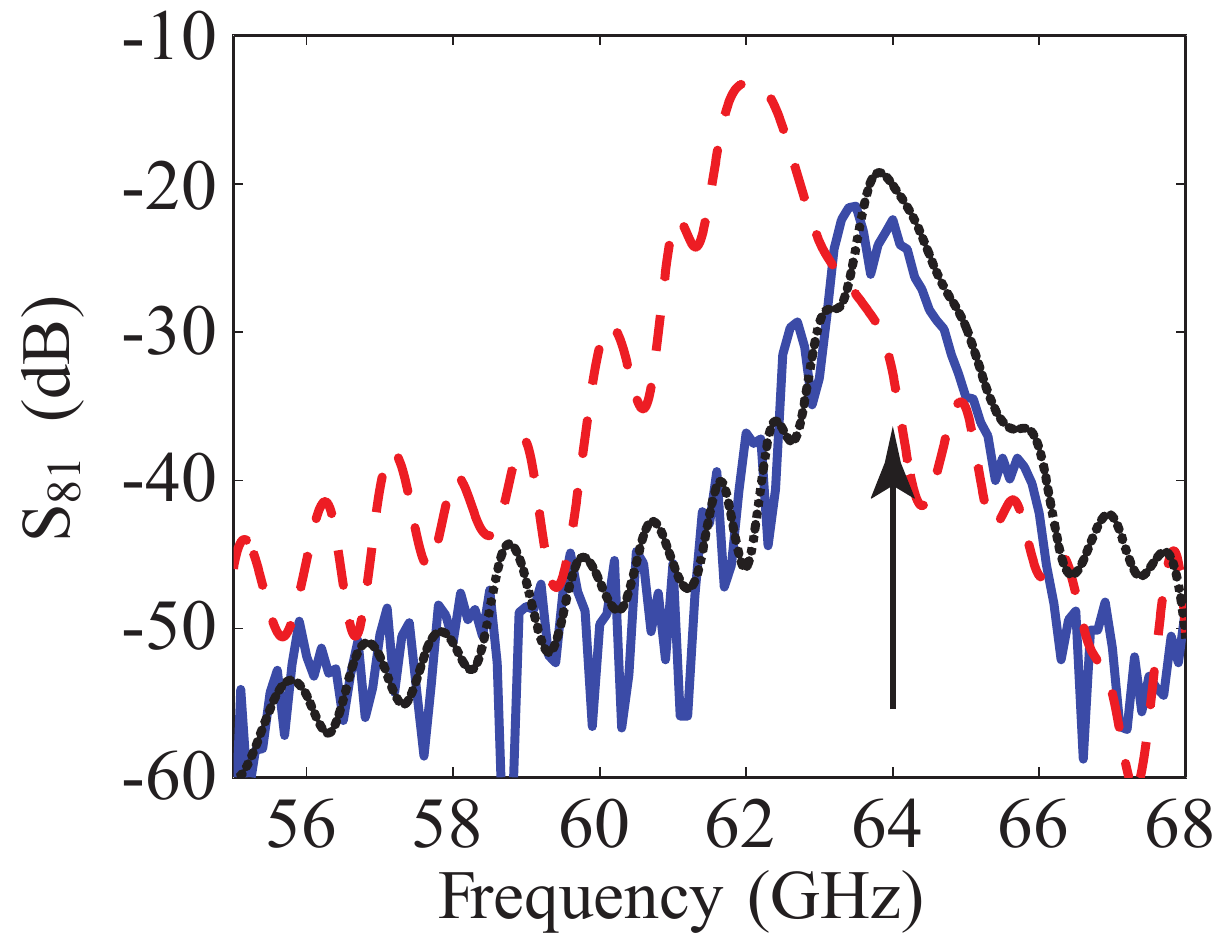}%
\caption{}%
\label{subfigb}%
\end{subfigure}\hfill%
\begin{subfigure}{.493\columnwidth}
\centering
\psfrag{a}[c][c][0.7]{Frequency (GHz)}
\psfrag{b}[c][c][0.7]{S$_{91}$ (dB)}
\includegraphics[width=\columnwidth]{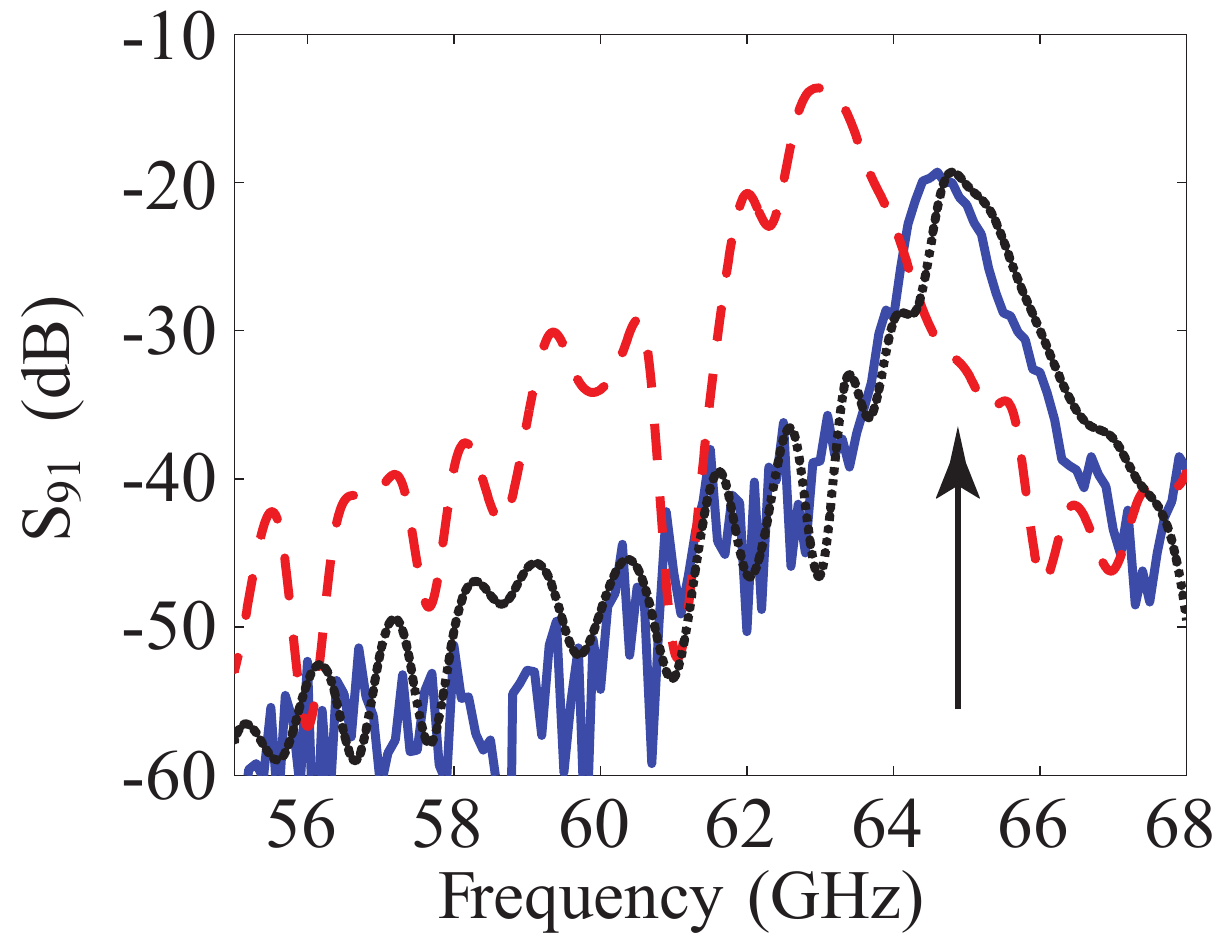}%
\caption{}%
\label{subfigb}%
\end{subfigure}\hfill%
\begin{subfigure}{.493\columnwidth}
\centering
\psfrag{a}[c][c][0.7]{Frequency (GHz)}
\psfrag{b}[c][c][0.7]{S$_{10,1}$ (dB)}
\includegraphics[width=\columnwidth]{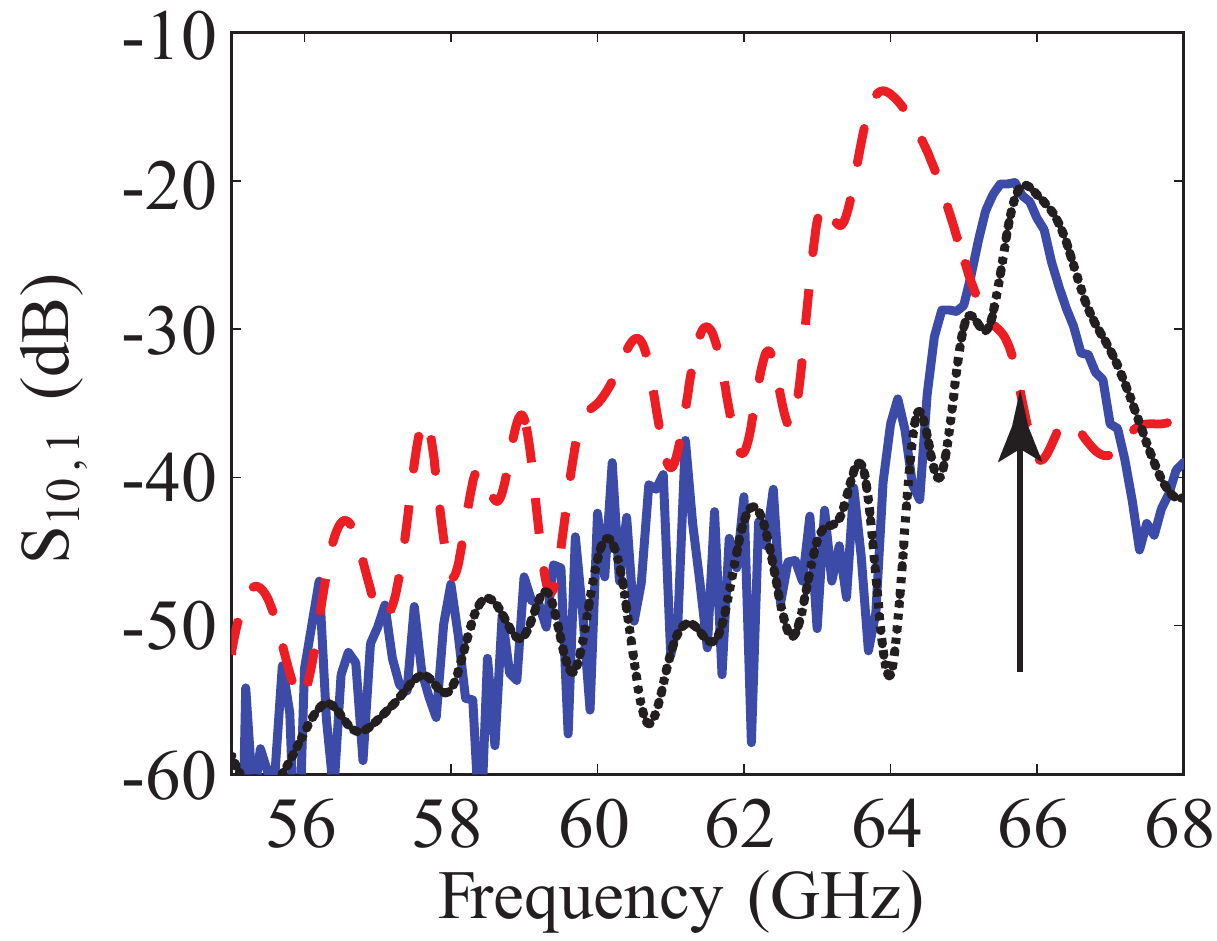}%
\caption{}%
\label{subfigb}%
\end{subfigure}\hfill%
\caption{Measured S-parameters of the integrated spectrum analyzer of Fig.~\ref{Fig:SAPrototype} demonstrating the spatial-spectral decomposition. Adjusted $\epsilon_r=2.08$ from nominal $\epsilon_r=2.2$ for Arlon DiClad 880. Adjusted $\tan\delta=0.003$ from nominal $\tan\delta=0.0009$ for Arlon DiClad 880 following the material characterization given in the Appendix. (a)-(f) Correspond to ports 3-10, respectively.}
\label{Fig:HornPorts}
\end{figure*}

The side-fire based integrated spectrum analyzer is next fabricated and some pictures of the prototype and the experimental setup is shown in Fig.~\ref{Fig:SAPrototype}(a-b). Fig.~\ref{Fig:SAPrototype}(a) shows the array of integrated horns on the receiver side connected to a standard WR-15 waveguide to a V-band coax cable. The integrated spectrum analyzer is next measured using a Vector Network Analyzer as shown in Fig.~\ref{Fig:SAPrototype}(b), where different measurements are taken across various horn ports. The S-parameters of the curved side-fire antenna is shown in Fig.~\ref{Fig:SAPrototype}(c,d) where \textcolor{black}{in spite} of an intended well-matched design in simulations, relatively larger reflections are observed, particularly around 62~GHz. This indicates a stop-band region and a high frequency shift in the response, as the broadside frequency was around 60 GHz. Transmission on the other hand stays around -15~dB. The spectrum analyzer was \textcolor{black}{then} re-simulated in HFSS using the newly characterized parameters of the dielectrics (provided in Appendix), and the resulting S-parameters are shown also in Fig.~\ref{Fig:SAPrototype}(c,d). The frequency shift is clearly captured as seen in $S_{11}$ and an increase in reflection is observed \textcolor{black}{in spite} of being practically acceptable. However an appreciable difference between measurements and simulations still remains. This suggests that small variations in tolerances and material properties may significantly accumulate to give rise to large errors, since this integrated spectrum analyzer is still electrically large \textcolor{black}{to} the order of about $120\lambda$.

Next, the transmission across various horn ports receivers from the input LWA port is measured and are shown in Fig.~\ref{Fig:HornPorts}. A clear shift in the transmission peak is observed across frequency as different horn receivers are measured, confirming the frequency scanning behaviour. This is attributed to the lower permittivity of the actual material. The measurements results are next compared with the characterized material response in FEM-HFSS and a very good agreement is observed where not only the frequency shift but the transmission levels are correctly reproduced. The spectrum analyzer thus operate between 59 GHz - 66 GHz (instead of 57 GHz - 64 GHz), with a resolution of approximately 1~GHz and an average transmission level of about -20 dB. The proposed side-fire LWA based integrated spectrum analyzer is thus successfully demonstrated.

\begin{table*}
\centering
\caption{Port and horn dimensions.}
\begin{tabularx}{\textwidth}{@{}l*{11}{C}c@{}}
\hline
& Dimension & Port 1 & Port 2 & Port 3 & Port 4 & Port 5 & Port 6 & Port 7 & Port 8 & Port 9 & Port 10 \\
\hline
& $a$ (mm)  & 3.7592      & 3.7592           & 3.7592    & 3.7592   & 3.7592     & 3.7592    & 3.7592     & 3.7592 & 3.7592    & 3.7592     \\
& $b$ (mm) & 1.8796        & 1.8796         & 1.8796     & 1.8796    & 1.8796     & 1.8796    & 1.8796     & 1.8796  & 1.8796    & 1.8796     \\ 
& $a_{ap}$ (mm)      & 3.3019       & 3.3019     & 3.2965  & 3.2965  & 3.2965   & 3.2965  & 3.2965   & 3.2700    & 3.2700    & 3.2700    \\ 
& $L_1$  (mm)     & 7.807       & 7.807          & 7.807   & 7.807 & 7.807   & 7.807 & 7.807   & 7.807    & 7.807    & 7.807    \\ 
& $L_2$ (mm)      & 5       & 5          & 5   & 25  & 5  & 25  & 5   & 25    & 5    & 25    \\ 
& $L_3$  (mm)      & 0        & 0          & 9.6   & 9 & 9.3   & 9 & 9.6    & 9.3    & 9.3    & 9    \\ 
& $L_4$ (mm)       & 4.2843        & 4.2843          & 4.2843   & 4.2843  & 4.2843    & 4.2843  & 4.2843  & 4.2843 & 4.2843    & 4.2843    \\ 
& $g$  (mm)     & 2.81       & 2.81          & 2.81   & 2.81  & 2.81   & 2.81  & 2.81    & 2.81   & 2.81    & 2.81 \\ 
& $\theta_h$  ($\degree$)      & 0       & 0         & 56    & 52   & 40    & 50   & 42    & 42     & 37  & 46    \\ 
\hline
\end{tabularx}
\end{table*}

\section{Conclusions \& Future Work}

An analog, low-profile and shielded spectrum analyzer has been proposed for operation at mm-wave frequencies around the 60 GHz band based on a novel side-fire LWA configuration. The proposed side-fire LWA has been systematically developed from a conventional 3-port waveguide T-junction which has been modified to a LWA unit cell with an internal matching mechanism to suppress stop-band to enable broadside radiation. The internal mechanism is achieved by introducing a transversal asymmetry of the unit cell, by placing a notch inside the waveguide, which results in a seamless leaky-wave radiation from backward to forward region including broadside. The proposed antenna is fully compatible with SIW technology resulting in a low-profile and is compatible with standard PCB fabrication. The resulting periodic side-fire antenna radiates in the plane of the antenna, whereby the leakage power be either be allowed to radiate in free-space or kept confined inside a PPW structure. The proposed side-fire structure thus can be completely shielded, which is particularly attractive for devising an analog broadband spectrum analyzer. Such a system has been demonstrated around the 60 GHz band, where a convex side-fire antenna has been used to focus the radiated beams in the near-field of the structure to make the entire system more compact, operating between \textcolor{black}{59 - 66} GHz in experiments with a 1 GHz resolution. Furthermore, a simple mathematical model consisting of an array of point sources has also been proposed which is found to faithfully reproduce the beam-scanning characteristics of the curved side-fire LWA in the near-field, providing a fast design tool to engineer such a system. 

The proposed side-fire antenna also provides a different interpretation of a LWA, where a periodic leaky-wave structure may be seen as a broadband multi-port power divider, whose outputs are connected to radiating apertures. The proposed spectrum analyzer furthermore offers several system features and benefits. The system is broadband in nature due to its wide-band radiating property of the stop-band suppressed side-fire LWA, directly operating at mm-wave frequencies based on SIW technology. It is a fast analog system minimizing digital computations compared to a fully digital system, except at the data acquisition stage. The frequency resolution of the system depends on the length of the side-fire antenna. This resolution can further be improved by optimizing the length of the antenna and controlling the per-unit-length leakage from the unit cell, \textcolor{black}{while} keeping the overall size of the analyzer the same. The low-profile and compact size of the spectrum analyzer represents a bench-top type system that can be placed on standard equipments with no electromagnetic interference due to its fully-shielded configuration. Thus the proposed integrated side-fire antenna and the associated spectrum analyzer system represent an attractive solution for next generation 5G mm-wave systems in the area of wireless communication and instrumentation, \textcolor{black}{for} instance.

\section{Appendix: Material Characterization}

\begin{figure}[htbp]
	\centering
	\psfrag{a}[c][c][0.8]{Frequency (GHz)}
	\psfrag{b}[c][c][0.8]{$S_{21}$ (dB)}
	\psfrag{c}[c][c][0.7]{Measured}
	\psfrag{d}[c][c][0.7]{FEM-HFSS (Fitted)}
	\psfrag{e}[c][c][0.7]{FEM-HFSS (Arlon)}
   	 \includegraphics[width=0.7\linewidth]{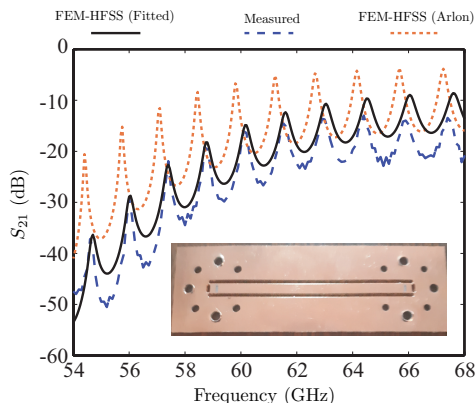}
 %
	\caption{Fabricated waveguide resonator for effective material characterization showing $S_{21}$. Arlon DiClad 880 ($\epsilon_r=2.2$ and $\tan\delta=0.0009$), and FEM-HFSS with fitted material ($\epsilon_r=2.08$ and $\tan\delta=0.003$). Resonator length and width are 50~mm and 2.5~mm, respectively.}\label{Fig:resonator}
\end{figure}

Given the sensitivity of the devices to electrical material parameters such as $\epsilon_r$ and $\tan\delta$ (particularly the beam location of the spectrum analyzer considering the large focal length) a material characterization of the panel used to fabricate the devices was performed. This was done by fabricating a resonator and measuring its frequency dependent transmission $S_{21}$, as shown in Fig.~\ref{Fig:resonator}. The location and magnitudes of the peaks can then be fitted by running additional simulations of the resonator with new values of $\epsilon_r$ and $\tan\delta$. These values which match measurement most closely provide a more accurate characterization of the panel material beyond the standard values of Arlon DiClad 880 provided in HFSS. However, given that the frequency dependence of the material parameters varies from what is predicted in software, over a large bandwidth the curves cannot be perfectly fit over the band, hence some mismatch will occur. The curves were fit best to match over the band of interest (57-64 GHz) and the extracted effective material parameters were found to be $\epsilon_r = 2.08$ and $\tan\delta = 0.003$.

\bibliographystyle{IEEEtran}
\bibliography{2020_SideFireSA_TAP_King}

\begin{thebibliography}{10}
\providecommand{\url}[1]{#1}
\csname url@samestyle\endcsname
\providecommand{\newblock}{\relax}
\providecommand{\bibinfo}[2]{#2}
\providecommand{\BIBentrySTDinterwordspacing}{\spaceskip=0pt\relax}
\providecommand{\BIBentryALTinterwordstretchfactor}{4}
\providecommand{\BIBentryALTinterwordspacing}{\spaceskip=\fontdimen2\font plus
\BIBentryALTinterwordstretchfactor\fontdimen3\font minus
  \fontdimen4\font\relax}
\providecommand{\BIBforeignlanguage}[2]{{%
\expandafter\ifx\csname l@#1\endcsname\relax
\typeout{** WARNING: IEEEtran.bst: No hyphenation pattern has been}%
\typeout{** loaded for the language `#1'. Using the pattern for}%
\typeout{** the default language instead.}%
\else
\language=\csname l@#1\endcsname
\fi
#2}}
\providecommand{\BIBdecl}{\relax}
\BIBdecl

\bibitem{farzaneh2009antenna}
S.~Farzaneh, A.~K. Ozturk, A.~R. Sebak, and R.~Paknys, ``Antenna-pattern
  measurement using spectrum analyzer for systems with frequency translation
  [measurements corner],'' \emph{IEEE Antennas Propag. Mag.}, vol.~51, no.~3,
  pp. 126--131, 2009.

\bibitem{sanchez2017millimeter}
M.~G. S{\'a}nchez, M.~P. T{\'a}boas, and E.~L. Cid, ``Millimeter wave radio
  channel characterization for 5{G} vehicle-to-vehicle communications,''
  \emph{Measurement}, vol.~95, pp. 223--229, 2017.

\bibitem{kawamura2014novel}
T.~Kawamura, H.~Shimotahira, and A.~Otani, ``Novel tunable filter for
  millimeter-wave spectrum analyzer over 100 {GHz},'' \emph{IEEE Trans.
  Instrum. Meas.}, vol.~63, no.~5, pp. 1320--1327, 2014.

\bibitem{bergland1969fast}
G.~Bergland, ``Fast {Fourier} transform hardware implementations--an
  overview,'' \emph{IEEE Trans. Audio Electroacoust.}, vol.~17, no.~2, pp.
  104--108, 1969.

\bibitem{rappaport2017overview}
T.~S. Rappaport, Y.~Xing, G.~R. MacCartney, A.~F. Molisch, E.~Mellios, and
  J.~Zhang, ``Overview of millimeter wave communications for fifth-generation
  (5{G}) wireless networks—with a focus on propagation models,'' \emph{{IEEE
  Trans. Antennas Propagat.}}, vol.~65, no.~12, pp. 6213--6230, 2017.

\bibitem{ghosh20165g}
A.~Ghosh, ``The 5{G} mmwave radio revolution,'' \emph{Microw. J.}, vol.~59,
  no.~9, pp. 22--36, 2016.

\bibitem{karjalainen2014challenges}
J.~Karjalainen, M.~Nekovee, H.~Benn, W.~Kim, J.~Park, and H.~Sungsoo,
  ``Challenges and opportunities of mm-wave communication in 5{G} networks,''
  in \emph{2014 9th international conference on cognitive radio oriented
  wireless networks and communications (CROWNCOM)}.\hskip 1em plus 0.5em minus
  0.4em\relax IEEE, 2014, pp. 372--376.

\bibitem{de2016optical}
H.~G. de~Chatellus, L.~R. Cort{\'e}s, and J.~Aza{\~n}a, ``Optical real-time
  {Fourier} transformation with kilohertz resolutions,'' \emph{Optica}, vol.~3,
  no.~1, pp. 1--8, 2016.

\bibitem{saleh2019fundamentals}
B.~E. Saleh and M.~C. Teich, \emph{Fundamentals of photonics}.\hskip 1em plus
  0.5em minus 0.4em\relax John Wiley \& Sons, 2019.

\bibitem{goodman2005introduction}
J.~W. Goodman, \emph{Introduction to {Fourier} optics}.\hskip 1em plus 0.5em
  minus 0.4em\relax Roberts and Company Publishers, 2005.

\bibitem{shirasaki1996large}
M.~Shirasaki, ``Large angular dispersion by a virtually imaged phased array and
  its application to a wavelength demultiplexer,'' \emph{Opt. Lett.}, vol.~21,
  no.~5, pp. 366--368, 1996.

\bibitem{laso2003real}
M.~A. Laso, T.~Lopetegi, M.~J. Erro, D.~Benito, M.~J. Garde, M.~A. Muriel,
  M.~Sorolla, and M.~Guglielmi, ``Real-time spectrum analysis in microstrip
  technology,'' \emph{{IEEE Trans. Microw. Theory Tech.}}, vol.~51, no.~3, pp.
  705--717, 2003.

\bibitem{wang2018real}
X.~Wang, A.~Akbarzadeh, L.~Zou, and C.~Caloz, ``Real-time spectrum sniffer for
  cognitive radio based on {Rotman} lens spectrum decomposer,'' \emph{IEEE
  Access}, vol.~6, pp. 52\,366--52\,373, 2018.

\bibitem{gupta2009microwave}
S.~Gupta, S.~Abielmona, and C.~Caloz, ``Microwave analog real-time spectrum
  analyzer ({RTSA}) based on the spectral--spatial decomposition property of
  leaky-wave structures,'' \emph{{IEEE Trans. Microw. Theory Tech.}}, vol.~57,
  no.~12, pp. 2989--2999, 2009.

\bibitem{gupta2008crlh}
S.~Gupta, C.~Caloz, and S.~Abielmona, ``{CRLH} leaky-wave real-time spectrum
  analyzer (rtsa) with unrestricted time-frequency resolution,'' in \emph{2008
  IEEE MTT-S International Microwave Symposium Digest}.\hskip 1em plus 0.5em
  minus 0.4em\relax IEEE, 2008, pp. 807--810.

\bibitem{gomez2010frequency}
J.~L. Gomez-Tornero, F.~Quesada-Pereira, A.~Alvarez-Melcon, G.~Goussetis, A.~R.
  Weily, and Y.~J. Guo, ``Frequency steerable two dimensional focusing using
  rectilinear leaky-wave lenses,'' \emph{{IEEE Trans. Antennas Propagat.}},
  vol.~59, no.~2, pp. 407--415, 2010.

\bibitem{garcia20111d}
M.~Garcia-Vigueras, J.~L. Gomez-Tornero, G.~Goussetis, A.~R. Weily, and Y.~J.
  Guo, ``{1-D}-leaky wave antenna employing parallel-plate waveguide loaded
  with {PRS} and {HIS},'' \emph{{IEEE Trans. Antennas Propagat.}}, vol.~59,
  no.~10, pp. 3687--3694, 2011.

\bibitem{martinez2011planar}
A.~J. Martinez-Ros, J.~L. Gomez-Tornero, and G.~Goussetis, ``Planar leaky-wave
  antenna with flexible control of the complex propagation constant,''
  \emph{{IEEE Trans. Antennas Propagat.}}, vol.~60, no.~3, pp. 1625--1630,
  2011.

\bibitem{king2019millimeter}
D.~J. King, M.~K. Emara, and S.~Gupta, ``Millimeter-wave near-field spectrum
  analyzer based on integrated side-fire antennas,'' in \emph{2019 13th
  European Conference on Antennas and Propagation (EuCAP)}.\hskip 1em plus
  0.5em minus 0.4em\relax IEEE, 2019, pp. 1--3.

\bibitem{jackson2012leaky}
D.~R. Jackson, C.~Caloz, and T.~Itoh, ``Leaky-wave antennas,'' \emph{Proc.
  IEEE}, vol. 100, no.~7, pp. 2194--2206, 2012.

\bibitem{yang2010full}
N.~Yang, C.~Caloz, and K.~Wu, ``Full-space scanning periodic phase-reversal
  leaky-wave antenna,'' \emph{{IEEE Trans. Microw. Theory Tech.}}, vol.~58,
  no.~10, pp. 2619--2632, 2010.

\bibitem{Sakakibara_slots}
K.~{Sakakibara}, J.~{Hirokawa}, M.~{Ando}, and N.~{Goto}, ``A
  linearly-polarized slotted waveguide array using reflection-cancelling slot
  pairs,'' \emph{IEICE Trans. Commun.}, vol. E77-B, no.~4, pp. 511--518, Apr.
  1994.

\bibitem{Caloz_CRLH}
C.~{Caloz}, T.~{Itoh}, and A.~{Rennings}, ``{CRLH} metamaterial leaky-wave and
  resonant antennas,'' \emph{{IEEE Antennas Propagat. Mag.}}, vol.~50, no.~5,
  pp. 25--39, 2008.

\bibitem{paulotto2008full}
S.~Paulotto, P.~Baccarelli, F.~Frezza, and D.~R. Jackson, ``Full-wave modal
  dispersion analysis and broadside optimization for a class of microstrip
  {CRLH} leaky-wave antennas,'' \emph{{IEEE Trans. Microw. Theory Tech.}},
  vol.~56, no.~12, pp. 2826--2837, 2008.

\bibitem{collin1969antenna}
R.~E. Collin and F.~J. Zucker, ``Antenna theory,'' 1969.

\bibitem{pozar2005microwave}
D.~M. Pozar, ``Microwave engineering 3e,'' \emph{Tramsmission Lines and
  Waveguides}, pp. 143--149, 2005.

\bibitem{arndt1987optimized}
F.~Arndt, I.~Ahrens, U.~Papziner, U.~Wiechmann, and R.~Wilkeit, ``Optimized
  {E}-plane {T}-junction series power dividers,'' \emph{{IEEE Trans. Microw.
  Theory Tech.}}, vol.~35, no.~11, pp. 1052--1059, 1987.

\bibitem{hirokawa1991analysis}
J.~Hirokawa, K.~Sakurai, M.~Ando, and N.~Goto, ``An analysis of a waveguide {T}
  junction with an inductive post,'' \emph{{IEEE Trans. Microw. Theory Tech.}},
  vol.~39, no.~3, pp. 563--566, 1991.

\bibitem{mallahzadeh2015periodic}
A.~Mallahzadeh and S.~Mohammad-Ali-Nezhad, ``Periodic collinear-slotted leaky
  wave antenna with open stopband elimination,'' \emph{{IEEE Trans. Antennas
  Propagat.}}, vol.~63, no.~12, pp. 5512--5521, 2015.

\bibitem{baccarelli2019open}
P.~Baccarelli, P.~Burghignoli, D.~Comite, W.~Fuscaldo, and A.~Galli,
  ``Open-stopband suppression via double asymmetric discontinuities in 1-{D}
  periodic 2-{D} leaky-wave structures,'' \emph{{IEEE Antennas Wirel. Propagat.
  Lett.}}, vol.~18, no.~10, pp. 2066--2070, 2019.

\bibitem{otto2013importance}
S.~Otto and C.~Caloz, ``Importance of transversal and longitudinal
  symmetry/asymmetry in the fundamental properties of periodic leaky-wave
  antennas,'' in \emph{2013 IEEE Antennas and Propagation Society International
  Symposium (APSURSI)}.\hskip 1em plus 0.5em minus 0.4em\relax IEEE, 2013, pp.
  240--241.

\bibitem{otto2014transversal}
S.~Otto, A.~Al-Bassam, A.~Rennings, K.~Solbach, and C.~Caloz, ``Transversal
  asymmetry in periodic leaky-wave antennas for {B}loch impedance and radiation
  efficiency equalization through broadside,'' \emph{{IEEE Trans. Antennas
  Propagat.}}, vol.~62, no.~10, pp. 5037--5054, 2014.

\bibitem{guglielmi1993broadside}
M.~Guglielmi and D.~Jackson, ``Broadside radiation from periodic leaky-wave
  antennas,'' \emph{{IEEE Trans. Antennas Propagat.}}, vol.~41, no.~1, pp.
  31--37, 1993.

\bibitem{josefsson2006conformal}
L.~Josefsson and P.~Persson, \emph{Conformal array antenna theory and
  design}.\hskip 1em plus 0.5em minus 0.4em\relax John Wiley \& Sons, 2006,
  vol.~29.

\bibitem{balanis2016antenna}
C.~A. Balanis, \emph{Antenna theory: analysis and design}.\hskip 1em plus 0.5em
  minus 0.4em\relax John Wiley \& Sons, 2016.

\bibitem{stutzman2012antenna}
W.~L. Stutzman and G.~A. Thiele, \emph{Antenna theory and design}.\hskip 1em
  plus 0.5em minus 0.4em\relax John Wiley \& Sons, 2012.

\end{thebibliography}

\end{document}